\documentclass[prx,twocolumn,superscriptaddress,showpacs,nobibnotes,notitlepage,nofootinbib]{revtex4-2}

\usepackage[colorlinks=true,linktoc=page,citecolor=red,linkcolor=blue]{hyperref}

\usepackage{bm} 
\usepackage{amsmath}
\usepackage{amssymb}
\usepackage{graphicx}
\usepackage{adjustbox}
\usepackage{color}
\usepackage{xcolor}
\usepackage{comment}
\definecolor{myRed}{HTML}{EE3377}   
\definecolor{myGreen}{HTML}{009E73} 
\definecolor{myBlue}{HTML}{0e35b3}

\usepackage{varwidth}
\usepackage{mathtools}
\usepackage{tikz}
\usetikzlibrary{tikzmark, arrows.meta}

\usepackage{soul}
\usepackage{enumerate}
\usepackage{mathrsfs}
\usepackage{tikz}
\usetikzlibrary{automata}
\usepackage{circuitikz}
\usetikzlibrary{patterns}
\usetikzlibrary{through,hobby}
\usetikzlibrary{decorations.markings}
\usetikzlibrary{fadings,shadings}
\usetikzlibrary{intersections}
\usetikzlibrary{plotmarks}
\usetikzlibrary{positioning}
\usetikzlibrary{decorations,arrows}
\usetikzlibrary{calc}
\usetikzlibrary{decorations.pathreplacing}
\usetikzlibrary{chains, scopes, positioning, backgrounds, shapes, fit, shadows, calc, arrows.meta}
\usepackage{braket}

\newcommand{\dec}{\text{dec}}
\usepackage{amsthm}

\usepackage{CJK}
\usepackage{booktabs}
\usepackage[utf8]{inputenc}
\setstcolor{myBlue}

\newcommand{\ee}{\mathsf{e}}
\newcommand{\mm}{\mathsf{m}}
\newcommand{\ff}{\mathsf{f}}
\newcommand{\s}{\mathsf{s}}

\newcommand{\DrawGateConnections}[1]{} 

\begin{document}

\title{Decohering Kitaev's Sixteenfold Way: A Holographic Approach}

\begin{abstract}
   We study anyon decoherence in 2D topological order, with the fermion decoherence in Kitaev's sixteenfold way as explicit examples. The sixteenfold way includes chiral topological phases, which are believed to admit no commuting projector Hamiltonian, making it challenging to study their decoherence directly. We further
   conjecture the density matrix produced by this decoherence induced phase transition can be obtained by a three-step procedure: first lift the topological order to its Walker-Wang model, followed by a transition to another Walker-Wang model, and finally trace over the bulk degrees of freedom. This establishes a connection between the decoherence of a topological order and a pure state phase transition in its 3D counterpart. Many examples are given. In particular, we show that decohering the fermion in each of the sixteen states always yields an intrinsically mixed-state topological order, a consequence of the fact the corresponding 3D bulk always transits into the 3D fermionic toric code (fTC). Our result suggests a systematic method for studying anyon decoherence in chiral topological orders at the level of commuting projector Hamiltonians.
\end{abstract}

\author{Shuyu Zhang}
\email{zhang.s.y.phys@gmail.com}
\affiliation{C. N. Yang Institute for Theoretical Physics and Department of Physics and Astronomy, Stony Brook University, Stony Brook, NY 11794, USA}

\author{Linhao Li}
\affiliation{Department of Physics, The Pennsylvania State University, University Park, Pennsylvania 16802, USA}

\author{Zhen Bi}
\affiliation{Department of Physics, The Pennsylvania State University, University Park, Pennsylvania 16802, USA}
\affiliation{Center for Theory of Emergent Quantum Matter, The Pennsylvania State University, University Park, Pennsylvania 16802, USA}
\author{Tzu-Chieh Wei}
\affiliation{C. N. Yang Institute for Theoretical Physics and Department of Physics and Astronomy, Stony Brook University, Stony Brook, NY 11794, USA}

\author{Zijian Song}
\email{zjsong.physics@gmail.com}
\affiliation{C. N. Yang Institute for Theoretical Physics and Department of Physics and Astronomy, Stony Brook University, Stony Brook, NY 11794, USA}

\maketitle

\tableofcontents

\section{Introduction}

Topological order and its phase transitions have been extensively studied over the past several decades, as they describe quantum phases of matter beyond the framework of local order parameters~\cite{wen1995topological,kitaev2003fault,kong2014braided,wen2016theory,wen2017colloquium,johnson2022classification}. In two dimensions, topological order is characterized by long-range entanglement and anyonic excitations with nontrivial fusion and braiding statistics~\cite{kitaev2006anyons,kitaev2006topological,levin2006detecting}. These properties connect the study of strongly correlated systems with quantum information, where nonlocal degrees of freedom can encode information protected against local perturbations~\cite{dennis2002topological,bonderson2012non}. Much of our understanding of topological order has been developed for pure ground states of gapped Hamiltonians. Extending this understanding to systems subject to environmental noise is essential for both the theory of quantum matter and its applications.

Interactions with an environment generally produce mixed states, raising questions not settled by the classification of pure ground states~\cite{ma2023average,lessa2025strong,ma2025topological,sang2025stability,sang2025mixed,ma2025circuit}. Which topological properties survive the loss of quantum coherence? How does the information stored in a topological state change under decoherence, and when does this change constitute a phase transition? Recent studies have shown that suitable noise channels can preserve nontrivial topological structure even after substantial quantum information has been lost~\cite{zhang2025strong,song2026strong,yang2025topological,wang2025fractional,bao2026mixed}. The resulting states can retain quantum or classical memory, and can exhibit intrinsic mixed-state topological order, for which no pure ground-state representative exists under the relevant notion of phase equivalence~\cite{ellison2025toward,sohal2025noisy,wang2025intrinsic}.

Characterizing these mixed-state phases requires careful treatment of both their correlations and the operations used to compare them. Different approaches emphasize local preparation and recovery, entanglement properties, or emergent generalized symmetries~\cite{coser2019classification,fan2024diagnostics,ellison2025toward,sang2025stability,sang2025mixed,weinstein2025efficient,zhang2025strong,PRXQuantum.6.010347,song2026strong,li2026generalized}. These perspectives reveal distinctions between the quantum and classical information retained by noisy states, while also raising questions about how the corresponding notions of order are related. Understanding when these approaches yield the same phase distinctions and when they capture different information remains an important open problem in the study of mixed-state quantum matter.

Commuting projector lattice models provide valuable settings for addressing these questions microscopically, in which density matrices, correlations, and the effects of noise can be examined directly~\cite{kitaev2003fault,lw05}. Chiral topological phases, however, present a particular challenge, since a nonzero chiral central charge obstructs a strictly two-dimensional commuting projector realization~\cite{kitaev2006anyons,kapustin2020local,kapustin2020thermal}, as well as the incompatibility of tensor-network description of gapped chiral topological order~\cite{wah2013projected,dubail2015tensor}. Systematic lattice approaches that accommodate chiral as well as nonchiral phases would therefore help connect general descriptions of mixed-state order with concrete physical realizations.

\subsection{Summary of results}

We propose a holographic lattice construction for anyonic decoherence and demonstrate its application to fermion decoherence throughout Kitaev's sixteenfold way. Related approaches in various dimensions have also been explored in the literature~\cite{ellison2025toward,sohal2025noisy,wang2025intrinsic,sun2025holographic,luo2025topological,qi2025symmetry,schafernameki2025symtft,mana2026mixed,lu2026holographic,mana2026holographic}. 

Our construction begins with an anyon theory described by a unitary modular tensor category (UMTC) $\mathcal{C}$~\cite{bakalov2001lectures,etingof2005fusion,drinfeld2010braided,davydov2011witt,etingof2015tensor} and its corresponding Walker-Wang model defined on a 3D slab with smooth boundaries~\cite{walker20123+,vonkeyserlingk2013three,burnell2013phase,burnell2014exactly}. Given a set of anyons $S$ closed under fusion, let $\mathcal{D}$ consist of the anyons in $\mathcal{C}$ that braid trivially with every anyon in $S$. We project the bulk ground state onto string-nets labeled only by $\mathcal{D}$, obtaining the Walker-Wang model with input $\mathcal{D}$. A spatial partial trace then produces a boundary density matrix, with a buffer retained which is necessary to support the boundary symmetry operators. We show that this density matrix realizes the mixed-state phase obtained by maximal decoherence of $S$ in $\mathcal{C}$.

We establish this picture explicitly for two decoherence channels of the two-dimensional toric code, corresponding to maximal bosonic and fermionic decoherence. The three-dimensional counterpart of the toric code is the Raussendorf-Bravyi-Harrington (RBH) cluster state~\cite{raussendorf2005long,raussendorf2006fault}. We define projections of the RBH state that correspond to maximal anyonic decoherence in 2D. The projection associated with bosonic decoherence produces the 3D toric code~\cite{tantivasadakarn2024long}, whose reduced boundary state is the 1-form strong-to-weak spontaneous symmetry breaking (SWSSB) state. The projection associated with fermionic decoherence instead produces the 3D fermionic toric code, whose reduced boundary state realizes intrinsic mixed-state topological order (imTO). These examples directly connect the decohered density matrices to bulk confinement-deconfinement transitions~\cite{burnell2013phase} and motivate the general construction.

An advantage of our construction is that it applies to decoherence in both nonchiral models, such as the toric code and other Kitaev quantum double models, and more significantly chiral topological orders. Our main family of examples is Kitaev's sixteenfold way, which consists of sixteen minimal modular extensions of the anyon theory $\{1,\ff\}$, where $\ff$ is a fermion~\cite{kitaev2006anyons}. We show that restricting each corresponding Walker-Wang model to the labels $\{1,\ff\}$ yields the 3D fermionic toric code. The fermion, initially confined in the bulk, becomes deconfined after such projection. Thus, all sixteen inputs, including both abelian and nonabelian chiral theories, share a common bulk endpoint. Applying the partial-trace construction then recovers the same intrinsically mixed-state topological order. This gives a unified lattice description of the common mixed-state outcome through explicit transitions between three-dimensional commuting projector models.

We identify these transitions using Hamiltonians, topological actions, and ground state wavefunctions, as appropriate to each example. For the abelian theories, the action formulation translates the projections into constraints on gauge fields and directly identifies the resulting bulk phase. By translating topological actions into entangling circuits, this method also provides a systematic construction of abelian Walker-Wang Hamiltonians with bosonic qudit degrees of freedom~\cite{chen2023higher,sun2026non}. For the Ising-type theories, projecting the string-net wavefunction leaves fermion loops with the exchange statistics of the 3D fermionic toric code. Beyond the sixteenfold way, semion decoherence in the double semion theory corresponds to a projection onto the anti-semion Walker-Wang model, illustrating how the construction can retain a chiral boundary sector. Together, these examples provide concrete lattice realizations of the proposed correspondence.

\subsection{Outline}

The remainder of the paper is organized as follows. In Sec.~\ref{sec:wu}, we review two decoherence channels on the toric code. One channel decoheres a boson while the other decoheres the fermion. We then present a holographic perspective in 3D by identifying phase transitions from the Walker-Wang model with toric code input. The bosonic case corresponds to a transition to the 3D toric code, while the fermionic case corresponds to a transition to the 3D fermionic toric code. In Sec.~\ref{sec:gc}, we give a categorical interpretation of the above two phase transitions. This motivates us to conjecture a procedure for studying general decoherence phase transitions in topological orders using the Walker-Wang construction. In Sec.~\ref{sec:examples}, we work out many examples in detail to validate the procedure. The main family of examples is Kitaev's sixteenfold way, in which we decohere the fermion. In addition, we also discuss decoherence in the double semion topological order. In Sec.~\ref{sec:conclusion}, we make concluding remarks and outline further directions. For completeness, in Appendix~\ref{app:ww}, we review the Walker-Wang model. In Appendix~\ref{app:ta}, we review the topological actions for abelian Walker-Wang models and provide a detailed derivation of the semion Walker-Wang model discussed in Sec.~\ref{sec:examples_sf}.

\section{Warm up: decohering the 2D toric code}\label{sec:wu}

Before discussing the general theory, in this section we first review two types of decoherence channels for the 2D toric code and their corresponding fixed-point mixed states, and then discuss their holographic correspondence in one dimension higher.

We first consider the decoherence channel given by Pauli $X$ as the Kraus operator, which incoherently proliferates the $\ee$ anyons (see Eq.~(\ref{eq:TC}) for the convention used in our toric-code Hamiltonian). Maximally decohering with this channel results in a 1-form strong-to-weak spontaneous symmetry breaking (SWSSB) phase with a strong bosonic 1-form symmetry and a weak bosonic 1-form symmetry. Interestingly, we demonstrate that this pure-state to mixed-state phase transition corresponds to a confinement-deconfinement transition in three dimensions, specifically from the 3D RBH cluster state to the 3D toric code. Tracing out the bulk of the resulting 3D bosonic toric code then yields the corresponding 2D SWSSB state. 

In the second part of this section, we consider the decoherence channel given by a combination of Pauli $X$ and $Z$ operators on nearest-neighboring sites, which incoherently proliferates the fermion $\ff$. Maximal decoherence under this channel results in the exotic intrinsic mixed-state topological order (imTO), and we demonstrate that this phase transition likewise corresponds to a confinement-deconfinement transition in one dimension higher, in this case from the 3D RBH cluster state to the 3D fermionic toric code. Tracing out the bulk of the resulting 3D fermionic toric code then yields the corresponding 2D imTO.

\subsection{Bosonic Decoherence}

We first consider the case where $\ee$ anyon is incoherently proliferated. The Hamiltonian of the 2D toric code is
\begin{equation}
    \label{eq:TC}\begin{aligned}
        H = - \sum_{v} B_v - \sum_{p} A_p,
    \end{aligned}
\end{equation}
where 
\begin{equation}
    \begin{aligned}
  B_v =    \begin{tikzpicture}
    [scale= 1, baseline={([yshift=-2.5pt]current bounding box.center)}]
        \draw[step = 2] (-1.2,-1.2) grid (1.2,1.2);
        \node[inner sep=0.5pt, fill=white] at (-0.7, 0) {\footnotesize $Z$};
        \node[inner sep=0.5pt, fill=white] at (0.7,0) {\footnotesize $Z$};
        \node[inner sep=0.5pt, fill=white] at (0, 0.7) {\footnotesize $Z$};
        \node[inner sep=0.5pt, fill=white] at (0,-0.7) {\footnotesize $Z$}; 
        \node[inner sep=0.5pt, fill=white] at (0.2,-0.2) {\footnotesize $v$};
    \end{tikzpicture}\ ,\quad
    A_p = \begin{tikzpicture}
    [scale= 0.8, baseline={([yshift=-2.5pt]current bounding box.center)}]
    \draw[step = 2] (0,0) grid (2,2);
        \node[inner sep=0.5pt, fill=white] at (0, 1) {\footnotesize $X$};
        \node[inner sep=0.5pt, fill=white] at (2,1) {\footnotesize $X$};
        \node[inner sep=0.5pt, fill=white] at (1, 0) {\footnotesize $X$};
        \node[inner sep=0.5pt, fill=white] at (1,2) {\footnotesize $X$};
        \node[inner sep=0.5pt, fill=white] at (1,1) {\footnotesize $p$}; 
    \end{tikzpicture}.
\end{aligned}
\end{equation}
Consider the toric code on an infinite plane, where the ground state is unique and its wavefunction is a coherent superposition of loops, up to normalization,
\begin{align}
    \ket{\psi} \sim \sum_{\gamma^*} \ket{\gamma^*},
\end{align}
where $\gamma^*$ is a loop on the dual lattice, and $\ket{\gamma^*}$ is the state  with $\ket{-}$ on all edges intersecting $\gamma^*$, and $\ket{+}$ otherwise. ($\ket{\pm}$ are eigenstates of the Pauli $X$ matrix: $X|\pm\rangle=\pm |\pm\rangle$.) The density matrix is given by a product of projectors
\begin{equation}
    \begin{aligned}
        \rho_{TC} = \ket{\psi
        }\bra{\psi} 
        &= \prod_v \frac{1+B_v}{2} \prod_p \frac{1 + A_p}{2}\\
        & \sim \sum_{\gamma^*, \gamma'^*} \ket{\gamma^*}\bra{\gamma'^*}. \label{eq:TC_density_matrix}
    \end{aligned} 
\end{equation}

\subsubsection{Decohered Density Matrix}

Consider a local decoherence channel acting on edge $e$
\begin{align}
    \mathcal{N}_e^p(\rho) = (1-p)\rho + p X_e \rho X_e,
\end{align}
where $X_e$ is the Pauli $X$ operator acting on the edge $e$. Physically, this local channel incoherently creates a pair of $\ee$ anyons at the two vertices bounding the edge $e$ with probability $p$. The global channel is given by the tensor product of local channels,
\begin{align}
    \mathcal{N}^p(\rho) := \bigotimes_e \mathcal{N}_e^p(\rho),
\end{align}
where $e$ runs over every edge. At the maximal decoherence limit $p = \frac{1}{2}$, the resulting mixed-state density matrix is a fixed point of the channel, satisfying
\begin{equation}
    \begin{aligned}
        \mathcal{N}^{1/2}\left(\mathcal{N}^{1/2}(\rho) \right) = \mathcal{N}^{1/2}(\rho).
    \end{aligned}
\end{equation}
To simplify notation, in the remainder of the paper we use $\mathcal{N}(\cdot)$ to denote $\mathcal{N}^p(\cdot)$ at its maximal decoherence limit.

Applying this maximal dephasing channel to the toric code density matrix in Eq.~\eqref{eq:TC_density_matrix} removes all non-diagonal entries of the density matrix in the loop basis, so that the decohered density matrix becomes a ``classical'' linear combination of loops,
\begin{align}
    \rho_{TC}^{\dec} = \mathcal{N}(\rho_{TC}) = \prod_p \frac{1 + A_p}{2} \sim \sum_{\gamma^*}\ket{\gamma^*}\bra{\gamma^*}. \label{eq:TC_charge_decoherence}
\end{align}
This mixed state is referred to as being in the 1-form SWSSB phase in the literature~\cite{zhang2025strong}. It has a strong bosonic 1-form symmetry given by $\ee$-loop operators, and a weak bosonic 1-form symmetry given by $\mm$-loop operators. Specifically, given loops $\gamma$ and $\gamma^*$ on the direct lattice and the dual lattice, we have,
\begin{equation}
    \begin{aligned}
        \bigg(\prod_{e \in \gamma} X_e\bigg) \rho_\ee = \rho_\ee, \\
        \bigg(\prod_{e \in \gamma^*} Z_e \bigg)\rho_\ee \bigg(\prod_{e \in \gamma^*} Z_e\bigg) = \rho_\ee.
    \end{aligned}
\end{equation}
Moreover, as discussed in Refs.~\cite{ma2025topological,sang2025mixed,song2026strong}, the endpoint of a truncated strong (weak) $1$-form symmetry operator has nontrivial braiding with a closed weak (strong) $1$-form symmetry operator encircling it, demonstrating the mixed 't Hooft anomaly between the strong and weak $1$-form symmetries.

On a nontrivial manifold or with a nontrivial boundary condition, the set of density matrices $\{\rho_{TC}^{\dec}\}$ forms an information convex set, and each extremal point is an eigenstate of the noncontractible strong 1-form symmetry, labeled by the charge of that strong symmetry~\cite{shi2020fusion,sang2025mixed,yang2025topological,song2026strong}. Consequently, this decohered density matrix encodes classical bits instead of quantum bits, and is therefore referred to as being in the classical memory phase.

In the past few years, extensive studies have sought to identify the phase transition between the quantum memory phase and the classical memory phase, as well as between the classical memory phase and the trivial mixed-state phase, under decoherence at either zero or finite temperature~\cite{sang2025stability,weinstein2025efficient,zhou2025finite,ma2025circuit}. Several candidate diagnostics have been proposed, among which two criteria are most widely recognized: two-way connectivity~\cite{coser2019classification} and the quantum Markov length~\cite{sang2025stability}.

Physically, two-way connectivity asks whether two mixed states can be converted into each other by quasi-local quantum channels. Since quantum channels need not be invertible, this requires a quasi-local channel in each direction, and two states belong to the same phase only if such channels exist in both directions. The quantum Markov length instead measures how locally reconstructible a state is: it is the decay length $\xi$ of the conditional mutual information $I(A{:}C|B)$ with respect to the width of a buffer region $B$ separating $A$ from $C$. The decay length $\xi$ plays the role analogous to the correlation length for a gapped pure ground state, and its divergence marks a mixed-state phase transition.

The bosonic decoherence of toric code illustrates that these two criteria are {\it inequivalent}. After $\ee$ has been incoherently proliferated by maximal dephasing, the state becomes a classical ensemble of closed $\ee$-strings, and the 1-form symmetry generated by the $\mm$-strings is broken explicitly from strong to weak. Under the criteria of two-way connectivity, this state is nevertheless trivial: it can be prepared from a product state by a finite-depth channel that incoherently proliferates the charge, and returned to a product state by a local replacement channel that discards each site and prepares a fixed product state, so quasi-local channels exist in both directions~\cite{sang2025mixed}. The key point is that these channels are not locally reversible and need not preserve the strong $1$-form symmetry or the associated topological memory, which the two-way connectivity criterion does not forbid.

The Markov length criterion is more restrictive, since it requires not only that the two states be connected, but that local reconstructibility be maintained along the entire path. The classical winding parity carried by the loop ensemble is a nonlocal label, and once a channel erases the information distinguishing different winding sectors, no state-independent local recovery channel can reconstruct the original sector. Consequently, any Lindbladian interpolation to a trivial state must encounter a point of diverging quantum Markov length~\cite{sang2025stability}.

\subsubsection{Bulk Phase Transition: a Holographic Perspective}

It has been suggested in the literature that decoherence, or more specifically, the incoherent proliferation of anyons, corresponds to anyon condensation in the doubled Hilbert space~\cite{sohal2025noisy,ellison2025toward,wang2025intrinsic}. It has further been pointed out that such a decohered density matrix can be obtained by decohering a corresponding Walker-Wang model~\cite{ellison2025toward}. In this section, we elaborate on this idea by constructing a three-dimensional phase transition through which we recover the density matrix $\rho_{TC}^{\dec}$.

The 2D toric code can naturally be regarded as a boundary of the 3D RBH state. Consider a 3D cubic lattice where two qubits are assigned to each edge. The RBH Hamiltonian is
\begin{equation}
    H_{\text{RBH}} = - \sum_{p} K_{p,1} - \sum_{p} K_{p,2}.
\end{equation}
Depending on the orientations of the plaquettes, the stabilizers are defined as
\begin{align*}
\hspace{-0.3cm}
K_{p,1}
=
    \begin{tikzpicture}[scale= 0.8, baseline={([yshift=-2.5pt]current bounding box.center)}]
            \coordinate (A) at (0,0,0);
            \coordinate (B) at (0,0,2);
            \coordinate (C) at (0,2,2);
            \coordinate (D) at (0,2,0);
%
            \coordinate (U) at (-2,0,0);
            \coordinate (O) at (2,2,2);
%
%
            %
            \draw (A) -- (B) --(C) -- (D) -- cycle;
            \draw (C) -- (O);
            \foreach \X/\Y in { A/B, B/C, C/D, D/A}
            \node[inner sep=0.5pt, fill=white] at ($( \X )!0.5!( \Y )$) {\footnotesize $X_1$}; 
            \node[inner sep=0.5pt, fill=white] at (1.3, 2, 2) {\footnotesize $Z_2$};
    \end{tikzpicture}
    ,\
    \begin{tikzpicture}[scale= 0.8, baseline={([yshift=-2.5pt]current bounding box.center)}]
%
            \coordinate (A) at (0,0,0);
            \coordinate (B) at (0,2,0);
            \coordinate (C) at (2,2,0);
            \coordinate (D) at (2,0,0);
%
            \coordinate (U) at (0,0,-2);
            \coordinate (O) at (2,2,2);
%
%
  %
            \draw (A) -- (B) --(C) -- (D) -- cycle;
            \draw (C) -- (O);
            \foreach \X/\Y in { A/B, B/C, C/D, D/A}
            \node[inner sep=0.5pt, fill=white] at ($( \X )!0.5!( \Y )$) {\footnotesize $X_1$}; 
            \node[inner sep=0.5pt, fill=white] at (2, 2, 1.1) {\footnotesize $Z_2$};
    \end{tikzpicture}
    ,\
    \begin{tikzpicture}[scale= 0.8, baseline={([yshift=-2.5pt]current bounding box.center)}]
        \coordinate (A) at (0,0,0);
        \coordinate (B) at (2,0,0);
        \coordinate (C) at (2,0,2);
        \coordinate (D) at (0,0,2);
%
        \coordinate (O) at (2,2,2);
        \coordinate (U) at (0,-2,0);
%
%
        \draw  (A) -- (B) -- (C) -- (D) -- cycle;
        \draw (C) -- (O);
%
 %
        \foreach \X/\Y in { A/B, B/C, C/D, D/A}
        \node[inner sep=0.5pt, fill=white] at ($( \X )!0.5!( \Y )$) {\footnotesize $X_1$};
        \node[inner sep=0.5pt, fill=white] at (2,1.3,2) {\footnotesize $Z_2$};        
    \end{tikzpicture},
\end{align*}
\begin{align*}
K_{p,2}
=
    \begin{tikzpicture}[scale= 0.8, baseline={([yshift=-2.5pt]current bounding box.center)}]
            \coordinate (A) at (0,0,0);
            \coordinate (B) at (0,0,2);
            \coordinate (C) at (0,2,2);
            \coordinate (D) at (0,2,0);
%
            \coordinate (U) at (-2,0,0);
            \coordinate (O) at (2,2,2);
%
%
            %
            \draw (A) -- (B) --(C) -- (D) -- cycle;
            \draw [dotted] (A) -- (U);
            \foreach \X/\Y in { A/B, B/C, C/D, D/A}
            \node[inner sep=0.5pt, fill=white] at ($( \X )!0.5!( \Y )$) {\footnotesize $X_2$}; 
            \node[inner sep=0.5pt, fill=white] at (-1.3, 0, 0) {\footnotesize $Z_1$};
    \end{tikzpicture}
    ,\
    \begin{tikzpicture}[scale= 0.8, baseline={([yshift=-2.5pt]current bounding box.center)}]
%
            \coordinate (A) at (0,0,0);
            \coordinate (B) at (0,2,0);
            \coordinate (C) at (2,2,0);
            \coordinate (D) at (2,0,0);
%
            \coordinate (U) at (0,0,-2);
            \coordinate (O) at (2,2,2);
%
%
  %
            \draw (A) -- (B) --(C) -- (D) -- cycle;
            \draw [dotted] (A) -- (U);
            \foreach \X/\Y in { A/B, B/C, C/D, D/A}
            \node[inner sep=0.5pt, fill=white] at ($( \X )!0.5!( \Y )$) {\footnotesize $X_2$}; 
            \node[inner sep=0.5pt, fill=white] at (0, 0, -1.1) {\footnotesize $Z_1$};
    \end{tikzpicture}
    ,\
    \begin{tikzpicture}[scale= 0.8, baseline={([yshift=-2.5pt]current bounding box.center)}]
        \coordinate (A) at (0,0,0);
        \coordinate (B) at (2,0,0);
        \coordinate (C) at (2,0,2);
        \coordinate (D) at (0,0,2);
%
        \coordinate (O) at (2,2,2);
        \coordinate (U) at (0,-2,0);
%
%
        \draw  (A) -- (B) -- (C) -- (D) -- cycle;
        \draw [dotted] (A) -- (U);
%
 %
        \foreach \X/\Y in { A/B, B/C, C/D, D/A}
        \node[inner sep=0.5pt, fill=white] at ($( \X )!0.5!( \Y )$) {\footnotesize $X_2$};
        \node[inner sep=0.5pt, fill=white] at (0,-1.3,0) {\footnotesize $Z_1$};        
    \end{tikzpicture}
\end{align*}
where dotted lines represent lines that are hidden behind the front faces after projecting the cubic lattice onto 2D. On the bottom and top boundaries, we further define the truncated vertex stabilizers, given by 
\begin{align*}
    \begin{tikzpicture}[scale= 0.8, baseline={([yshift=-2.5pt]current bounding box.center)}]
    %
        \coordinate (O) at (0,0,0);
        \coordinate (A) at (-1.5,0,0);
        \coordinate (B) at (1.5,0,0);
        \coordinate (C) at (0,-1.5,0);
        \coordinate (D) at (0,1.5,0);
        \coordinate (E) at (0,0,1.5);
        \coordinate (F) at (0,0,-1.5);
        %
        \draw (A) -- (B);
        \draw (O) -- (D); 
        \draw (E) -- (F);
        \foreach \X/\Y in { O/E, O/F}
        \node[inner sep=0.5pt, fill=white] at ($( \X )!0.5!( \Y )$) {\footnotesize $Z_i$};
        \foreach \X/\Y in {O/D}
        \node[inner sep=0.5pt, fill=white] at ($( \X )!0.5!( \Y )$) {\footnotesize $Z_i$};
        \foreach \X/\Y in {O/A, O/B}
        \node[inner sep=0.5pt, fill=white] at ($( \X )!0.5!( \Y )$) {\footnotesize $Z_i$}; 
    \end{tikzpicture}, \quad  
    \begin{tikzpicture}[scale= 0.8, baseline={([yshift=-2.5pt]current bounding box.center)}]
    %
        \coordinate (O) at (0,0,0);
        \coordinate (A) at (-1.5,0,0);
        \coordinate (B) at (1.5,0,0);
        \coordinate (C) at (0,-1.5,0);
        \coordinate (D) at (0,1.5,0);
        \coordinate (E) at (0,0,1.5);
        \coordinate (F) at (0,0,-1.5);
        %
        \draw (A) -- (B);
        \draw (C) -- (O); 
        \draw (E) -- (F);
        \foreach \X/\Y in { O/E, O/F}
        \node[inner sep=0.5pt, fill=white] at ($( \X )!0.5!( \Y )$) {\footnotesize $Z_i$};
        \foreach \X/\Y in {O/C}
        \node[inner sep=0.5pt, fill=white] at ($( \X )!0.5!( \Y )$) {\footnotesize $Z_i$};
        \foreach \X/\Y in {O/A, O/B}
        \node[inner sep=0.5pt, fill=white] at ($( \X )!0.5!( \Y )$) {\footnotesize $Z_i$}; 
    \end{tikzpicture},\quad 
    i = 1,2,
\end{align*}
which can be written as products of truncated plaquette stabilizers (i.e., truncated $K_p$ operators). By projecting the second qubit on each edge onto the $Z_2 = +1$ subspace, one can drive a phase transition from the RBH cluster state to the 3D toric code, as studied in Ref.~\cite{tantivasadakarn2024long}. Equivalently, one can add the $-\Lambda \sum_{e}Z_{e,2}$ term to the Hamiltonian $H_{\text{RBH}}$ and take the limit $\Lambda \rightarrow \infty$. Under this projection, only the commuting terms in $H_{\text{RBH}}$ survive. The terms $K_{p,1}$ reduce to the plaquette terms of the toric code
\begin{equation}
    \begin{aligned}
        \begin{tikzpicture}[scale= 0.8, baseline={([yshift=-2.5pt]current bounding box.center)}]
            \coordinate (A) at (0,0,0);
            \coordinate (B) at (0,0,2);
            \coordinate (C) at (0,2,2);
            \coordinate (D) at (0,2,0);
%
            \coordinate (U) at (-2,0,0);
            \coordinate (O) at (2,2,2);
%
%
            %
            \draw (A) -- (B) --(C) -- (D) -- cycle;
            \foreach \X/\Y in { A/B, B/C, C/D, D/A}
            \node[inner sep=0.5pt, fill=white] at ($( \X )!0.5!( \Y )$) {\footnotesize $X_1$}; 
    \end{tikzpicture}, \quad
    \begin{tikzpicture}[scale= 0.8, baseline={([yshift=-2.5pt]current bounding box.center)}]
%
            \coordinate (A) at (0,0,0);
            \coordinate (B) at (0,2,0);
            \coordinate (C) at (2,2,0);
            \coordinate (D) at (2,0,0);
%
            \coordinate (U) at (0,0,-2);
            \coordinate (O) at (2,2,2);
%
%
  %
            \draw (A) -- (B) --(C) -- (D) -- cycle;
            \foreach \X/\Y in { A/B, B/C, C/D, D/A}
            \node[inner sep=0.5pt, fill=white] at ($( \X )!0.5!( \Y )$) {\footnotesize $X_1$}; 
    \end{tikzpicture}, \quad
    \begin{tikzpicture}[scale= 0.8, baseline={([yshift=-2.5pt]current bounding box.center)}]
        \coordinate (A) at (0,0,0);
        \coordinate (B) at (2,0,0);
        \coordinate (C) at (2,0,2);
        \coordinate (D) at (0,0,2);
%
        \coordinate (O) at (2,2,2);
        \coordinate (U) at (0,-2,0);
%
%
        \draw  (A) -- (B) -- (C) -- (D) -- cycle;
%
 %
        \foreach \X/\Y in { A/B, B/C, C/D, D/A}
        \node[inner sep=0.5pt, fill=white] at ($( \X )!0.5!( \Y )$) {\footnotesize $X_1$};
    \end{tikzpicture}. \label{eq:3DTC_plaquette}
    \end{aligned}
\end{equation}
On the other hand, $X_2$ in $K_{p,2}$ anti-commutes with the projection, so $X_2$ must be canceled by taking the product of these plaquette terms around a cube, resulting in the vertex term of the toric code,
\begin{align}
    B_v^{3D} = \prod_{p\in c} K_{p,2} 
    \hspace{5pt}
    =
    \hspace{5pt}
    \begin{tikzpicture}[scale= 0.8, baseline={([yshift=-2.5pt]current bounding box.center)}]
    %
        \coordinate (O) at (0,0,0);
        \coordinate (A) at (-1.5,0,0);
        \coordinate (B) at (1.5,0,0);
        \coordinate (C) at (0,-1.5,0);
        \coordinate (D) at (0,1.5,0);
        \coordinate (E) at (0,0,1.5);
        \coordinate (F) at (0,0,-1.5);
        %
        \draw (A) -- (B);
        \draw (C) -- (D); 
        \draw (E) -- (F);
        \foreach \X/\Y in { O/E, O/F}
        \node[inner sep=0.5pt, fill=white] at ($( \X )!0.5!( \Y )$) {\footnotesize $Z_1$};
        \foreach \X/\Y in {O/C, O/D}
        \node[inner sep=0.5pt, fill=white] at ($( \X )!0.5!( \Y )$) {\footnotesize $Z_1$};
        \foreach \X/\Y in {O/A, O/B}
        \node[inner sep=0.5pt, fill=white] at ($( \X )!0.5!( \Y )$) {\footnotesize $Z_1$}; 
    \end{tikzpicture}. \label{eq:3DTC_vertex}
\end{align}
Finally, the $Z_2$ boundary terms becomes identity, and the $Z_1$ boundary terms remain invariant. The collection of vertex and plaquette terms above is exactly the 3D toric code with smooth boundaries on both the top and bottom.

The above phase transition can also be interpreted as gauging a 1-form symmetry of the RBH cluster state~\cite{tantivasadakarn2024long}, and, specifically, the RBH cluster state has two $\mathbb{Z}_2$ 1-from symmetries generated by the vertex terms
\begin{align*}
    \begin{tikzpicture}[scale= 0.8, baseline={([yshift=-2.5pt]current bounding box.center)}]
    %
        \coordinate (O) at (0,0,0);
        \coordinate (A) at (-1.5,0,0);
        \coordinate (B) at (1.5,0,0);
        \coordinate (C) at (0,-1.5,0);
        \coordinate (D) at (0,1.5,0);
        \coordinate (E) at (0,0,1.5);
        \coordinate (F) at (0,0,-1.5);
        %
        \draw (A) -- (B);
        \draw (C) -- (D); 
        \draw (E) -- (F);
        \foreach \X/\Y in { O/E, O/F}
        \node[inner sep=0.5pt, fill=white] at ($( \X )!0.5!( \Y )$) {\footnotesize $Z_i$};
        \foreach \X/\Y in {O/C, O/D}
        \node[inner sep=0.5pt, fill=white] at ($( \X )!0.5!( \Y )$) {\footnotesize $Z_i$};
        \foreach \X/\Y in {O/A, O/B}
        \node[inner sep=0.5pt, fill=white] at ($( \X )!0.5!( \Y )$) {\footnotesize $Z_i$}; 
    \end{tikzpicture}, \quad
    i = 1,2
\end{align*}
The stabilizers $K_p$'s decorate the domain walls of each $1$-form symmetry with the $1$-form charge of the other. Therefore, it is an SPT state protected by the two $\mathbb{Z}_2$ 1-form symmetries. The degenerate ground states of the toric code supported on the boundary are the higher-dimensional analog of the degenerate edge modes of the 1D cluster state. See~\cite{chen2014symmetry} for the SPT construction using decorated domain walls. It is well-known that, given a state which is symmetric under a $\mathbb{Z}_2$ 1-form symmetry, the procedure of entangling the state using the RBH entangler followed by projection effectively gauges the 1-form symmetry.

The density matrix of the 3D toric code can be written as
\begin{equation}
    \begin{aligned}
        \rho_{3DTC} = \prod_v \frac{1 + B_v}{2} \prod_{p} \frac{1 + A_p}{2},
    \end{aligned}
\end{equation}
where $A_p$ and $B_v$ are defined in Eqs.~\eqref{eq:3DTC_plaquette} and \eqref{eq:3DTC_vertex} as well as the truncated vertex terms on the boundary, respectively. The smooth boundary of the 3D toric code can no longer be regarded as a 2D toric code. In fact, by tracing out the bulk and retaining only the top boundary, one can show that the reduced density matrix takes the form
\begin{equation}
    \begin{aligned}
        \operatorname{Tr}_{\rm bulk} (\rho_{3DTC}) = \prod_{p} \frac{1 + A_p}{2},
    \end{aligned}
\end{equation}
which is exactly $\rho_{TC}^{\dec}$ as defined in Eq.~\eqref{eq:TC_charge_decoherence}.

It is not a coincidence that starting from the 2D toric code, constructing the 3D RBH model, applying the $Z_2$ projection, and tracing out the bulk yields the same density matrix as the maximally charge-decohered toric code. This example motivates us to interpret the strong and weak 1-form symmetries of the decohered state as boundary manifestations of extended operators in a three-dimensional bulk.

As discussed above, the RBH cluster state has two 1-form symmetries. Consider tracing out the bulk of the RBH state while retaining only the top boundary without applying projections in the bulk, the reduced density matrix can be written as
\begin{equation}
    \begin{aligned}
        \operatorname{Tr}_{\text{Bulk}}(\rho_{\text{RBH}}) = \prod_{v} \frac{1 + \widetilde{B}_v}{2} \prod_p \frac{1 + \widetilde{A}_p}{2},\label{eq:boundary_RBH}
    \end{aligned}
\end{equation}
where
\begin{equation}
    \begin{aligned}
        \widetilde{B}_v &= \begin{tikzpicture}[scale= 1, baseline={([yshift=-2.5pt]current bounding box.center)}]
        \def\L{2}
        \draw (0,0) -- (\L,0);
        \draw (\L, 0) -- (2*\L, 0);
        \draw (0.5*\L, -0.5*\L) -- (1.5*\L, -0.5*\L) -- (2*\L, 0);
        \draw (0.5*\L,-0.5*\L) -- (1.5*\L,0.5*\L);

        \draw (\L*0.5,0) node[inner sep=0.5pt, fill=white] {\footnotesize $Z_1$};
        \draw (\L*1.5, 0) node[inner sep=0.5pt, fill=white] {\footnotesize $Z_1 X_2$};
        \draw (0.75*\L, -0.25*\L) node[inner sep=0.5pt, fill=white] {\footnotesize $Z_1 X_2$};
        \draw (1.25*\L, 0.25*\L) node[inner sep=0.5pt, fill=white] {\footnotesize $Z_1$};
        \draw (\L, -0.5*\L) node[inner sep=0.5pt, fill=white] {\footnotesize $X_2$};
        \draw (1.75*\L, -0.25*\L) node[inner sep=0.5pt, fill=white] {\footnotesize $X_2$};
    \end{tikzpicture}, \\
    \widetilde{A}_p &= \quad\ \begin{tikzpicture}[scale= 1, baseline={([yshift=-2.5pt]current bounding box.center)}]
        \coordinate (A) at (0,0,0);
        \coordinate (B) at (2,0,0);
        \coordinate (C) at (2,0,2);
        \coordinate (D) at (0,0,2);
%
        \coordinate (O) at (2,2,2);
        \coordinate (U) at (0,-2,0);
%
%
        \draw  (A) -- (B) -- (C) -- (D) -- cycle;
%
 %
        \foreach \X/\Y in { A/B, B/C, C/D, D/A}
        \node[inner sep=0.5pt, fill=white] at ($( \X )!0.5!( \Y )$) {\footnotesize $X_1$};
    \end{tikzpicture}. 
    \end{aligned}
\end{equation}
By applying a Hadamard gate to all second qubits, followed by a controlled-$X$ gate between each second qubit and its corresponding first qubit along the diagonal direction, the density matrix in Eq.~\eqref{eq:boundary_RBH} transforms into the density matrix of a pure-state 2D toric code defined on the first qubits, tensored with a maximally mixed state on the second qubits. Consequently, tracing out the bulk of the RBH state while retaining only the top boundary yields a density matrix in the same phase as the pure-state 2D toric code. From the perspective of symmetries, this is equivalent to saying that the reduced density matrix has the same strong 1-form symmetries as the pure-state 2D toric code, and is therefore in the same phase.

On the other hand, after the phase transition induced by the projection, performing the same tracing procedure shows that the reduced density matrix is exactly the charge-decohered density matrix in Eq.~\eqref{eq:TC_charge_decoherence}. Consequently, the electric 1-form symmetry remains strong, while the magnetic 1-form symmetry becomes weak. As discussed in Ref.~\cite{ellison2025toward}, the magnetic membrane operator cannot be purely supported on the smooth boundary; for the logical symmetry operator, it must terminate on both the top and bottom boundaries. Consequently, after tracing, the symmetry associated with the magnetic membrane becomes weak. The electric symmetry of the 3D toric code, however, is a 2-form symmetry whose symmetry operator is line-like, and can therefore be purely supported on the top boundary alone. After tracing, this symmetry remains strong, and the reduced density matrix is thus in the SWSSB phase.

The above discussion of 3D phase transitions and their corresponding 2D reduced density matrices reveals a holographic correspondence between 2D decoherence and 3D phase transitions. A detailed general discussion is provided in Sec.~\ref{sec:gc}. Before turning to the general construction, we present another exotic example in the next subsection, showing that intrinsic mixed-state topological phases can also be understood holographically, where the corresponding phase transition drives the system to the 3D fermionic toric code.

\subsection{Fermionic Decoherence}

Next we consider the incoherent proliferation of the fermionic anyons, which leads to a SWSSB phase with a strong fermionic 1-form symmetry and weak bosonic 1-form symmetries. Specifically, the local channel which incoherently proliferates a pair of $\ff$ is given by
\begin{equation}
    \mathcal{N}_e(\rho) = \frac{1}{2} \rho + \frac{1}{2} W_e \rho W_e,
\end{equation}
where $W_e$ is a short ribbon operator creating a pair of fermion. Depending on the orientation of the edge, $W_e$ is given by
\begin{equation}
    \begin{aligned}
    \begin{tikzpicture}[scale= 0.8, baseline={([yshift=-2.5pt]current bounding box.center)}]
        \def\L{2}
        \draw (0,0) -- (\L,0);
        \draw (\L, -\L) -- (\L, 0);
        \draw (\L*0.5,0) node[inner sep=0.5pt, fill=white] {\footnotesize $X_e$};
        \draw (\L, -\L*0.5) node[inner sep=0.5pt, fill=white] {\footnotesize $Z$};
    \end{tikzpicture}, \quad 
    \begin{tikzpicture}[scale= 0.8, baseline={([yshift=-2.5pt]current bounding box.center)}]
        \def\L{2}
        \draw (0,0) -- (\L,0);
        \draw (0, 0) -- (0, \L);
        \draw (\L*0.5,0) node[inner sep=0.5pt, fill=white] {\footnotesize $Z$};
        \draw (0, \L*0.5) node[inner sep=0.5pt, fill=white] {\footnotesize $X_e$};
    \end{tikzpicture},
    \end{aligned}
\end{equation}
where $X_e$ labels the edge corresponding to $W_e$. Applying the maximal fermion decoherence channel to the ground state of the 2D toric code yields the density matrix
\begin{equation}
    \rho_{TC}^{\dec,\ff}  = \prod_v \frac{1 + \widetilde{A_v}}{2}, \label{eq:fermion_decoherence}
\end{equation}
where $\widetilde{A}_v$ is given as follows:
\begin{equation}
    \begin{aligned}
        \widetilde{A}_v = \begin{tikzpicture}
    [scale= 0.8, baseline={([yshift=-2.5pt]current bounding box.center)}]
        \draw[step = 2] (0,0) grid (2,-2);
        \draw[step = 2] (-1.5,-1.5) grid (1.5,1.5);
        \node[inner sep=0.5pt, fill=white] at (0,0) {};
        \node[inner sep=0.5pt, fill=white] at (-1, 0) {\footnotesize $Z$};
        \node[inner sep=0.5pt, fill=white] at (0,-1) {\footnotesize $ZX$};
        
        \node[inner sep=0.5pt, fill=white] at (0, 1) {\footnotesize $Z$};
        \node[inner sep=0.5pt, fill=white] at (2,-1) {\footnotesize $X$};
        \node[inner sep=0.5pt, fill=white] at (1, 0) {\footnotesize$ZX$};
        \node[inner sep=0.5pt, fill=white] at (1,-2) {\footnotesize $X$};
        
        \node[inner sep=0.5pt, fill=white] at (0.2,0.2) {\footnotesize $v$};
    \end{tikzpicture}.
    \end{aligned}
\end{equation}
Note that the operator $\widetilde{A}_v$ represents the smallest fermion loop.

This density matrix likewise lies in the SWSSB phase. However, the strong 1-form symmetry is fermionic rather than bosonic, making it anomalous, while the two weak 1-form symmetries remain bosonic. It has further been shown that $\rho_{TC}^{\dec,\ff}$ cannot be connected to the trivial mixed state via two-way connectivity. Hence, any path connecting this state to the trivial state must encounter a point of diverging Markov length, and it is for this reason that the phase is referred to as an {\it intrinsic mixed-state topological order} (imTO)~\cite{sohal2025noisy,ellison2025toward,wang2025intrinsic}.

Recall that the 2D bosonic decoherence corresponds to a 3D phase transition from the RBH cluster state to the 3D toric code. In the rest of this subsection, we demonstrate that the 2D fermionic decoherence corresponds analogously to a 3D phase transition from the RBH cluster state to the 3D fermionic toric code. Starting from $H_{\text{RBH}}$, instead of projecting onto the $Z_2 = 1$ subspace, we project onto the $Z_1Z_2 = 1$ subspace. This reduces the Hilbert space on each edge to a single effective qubit, and we further make the following identification
\begin{align}
    X_1X_2 \mapsto X,
    \hspace{25pt}
    Z_1 = Z_2 \mapsto Z.
\end{align}
In this case, both $K_{p,1}$ and $K_{p,2}$ anti-commute with the projection. There are two ways to form commuting terms. First, as before, we can form a product of the plaquette stabilizers around a cube to cancel the Pauli $X$, giving vertex terms
\begin{equation}
\begin{aligned}
    \begin{tikzpicture}[scale= 0.8, baseline={([yshift=-2.5pt]current bounding box.center)}]
    %
        \coordinate (O) at (0,0,0);
        \coordinate (A) at (-1.5,0,0);
        \coordinate (B) at (1.5,0,0);
        \coordinate (C) at (0,-1.5,0);
        \coordinate (D) at (0,1.5,0);
        \coordinate (E) at (0,0,1.5);
        \coordinate (F) at (0,0,-1.5);
        %
        \draw (A) -- (B);
        \draw (C) -- (D); 
        \draw (E) -- (F);
        \foreach \X/\Y in { O/E, O/F}
        \node[inner sep=0.5pt, fill=white] at ($( \X )!0.5!( \Y )$) {\footnotesize $Z_i$};
        \foreach \X/\Y in {O/C, O/D}
        \node[inner sep=0.5pt, fill=white] at ($( \X )!0.5!( \Y )$) {\footnotesize $Z_i$};
        \foreach \X/\Y in {O/A, O/B}
        \node[inner sep=0.5pt, fill=white] at ($( \X )!0.5!( \Y )$) {\footnotesize $Z_i$}; 
    \end{tikzpicture}
    \hspace{10pt}
    \mapsto
    \hspace{10pt}
    \begin{tikzpicture}[scale= 0.8, baseline={([yshift=-2.5pt]current bounding box.center)}]
    %
        \coordinate (O) at (0,0,0);
        \coordinate (A) at (-1.5,0,0);
        \coordinate (B) at (1.5,0,0);
        \coordinate (C) at (0,-1.5,0);
        \coordinate (D) at (0,1.5,0);
        \coordinate (E) at (0,0,1.5);
        \coordinate (F) at (0,0,-1.5);
        %
        \draw (A) -- (B);
        \draw (C) -- (D); 
        \draw (E) -- (F);
        \foreach \X/\Y in { O/E, O/F}
        \node[inner sep=0.5pt, fill=white] at ($( \X )!0.5!( \Y )$) {\footnotesize $Z$};
        \foreach \X/\Y in {O/C, O/D}
        \node[inner sep=0.5pt, fill=white] at ($( \X )!0.5!( \Y )$) {\footnotesize $Z$};
        \foreach \X/\Y in {O/A, O/B}
        \node[inner sep=0.5pt, fill=white] at ($( \X )!0.5!( \Y )$) {\footnotesize $Z$}; 
    \end{tikzpicture}, \label{eq:fTC_Av}
\end{aligned}
\end{equation}
where $i = 1, 2$. Second, we can take the product $K_{p,1}K_{p,2}$. After the substitution $X_1X_2 \mapsto X$, we obtain the plaquette terms
\begin{align}\label{eq:ferm_tc_plaq}
    \begin{tikzpicture}[scale= 0.8, baseline={([yshift=-2.5pt]current bounding box.center)}]
            \coordinate (A) at (0,0,0);
            \coordinate (B) at (0,0,2);
            \coordinate (C) at (0,2,2);
            \coordinate (D) at (0,2,0);
%
            \coordinate (U) at (-2,0,0);
            \coordinate (O) at (2,2,2);
%
%
            %
            \draw (A) -- (B) --(C) -- (D) -- cycle;
            \draw [dotted] (A) -- (U);
            \draw (C) -- (O);
            \foreach \X/\Y in { A/B, B/C, C/D, D/A}
            \node[inner sep=0.5pt, fill=white] at ($( \X )!0.5!( \Y )$) {\footnotesize $X$}; 
            \node[inner sep=0.5pt, fill=white] at (-1.1, 0, 0) {\footnotesize $Z$};
            \node[inner sep=0.5pt, fill=white] at (1.1, 2, 2) {\footnotesize $Z$};
    \end{tikzpicture},
    \begin{tikzpicture}[scale= 0.8, baseline={([yshift=-2.5pt]current bounding box.center)}]
%
            \coordinate (A) at (0,0,0);
            \coordinate (B) at (0,2,0);
            \coordinate (C) at (2,2,0);
            \coordinate (D) at (2,0,0);
%
            \coordinate (U) at (0,0,-2);
            \coordinate (O) at (2,2,2);
%
%
  %
            \draw (A) -- (B) --(C) -- (D) -- cycle;
            \draw [dotted] (A) -- (U);
            \draw (C) -- (O);
            \foreach \X/\Y in { A/B, B/C, C/D, D/A}
            \node[inner sep=0.5pt, fill=white] at ($( \X )!0.5!( \Y )$) {\footnotesize $X$}; 
            \node[inner sep=0.5pt, fill=white] at (0, 0, -1.1) {\footnotesize $Z$};
            \node[inner sep=0.5pt, fill=white] at (2, 2, 1.1) {\footnotesize $Z$};
    \end{tikzpicture},
    \begin{tikzpicture}[scale= 0.8, baseline={([yshift=-2.5pt]current bounding box.center)}]
        \coordinate (A) at (0,0,0);
        \coordinate (B) at (2,0,0);
        \coordinate (C) at (2,0,2);
        \coordinate (D) at (0,0,2);
%
        \coordinate (O) at (2,2,2);
        \coordinate (U) at (0,-2,0);
%
%
        \draw  (A) -- (B) -- (C) -- (D) -- cycle;
        \draw [dotted] (A) -- (U);
        \draw (C) -- (O);
%
 %
        \foreach \X/\Y in { A/B, B/C, C/D, D/A}
        \node[inner sep=0.5pt, fill=white] at ($( \X )!0.5!( \Y )$) {\footnotesize $X$};
        \node[inner sep=0.5pt, fill=white] at (2,1.1,2) {\footnotesize $Z$};
        \node[inner sep=0.5pt, fill=white] at (0,-1.1,0) {\footnotesize $Z$};        
    \end{tikzpicture}
\end{align}
Lastly, the boundary terms are identified as follows,
\begin{equation}
\begin{aligned}
    &\begin{tikzpicture}[scale= 0.8, baseline={([yshift=-2.5pt]current bounding box.center)}]
    %
        \coordinate (O) at (0,0,0);
        \coordinate (A) at (-1.5,0,0);
        \coordinate (B) at (1.5,0,0);
        \coordinate (C) at (0,-1.5,0);
        \coordinate (D) at (0,1.5,0);
        \coordinate (E) at (0,0,1.5);
        \coordinate (F) at (0,0,-1.5);
        %
        \draw (A) -- (B);
        \draw (O) -- (D); 
        \draw (E) -- (F);
        \foreach \X/\Y in { O/E, O/F}
        \node[inner sep=0.5pt, fill=white] at ($( \X )!0.5!( \Y )$) {\footnotesize $Z_i$};
        \foreach \X/\Y in {O/D}
        \node[inner sep=0.5pt, fill=white] at ($( \X )!0.5!( \Y )$) {\footnotesize $Z_i$};
        \foreach \X/\Y in {O/A, O/B}
        \node[inner sep=0.5pt, fill=white] at ($( \X )!0.5!( \Y )$) {\footnotesize $Z_i$}; 
    \end{tikzpicture}
    \hspace{10pt}
    \mapsto
    \hspace{10pt}
    \begin{tikzpicture}[scale= 0.8, baseline={([yshift=-2.5pt]current bounding box.center)}]
    %
        \coordinate (O) at (0,0,0);
        \coordinate (A) at (-1.5,0,0);
        \coordinate (B) at (1.5,0,0);
        \coordinate (C) at (0,-1.5,0);
        \coordinate (D) at (0,1.5,0);
        \coordinate (E) at (0,0,1.5);
        \coordinate (F) at (0,0,-1.5);
        %
        \draw (A) -- (B);
        \draw (O) -- (D); 
        \draw (E) -- (F);
        \foreach \X/\Y in { O/E, O/F}
        \node[inner sep=0.5pt, fill=white] at ($( \X )!0.5!( \Y )$) {\footnotesize $Z$};
        \foreach \X/\Y in {O/D}
        \node[inner sep=0.5pt, fill=white] at ($( \X )!0.5!( \Y )$) {\footnotesize $Z$};
        \foreach \X/\Y in {O/A, O/B}
        \node[inner sep=0.5pt, fill=white] at ($( \X )!0.5!( \Y )$) {\footnotesize $Z$}; 
    \end{tikzpicture}\\
    &\begin{tikzpicture}[scale= 0.8, baseline={([yshift=-2.5pt]current bounding box.center)}]
    %
        \coordinate (O) at (0,0,0);
        \coordinate (A) at (-1.5,0,0);
        \coordinate (B) at (1.5,0,0);
        \coordinate (C) at (0,-1.5,0);
        \coordinate (D) at (0,1.5,0);
        \coordinate (E) at (0,0,1.5);
        \coordinate (F) at (0,0,-1.5);
        %
        \draw (A) -- (B);
        \draw (C) -- (O); 
        \draw (E) -- (F);
        \foreach \X/\Y in { O/E, O/F}
        \node[inner sep=0.5pt, fill=white] at ($( \X )!0.5!( \Y )$) {\footnotesize $Z_i$};
        \foreach \X/\Y in {O/C}
        \node[inner sep=0.5pt, fill=white] at ($( \X )!0.5!( \Y )$) {\footnotesize $Z_i$};
        \foreach \X/\Y in {O/A, O/B}
        \node[inner sep=0.5pt, fill=white] at ($( \X )!0.5!( \Y )$) {\footnotesize $Z_i$}; 
    \end{tikzpicture}
    \hspace{10pt}
    \mapsto
    \hspace{10pt}
    \begin{tikzpicture}[scale= 0.8, baseline={([yshift=-2.5pt]current bounding box.center)}]
    %
        \coordinate (O) at (0,0,0);
        \coordinate (A) at (-1.5,0,0);
        \coordinate (B) at (1.5,0,0);
        \coordinate (C) at (0,-1.5,0);
        \coordinate (D) at (0,1.5,0);
        \coordinate (E) at (0,0,1.5);
        \coordinate (F) at (0,0,-1.5);
        %
        \draw (A) -- (B);
        \draw (C) -- (O); 
        \draw (E) -- (F);
        \foreach \X/\Y in { O/E, O/F}
        \node[inner sep=0.5pt, fill=white] at ($( \X )!0.5!( \Y )$) {\footnotesize $Z$};
        \foreach \X/\Y in {O/C}
        \node[inner sep=0.5pt, fill=white] at ($( \X )!0.5!( \Y )$) {\footnotesize $Z$};
        \foreach \X/\Y in {O/A, O/B}
        \node[inner sep=0.5pt, fill=white] at ($( \X )!0.5!( \Y )$) {\footnotesize $Z$}; 
    \end{tikzpicture}, \label{eq:FTC_boundary}
\end{aligned}
\end{equation}
where $i = 1,2$. These stabilizers define the 3D fermionic TC with a smooth boundary condition~\cite{bchktz23}.

Unlike the bosonic toric code in 3D, the point excitations in the fermionic TC are fermions, thus generating a fermionic $\mathbb{Z}_2$ 2-form symmetry, while the $\mathbb{Z}_2$ 1-form symmetry generated by flux loops is still bosonic. The smooth boundaries again allow the loop excitations to condense. Following a similar argument, when tracing out the bottom half of the system, the fermionic loops can be deformed to the top boundary, hence remain a strong symmetry, while the flux membranes terminating on the top boundary must also have support in the bulk, generating a weak 1-form symmetry after tracing. One can verify that this yields the density matrix $\rho_{TC}^{\dec,\ff}$.

The density matrix of the 3D fermionic toric code can be written as
\begin{equation}
    \begin{aligned}
        \rho_{3DFTC} = \prod_v \frac{1 + A_v}{2} \prod_p \frac{1 + B_p}{2}, 
    \end{aligned}
\end{equation}
where $A_v$ and $B_p$ are defined in Eqs.~\eqref{eq:fTC_Av} and \eqref{eq:ferm_tc_plaq}, and the boundary vertex terms Eq.~\eqref{eq:FTC_boundary}, respectively. Tracing out the bulk while retaining only the top boundary, the reduced density matrix can be written as
\begin{equation}
    \begin{aligned}
        \operatorname{Tr}_{\text{bulk}}(\rho_{3DFTC}) = \prod_v \frac{1 + \widetilde{A}_v}{2},
    \end{aligned}
\end{equation}
which is exactly the maximally fermion-decohered toric code in Eq.~\eqref{eq:fermion_decoherence}.

In summary, we have shown that both the bosonic and fermionic decoherence can be related to a phase transition in their 3D counterparts. Specifically, our examples suggest that two distinct paths lead to the same maximally decohered density matrix of the original 2D toric code. First, one can directly decohere the 2D toric code using Kraus operators given by the short anyon creation operators. Second, one can construct the 3D RBH cluster state, drive it through a phase transition via projection, and then trace out the bulk while retaining only the top boundary. The density matrices obtained from both paths are identical, and different pure-state phase transitions in the 3D counterpart correspond to different types of decoherence in 2D.

As we will show in the next section, the original 2D anyon theory can be generalized to any unitary modular tensor category (UMTC), and the 3D RBH cluster state can be generalized to its corresponding Walker-Wang model~\cite{walker20123+}. After nontrivial decoherence, the original anyon theory becomes another category, possibly premodular, while the corresponding 3D bulk transitions to a different Walker-Wang model.

This construction is particularly powerful for chiral anyon theories, i.e., those with non-zero chiral central charge. Such theories admit no commuting projector Hamiltonian in 2D, making it difficult to study their decoherence directly at the lattice level. Consequently, previous studies have addressed the decoherence of chiral phases only from a categorical perspective~\cite{wang2025intrinsic,ellison2025toward,sohal2025noisy}. Our construction circumvents this obstruction: since a commuting projector Hamiltonian always exists for Walker-Wang models, one can instead study the decoherence of a chiral anyon theory through the phase transition of its associated 3D bulk. This provides a concrete, rigorous lattice treatment of decoherence for chiral topological phases.

\section{General Construction: A 3D Phase Transition}\label{sec:gc}

The examples in the previous section serve as toy models for the general theory introduced in this section. In this section, we first give a categorical interpretation of these transitions, and then state a general conjecture relating Walker-Wang models in 3D to mixed-state phase transitions in 2D.

The input of a Walker-Wang model~\cite{walker20123+} is a unitary braided fusion category $\mathcal{C}$, or premodular category as referred to in the literature, whose $R$-symbols and $F$-symbols satisfy the pentagon and hexagon equations, while the $S$-matrix can be degenerate. The Walker-Wang model is defined on a 3D cubic lattice (with vertices resolved into trivalent junctions), and the Hilbert space consists of one $|\mathcal{C}|$-dimensional qudit per edge, spanned by the simple objects in $\mathcal{C}$. The Walker-Wang Hamiltonian is given by a sum of commuting projectors
\begin{align}
    H = -\sum_{v}B_v - \sum_p K_p,
\end{align}
where the vertex operators enforce fusion at each trivalent vertex. The subspace defined by $B_v = 1$ for all vertices $v$ is the subspace of string-nets. The plaquette operator is defined by
\begin{align}
    K_p = \frac{1}{D^2}\sum_{a\in \mathcal{C}} d_a K_p^a,
\end{align}
where $d_a$ is the quantum dimension of the anyon $a$, and $D = \sqrt{\sum_a d_a^2}$ is the total quantum dimension. The operator $K_p^a$ is defined by inserting a loop of anyon $a$ around the plaquette $p$, and resolving it into the edges using $R$-symbols and $F$-symbols. We refer the reader to the orginal paper~\cite{walker20123+}or Appendix~\ref{app:ww} for more details.

If the input category is modular, meaning that no transparent anyon exists, then the Walker-Wang model is an invertible phase, and no deconfined excitations exist. However, if the category is not modular, and transparent anyons exist, then such anyons become deconfined, and must be either a boson or a fermion. More concretely, for each anyon $a\in\mathcal{C}$, one can define a string operator by inserting a string of anyon $a$, and resolve it onto the edges using $R$-symbols and $F$-symbols~\cite{vonkeyserlingk2013three}. Such operators trivially commute with the vertex operators. However, if $a$ is not transparent, the string operator anti-commutes with all plaquette operators it threads through. If $a$ is transparent, it also commutes with all plaquette operators as well, hence is a logical operator.

For considerations of mixed states, motivated by the toric code example we study in the previous section, we define the Walker-Wang model on a slab. The boundary condition is chosen such that the top boundary hosts the 2D anyon theory $\mathcal{C}$, and the bottom boundary hosts the corresponding $\overline{\mathcal{C}}$. Such boundaries are smooth boundaries~\cite{vonkeyserlingk2013three}, which are boundaries of the cubic lattice with no edges sticking out. The boundary terms are defined in the same way as the bulk terms, with each vertex term enforcing fusion, and each plaquette term being a sum of anyon loop fusion operators, enforcing zero flux. The existence of boundaries also changes ground space properties of the WW model. On each boundary, we can define the closed string operators by inserting anyon strings for each $a\in \mathcal{C}$. Since such strings do not thread through any plaquettes, they commute with all Hamiltonian terms. As a result, they can act as logical operators on non-contractible boundary cycles even if $\mathcal{C}$ is modular. However, for non-transparent $a$, such string operators are strictly localized near the boundary, and cannot be deformed into the bulk without creating excitations. By resolving the strings into the edges of the lattice, we obtain operators acting strictly on the boundary edges, and possibly some edges connected to the boundary edges.

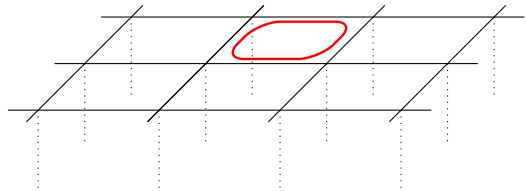
\begin{figure}
    \begin{tikzpicture}[scale = 0.8, baseline = -0.4cm]
        \coordinate (A) at (0,0,0);
        \coordinate (B) at (2,0,0);
        \coordinate (C) at (2,0,2);
        \coordinate (D) at (0,0,2);
%
        \coordinate (O) at (2,2,2);
        \coordinate (U) at (0,-2,0);
%
%
        \draw (-0.5,0,0) -- (6.5,0,0);
        \draw (-0.5,0,2) -- (6.5,0,2);
        \draw (-0.5,0,4) -- (6.5,0,4);
        \draw (0,0,-0.5) -- (0,0,4.5);
        \draw (2,0,-0.5) -- (2,0,4.5);
        \draw (2,0,-0.5) -- (2,0,4.5);
        \draw (4,0,-0.5) -- (4,0,4.5);
        \draw (6,0,-0.5) -- (6,0,4.5);
        \draw [dotted] (0,0,0) -- (0,-1.3,0);    
        \draw [dotted] (2,0,0) -- (2,-1.3,0);    
        \draw [dotted] (4,0,0) -- (4,-1.3,0);    
        \draw [dotted] (6,0,0) -- (6,-1.3,0);   
        \draw [dotted] (0,0,2) -- (0,-1.3,2);    
        \draw [dotted] (2,0,2) -- (2,-1.3,2);    
        \draw [dotted] (4,0,2) -- (4,-1.3,2);    
        \draw [dotted] (6,0,2) -- (6,-1.3,2);
        \draw [dotted] (0,0,4) -- (0,-1.3,4);    
        \draw [dotted] (2,0,4) -- (2,-1.3,4);    
        \draw [dotted] (4,0,4) -- (4,-1.3,4);    
        \draw [dotted] (6,0,4) -- (6,-1.3,4);   
        \draw[red, line width=0.9pt, rounded corners=8pt] (2.2,0,0.2) -- (3.8,0,0.2) -- (3.8,0,1.8) -- (2.2,0,1.8) -- cycle;
    \end{tikzpicture}
    \caption{Illustration of the smooth boundary of the coarse-grained lattice of the Walker-Wang model. The Walker-Wang model is defined on a 3D cubic lattice, with each vertex subsequently resolved into trivalent junctions. The detailed lattice construction is given in Appendix~\ref{app:ww}. Here we show a coarse-grained version, where the red loop denotes a loop insertion operation.}
\end{figure}

Having reviewed the essentials of Walker-Wang construction, we now reconsider the RBH cluster state as an illustrative example. The RBH cluster state is a ground state of the Walker-Wang model with input the toric code UMTC $\{1, \ee, \mm, \ff\}$, hence two qubits per edge. The state of the first qubit $\ket{0}$ or $\ket{1}$ labels whether or not $\ee$ exists on the edge, and the state of the second qubit labels whether $\mm$ exists. If both qubits are in the $\ket{11}$ state, then $\ff$ labels the edge. Specifically, the four states $\ket{00}$, $\ket{01}$, $\ket{10}$ and $\ket{11}$ respectively label $1$, $\mm$, $\ee$ and $\ff$~\cite{roberts20243fermion}.

First, consider the transition to the 3D toric code with smooth boundaries, which we carried out by projecting onto the $Z_2 = 1$ subspace. From the categorical perspective, this annihilates all string-nets with $\mm$ and $\ff$ labels, leaving only the vacuum and $\ee$ labels. Therefore, the $\ee$ anyon, which previously only exists on the boundary, now becomes deconfined in the bulk. By definition, this gives the Walker-Wang model on the input premodular category $\{1, \ee\}$ with a smooth boundary. Indeed, the 3D toric code is precisely the Walker-Wang model of $\{1, \ee\}$.

Secondly, consider the transition to the 3D fermionic toric code, which we carried out by projecting onto the $Z_1Z_2 = 1$ subspace. Under the projection, the basis state on each edge can either be $\ket{00}$ or $\ket{11}$. That is, only the $\ff$ label is allowed apart from the vacuum, and all string-nets with $\ee$ and $\mm$ labels are annihilated. The outcome is the Walker-Wang model with input the premodular category $\{1, \ff\}$. This is precisely the 3D fermionic toric code.

These two toric code examples suggest the commutative diagram in Fig.~\ref{fig:commutitive_diagram_TC}. More generally, we conjecture that a similar statement holds for decoherence of a general category. Given a general UMTC $\mathcal{C}$, we choose a set of anyons $S\subseteq \mathcal{C}$ that is closed under fusion. Let $\mathcal{D}\subseteq \mathcal{C}$ be the subcategory of $\mathcal{C}$ which braids trivially with all anyons in $S$. Then the density matrix under maximal decoherence of $S$ corresponds to the transition from the Walker-Wang model of $\mathcal{C}$ to the Walker-Wang model of $\mathcal{D}$ by only keeping string-nets containing anyon labels that are in $\mathcal{D}$. The commutative diagram is given in Fig.~\ref{fig:commutitive_diagram_category}.

\begin{figure}[h]
    \centering
    \begin{tikzpicture}[
  >=Latex,
  object/.style={
    draw,
    rounded corners=2pt,
    minimum width=4.35cm,
    minimum height=1.25cm,
    inner sep=7pt,
    align=center
  },
  map/.style={
    ->,
    line width=0.8pt,
    shorten <=9pt,
    shorten >=9pt
  },
  arrowlabel/.style={
    fill=white,
    inner xsep=4pt,
    inner ysep=2pt,
    font=\small,
    align=center
  },
  scale = 0.65
]
\node (C) at (0,0) {
  2D TC
};

\node (rho) at (8.3,0) {
  {\small Density matrix $\rho_{TC}^{\dec}$/$\rho_{TC}^{\dec,\ff}$}
};

\node (WW) at (0,-4.1) {
   RBH
};

\node (T3d) at (8.3,-4.1) {
  {\small 3D TC/fermionic TC}
};

\draw[map] (C.east) -- (rho.west)
  node[midway, above=5pt, arrowlabel] {Decohere $\ee$/$\ff$};

\draw[map] (C.south) -- (WW.north)
  node[midway, left=7pt, arrowlabel] {WW};

\draw[map] (WW.east) -- (T3d.west)
  node[midway, above=5pt, arrowlabel]
  {Projection\\$Z_2=1$/$Z_1Z_2=1$};

\draw[map] (T3d.north) -- (rho.south)
  node[midway, right=0cm, arrowlabel]
  {Tracing};
\end{tikzpicture}
    \caption{The commutative diagram of toric code decoherence. Decoherence can be carried out by first lifting to the RBH cluster state, making measurements, then trace out the bottom half.}
    \label{fig:commutitive_diagram_TC}
\end{figure}
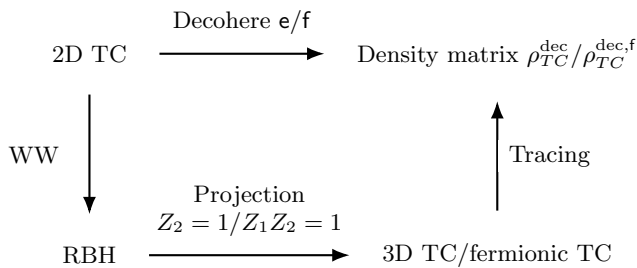
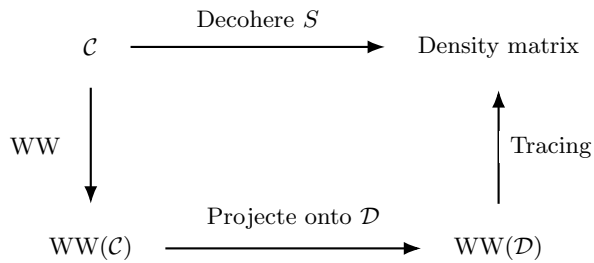
\begin{figure}[h]
    \centering
    \begin{tikzpicture}[
  >=Latex,
  object/.style={
    draw,
    rounded corners=2pt,
    minimum width=4.35cm,
    minimum height=1.25cm,
    inner sep=7pt,
    align=center
  },
  map/.style={
    ->,
    line width=0.8pt,
    shorten <=9pt,
    shorten >=9pt
  },
  arrowlabel/.style={
    fill=white,
    inner xsep=4pt,
    inner ysep=2pt,
    font=\small,
    align=center
  },
  scale = 0.65
]
\node (C) at (0,0) {
  $\mathcal{C}$
};

\node (rho) at (8.3,0) {
  {\small Density matrix}
};

\node (WW) at (0,-4.1) {
   WW($\mathcal{C}$)
};

\node (T3d) at (8.3,-4.1) {
  {WW($\mathcal{D}$)}
};

\draw[map] (C.east) -- (rho.west)
  node[midway, above=5pt, arrowlabel] {Decohere $S$};

\draw[map] (C.south) -- (WW.north)
  node[midway, left=7pt, arrowlabel] {WW};

\draw[map] (WW.east) -- (T3d.west)
  node[midway, above=5pt, arrowlabel]
  {Projecte onto $\mathcal{D}$};

\draw[map] (T3d.north) -- (rho.south)
  node[midway, right=0cm, arrowlabel]
  {Tracing};
\end{tikzpicture}
    \caption{Commutative diagram corresponding to decohering $S$ in the topological order $\mathcal{C}$. We first lift to the 3D WW model of $\mathcal{C}$, keeping only string-nets labeled by $\mathcal{D}$, then trace out the bottom half.  }
    \label{fig:commutitive_diagram_category}
\end{figure}

It is worth noting that a completely analogous transition can be defined in two dimensions on the Levin-Wen model~\cite{lw05}. That is, given a subcategory $\mathcal{D}$ of $\mathcal{C}$, we project out all the string-nets with an anyon label in $\mathcal{C}\backslash\mathcal{D}$. This induces a transition from the Levin-Wen model on $\mathcal{C}$ to that on $\mathcal{D}$.

An advantage of our construction is that it only requires the input category $\mathcal{C}$ to be a UMTC, allowing chiral topological orders to be studied within the same framework. For phases with nonzero chiral central charge, the absence of a strictly two-dimensional commuting-projector realization makes a direct fixed-point lattice treatment difficult. The corresponding Walker-Wang models nevertheless admit commuting-projector Hamiltonians in three dimensions. Our proposed correspondence therefore provides a systematic route to investigating decoherence of these phases through explicit projections and bulk phase transitions, complementing existing categorical descriptions with concrete lattice constructions.

Various examples of this commutative diagram are provided in Sec.~\ref{sec:examples}. This construction is particularly powerful for chiral anyon theories, which admit no commuting projector Hamiltonian in 2D, while their Walker-Wang models do in 3D.

\section{Examples}\label{sec:examples}

In this section we discuss several examples involving decoherence in the 3-fermion anyon theory, the semion-fermion anyon theory, the double semion anyon theory, as well as the Ising anyon theory. Most of these categories have a non-zero chiral central charge, making it hard to study their decoherence directly.

Given an abelian topological order $\mathcal{C}$, there are a couple of ways to write down the Hamiltonian of the Walker-Wang model. One approach, taken by Ref.~\cite{sun2026non}, is to find a suitable anyon theory $\widetilde{\mathcal{C}}$ which maps to $\mathcal{C}$ by condensing transparent bosons, and whose Walker-Wang model is easier to write down. In the Walker-Wang model of $\widetilde{\mathcal{C}}$, the transparent bosons are deconfined in the bulk. If $\mathcal{C}$ is modular, condensation will confine the bulk, leaving only $\mathcal{C}$ on a smooth boundary. The semion Walker-Wang model is derived in this way in~\cite{sun2026non}.

A more systematic way is to use the topological action~\cite{chen2023higher}. We mainly adopt this approach, which we review in Appendix~\ref{app:ta}. Upon obtaining the Hamiltonian, we mainly focus on the bulk and identify the Walker-Wang models before and after the transition. As described in Sec.~\ref{sec:gc}, it follows that the transition preserves smooth boundaries.

\subsection{Decohereing Vacuum}

In this section, we first consider the vacuous case, that is, decohering the vacuum in the topological order $\mathcal{C}$. This is a trivial channel which has no effect on the state. On the other hand, this corresponds to taking $S = \{1\}$, hence $\mathcal{D} = \mathcal{C}$, and the projection step is also trivial. Indeed, since the Walker-Wang model of $\mathcal{C}$ supports $\mathcal{C}$ on its top smooth boundary, the density matrix obtained by tracing out the bottom half of the system inherits the strong 1-form symmetries associated with $\mathcal{C}$. When the bulk is confined, we expect the mixed state to be in the same phase as a pure ground state realizing $\mathcal{C}$. However, the partial trace needs to be defined carefully. As mentioned in Sec.~\ref{sec:gc}, the boundary symmetry operators may act on a thin layer of qudits in the bulk. This layer must not be traced out, and it serves as a buffer region thick enough to retain the microscopic bulk support of the boundary symmetry operators. This accounts for the discrete lattice structure absent from the continuum topological description and ensures that the associated strong 1-form symmetries survive the partial trace.

If the anyon theory $\mathcal{C}$ is chiral, then it is not clear how to specify a pure reference ground state within this construction, as there is no commuting projector realization of this phase in 2D~\cite{kitaev2006anyons,kapustin2020local}. A necessary condition is that it must have the anyon theory $\mathcal{C}$ as its emergent strong 1-form symmetry, which are the 1-form symmetries of a pure ground state realizing $\mathcal{C}$. As discussed above, when $\mathcal{C}$ is a UMTC, the nontrivial anyons of $\mathcal{C}$ are deconfined only on the boundaries. Retaining the top boundary together with a sufficiently thick buffer region ensures that the corresponding boundary string operators remain fully supported in the retained subsystem. The reduced density matrix therefore inherits all the strong 1-form symmetries associated with $\mathcal{C}$. Consistent with the conjectured classification of locally correlated mixed-state topological orders by their strong 1-form symmetries~\cite{ellison2025toward}, we conjecture that these two states belong to the same mixed-state phase. Whether quasilocal channels connecting them in both directions exist remains an open question.

However, the above argument can be made explicit for the non-chiral toric code model. Specifically, in the example discussed in Sec.~\ref{sec:wu}, we show that, after tracing out the bulk of the RBH cluster state and retaining only the top boundary, the reduced density matrix can be transformed into a pure-state density matrix of the 2D toric code tensored with a trivial maximally mixed state, as shown in Eq.~\eqref{eq:boundary_RBH}.

\subsection{Kitaev's sixteenfold Way}

In this section, we consider Kitaev's sixteenfold way~\cite{kitaev2006anyons}. These are the sixteen minimal modular extensions of the premodular anyon theory $\{1,\ff\}$, where $\ff$ is a fermion. Among the sixteen modular anyon theories, there are four $\mathbb{Z}_2 \times \mathbb{Z}_2$ types, four $\mathbb{Z}_4$ types, and eight Ising types. Categorical data of these sixteenfold ways are listed in Table.~\ref{tab:sixteenfold}. In all cases, we consider the decoherence of the fermion $\ff$. A construction based on fermionic ground states and bosonization establishes that a mixed state characterized by $\{1,\ff\}$ admits purifications in each of the sixteen topological phases~\cite{ellison2025toward}. Our holographic construction instead maps decoherence/purification between the mixed-state order characterized by $\{1,\ff\}$ and Kitaev's sixteenfold way to phase transitions in 3D. Specifically, fermion decoherence corresponds to transitions from Walker-Wang models with these sixteen anyon theories as input to the 3D fermionic toric code.

\begin{table*}[t]
  \centering
  \small
  \setlength{\tabcolsep}{6pt}
  \renewcommand{\arraystretch}{1.25}
  \resizebox{0.8\linewidth}{!}{
  \begin{tabular}{ c  l  l  l  c }
    \toprule
    $\nu \bmod 16$ & Type & Simple objects & Topological spin(s) & $c_-\bmod 8$ \\
    \midrule
    $0$  & abelian, toric-code type                       & $\{1,e,m,\psi=em\}$ & $\theta_e=\theta_m=1$                & $0$    \\
    $1$  & Ising type                                     & $\{1,\psi,\sigma\}$                       & $\theta_\sigma=e^{\pi i/8}$          & $1/2$  \\
    $2$  & abelian, $\mathbb{Z}_4$ type                   & $\{1,a,\psi=a^2,a^3\}$                                       & $\theta_a=\theta_{a^3}=e^{\pi i/4}$  & $1$    \\
    $3$  & Ising type                                     & $\{1,\psi,\sigma\}$                                          & $\theta_\sigma=e^{3\pi i/8}$         & $3/2$  \\
    $4$  & abelian, $\mathbb{Z}_2\times\mathbb{Z}_2$ type & $\{1,x,y,\psi\}$                                            & $\theta_x=\theta_y=i$                & $2$    \\
    $5$  & Ising type                                     & $\{1,\psi,\sigma\}$                                          & $\theta_\sigma=e^{5\pi i/8}$         & $5/2$  \\
    $6$  & abelian, $\mathbb{Z}_4$ type                   & $\{1,a,\psi=a^2,a^3\}$                                       & $\theta_a=\theta_{a^3}=e^{3\pi i/4}$ & $3$    \\
    $7$  & Ising type                                     & $\{1,\psi,\sigma\}$                                          & $\theta_\sigma=e^{7\pi i/8}$         & $7/2$  \\
    $8$  & abelian, three-fermion type                    & $\{1,x,y,\psi\}$                                            & $\theta_x=\theta_y=-1$               & $4$    \\
    $9$  & Ising type                                     & $\{1,\psi,\sigma\}$                                          & $\theta_\sigma=e^{9\pi i/8}$         & $9/2$  \\
    $10$ & abelian, $\mathbb{Z}_4$ type                   & $\{1,a,\psi=a^2,a^3\}$                                       & $\theta_a=\theta_{a^3}=e^{5\pi i/4}$ & $5$    \\
    $11$ & Ising type                                     & $\{1,\psi,\sigma\}$                                          & $\theta_\sigma=e^{11\pi i/8}$        & $11/2$ \\
    $12$ & abelian, $\mathbb{Z}_2\times\mathbb{Z}_2$ type & $\{1,x,y,\psi\}$                                            & $\theta_x=\theta_y=-i$               & $6$    \\
    $13$ & Ising type                                     & $\{1,\psi,\sigma\}$                                          & $\theta_\sigma=e^{13\pi i/8}$        & $13/2$ \\
    $14$ & abelian, $\mathbb{Z}_4$ type                   & $\{1,a,\psi=a^2,a^3\}$                                       & $\theta_a=\theta_{a^3}=e^{7\pi i/4}$ & $7$    \\
    $15$ & Ising type                                     & $\{1,\psi,\sigma\}$                                          & $\theta_\sigma=e^{15\pi i/8}$        & $15/2$ \\
    \bottomrule
  \end{tabular}}
  \caption{Kitaev's sixteenfold way: the modular tensor category $\mathcal{C}_\nu\simeq SO(\nu)_1$ obtained by gauging $\mathbb{Z}_2^f$ of a class-D superconductor with BdG Chern number $\nu$, as a function of $\nu\bmod 16$. Every theory contains the distinguished fermion $\psi$ ($\theta_\psi=-1$), and the rest of objects are fermion-parity fluxes with monodromy $-1$ against $\psi$.}
  \label{tab:sixteenfold}
\end{table*}

\subsubsection{Decohering the 3-Fermion Topological Order}

The 3-fermion anyon theory~\cite{rowell2009classification,bombin2009interacting,bombin2012universal,walker20123+,vonkeyserlingk2013three,burnell2014exactly,teo2015theory,haah2023nontrivial}, $\nu=8$ in Table~\ref{tab:sixteenfold}, has four types of anyons, labeled by $\{1, \ff_1, \ff_2, \ff_3\}$, each of which is a fermion except the vacuum. This anyon theory has $\mathbb{Z}_2\times \mathbb{Z}_2$ fusion rules. More specifically, we have
\begin{equation}
\begin{aligned}
\ff_i \times \ff_i = 1, \quad \ff_i \times \ff_j = \ff_k,
\end{aligned}
\end{equation}
where $i, j, k \in \{1, 2, 3\}$ and $i \neq j \neq k$. The $S$-matrix of the 3-fermion model is identical to that of the toric code, while all nontrivial anyons are fermions. This anyon theory is chiral, since
\begin{equation}
\begin{aligned}
e^{2\pi i c_-/8} = \frac{1}{\mathcal{D}} \sum_{a \in \mathcal{C}} d_a^2\text{} \theta_a = \frac{1}{2}(1 - 1 - 1 - 1) = -1,
\end{aligned}
\end{equation}
where $d_a$ is the {\it quantum dimension} of the anyon $a$ and $\theta_a$ is its {\it topological spin}. Consequently, the chiral central charge of this model is $c_- = 4 \mod 8$. Since the anyon theory of the 3-fermion model is a UMTC with non-zero chiral central charge, no commuting projector Hamiltonian realization of the 3-fermion model can exist in a purely 2D lattice system. However, by regarding it as the boundary of a 3D Walker-Wang model, such a 3D commuting projector Hamiltonian can be defined.

We study the decoherence of the fermion $\ff_1$ by considering the density matrix resulting from the incoherent proliferation of $\ff_1$. Although the 3-fermion topological order has no commuting projector realization in 2D, we can carry out the decoherence according to the commutative diagram in Sec.~\ref{sec:gc}. First, we lift the 3-fermion anyon theory to 3D using the Walker-Wang construction and derive the Walker-Wang Hamiltonian using the topological action~\cite{chen2023higher,li2023measuring,li2024domain}, which we review in Appendix~\ref{app:ta}. This formulation allows the entangling circuits and stabilizers to be constructed systematically from the action. The projections discussed above translate directly into constraints on the gauge fields, and the resulting action identifies the phase after projection.

The action for the 3-fermion Walker-Wang model is
\begin{align}
\mathcal{L}[A_1, A_2] = \frac{1}{2}(A_1\cup A_1 + A_2\cup A_2 + A_1\cup A_2 ),
\end{align}
where $A_1$ and $A_2$ are 2-form $\mathbb{Z}_2$ gauge fields, which are cocycles $A_i \in $Ker$\delta^2$. Therefore, the Hilbert space is given by two qubits per edge (of the dual lattice), and the stabilizers are given by two terms per vertex and two per plaquette, as shown below:
\begin{align*}
    B_{v,i} = 
    \begin{tikzpicture}[scale= 0.8, baseline={([yshift=-2.5pt]current bounding box.center)}]
    %
        \coordinate (O) at (0,0,0);
        \coordinate (A) at (-1.5,0,0);
        \coordinate (B) at (1.5,0,0);
        \coordinate (C) at (0,-1.5,0);
        \coordinate (D) at (0,1.5,0);
        \coordinate (E) at (0,0,1.5);
        \coordinate (F) at (0,0,-1.5);
        %
        \draw (A) -- (B);
        \draw (C) -- (D); 
        \draw (E) -- (F);
        \foreach \X/\Y in { O/E, O/F}
        \node[inner sep=0.5pt, fill=white] at ($( \X )!0.5!( \Y )$) {\footnotesize $Z_i$};
        \foreach \X/\Y in {O/C, O/D}
        \node[inner sep=0.5pt, fill=white] at ($( \X )!0.5!( \Y )$) {\footnotesize $Z_i$};
        \foreach \X/\Y in {O/A, O/B}
        \node[inner sep=0.5pt, fill=white] at ($( \X )!0.5!( \Y )$) {\footnotesize $Z_i$}; 
    \end{tikzpicture}, \quad
    i = 1,2
\end{align*}
\begin{align*}
    K_{p,1}=
    \begin{tikzpicture}[scale= 0.75, baseline={([yshift=-2.5pt]current bounding box.center)}]
            \coordinate (A) at (0,0,0);
            \coordinate (B) at (0,0,2);
            \coordinate (C) at (0,2,2);
            \coordinate (D) at (0,2,0);
%
            \coordinate (U) at (-2,0,0);
            \coordinate (O) at (2,2,2);
%
%
            %
            \draw (A) -- (B) --(C) -- (D) -- cycle;
            \draw [dotted] (A) -- (U);
            \draw (C) -- (O);
            \foreach \X/\Y in { A/B, B/C, C/D, D/A}
            \node[inner sep=0.5pt, fill=white] at ($( \X )!0.5!( \Y )$) {\footnotesize $X_1$}; 
            \node[inner sep=0.5pt, fill=white] at (-1.4, 0, 0) {\footnotesize $Z_1$};
            \node[inner sep=0.5pt, fill=white] at (1.4, 2, 2) {\footnotesize $Z_1Z_2$};
    \end{tikzpicture},
    \begin{tikzpicture}[scale= 0.75, baseline={([yshift=-2.5pt]current bounding box.center)}]
%
            \coordinate (A) at (0,0,0);
            \coordinate (B) at (0,2,0);
            \coordinate (C) at (2,2,0);
            \coordinate (D) at (2,0,0);
%
            \coordinate (U) at (0,0,-2);
            \coordinate (O) at (2,2,2);
%
%
  %
            \draw (A) -- (B) --(C) -- (D) -- cycle;
            \draw [dotted] (A) -- (U);
            \draw (C) -- (O);
            \foreach \X/\Y in { A/B, B/C, C/D, D/A}
            \node[inner sep=0.5pt, fill=white] at ($( \X )!0.5!( \Y )$) {\footnotesize $X_1$}; 
            \node[inner sep=0.5pt, fill=white] at (0, 0, -1.4) {\footnotesize $Z_1$};
            \node[inner sep=0.5pt, fill=white] at (2, 2, 1.4) {\footnotesize $Z_1Z_2$};
    \end{tikzpicture},
    \begin{tikzpicture}[scale= 0.75, baseline={([yshift=-2.5pt]current bounding box.center)}]
        \coordinate (A) at (0,0,0);
        \coordinate (B) at (2,0,0);
        \coordinate (C) at (2,0,2);
        \coordinate (D) at (0,0,2);
%
        \coordinate (O) at (2,2,2);
        \coordinate (U) at (0,-2,0);
%
%
        \draw  (A) -- (B) -- (C) -- (D) -- cycle;
        \draw [dotted] (A) -- (U);
        \draw (C) -- (O);
%
 %
        \foreach \X/\Y in { A/B, B/C, C/D, D/A}
        \node[inner sep=0.5pt, fill=white] at ($( \X )!0.5!( \Y )$) {\footnotesize $X_1$};
        \node[inner sep=0.5pt, fill=white] at (2,1.4,2) {\footnotesize $Z_1Z_2$};
        \node[inner sep=0.5pt, fill=white] at (0,-1.4,0) {\footnotesize $Z_1$};        
    \end{tikzpicture}.
\end{align*}
\begin{align*}
    K_{p,2} =
    \begin{tikzpicture}[scale= 0.75, baseline={([yshift=-2.5pt]current bounding box.center)}]
            \coordinate (A) at (0,0,0);
            \coordinate (B) at (0,0,2);
            \coordinate (C) at (0,2,2);
            \coordinate (D) at (0,2,0);
%
            \coordinate (U) at (-2,0,0);
            \coordinate (O) at (2,2,2);
%
%
            %
            \draw (A) -- (B) --(C) -- (D) -- cycle;
            \draw [dotted] (A) -- (U);
            \draw (C) -- (O);
            \foreach \X/\Y in { A/B, B/C, C/D, D/A}
            \node[inner sep=0.5pt, fill=white] at ($( \X )!0.5!( \Y )$) {\footnotesize $X_2$}; 
            \node[inner sep=0.5pt, fill=white] at (-1.4, 0, 0) {\footnotesize $Z_1Z_2$};
            \node[inner sep=0.5pt, fill=white] at (1.4, 2, 2) {\footnotesize $Z_2$};
    \end{tikzpicture},
    \begin{tikzpicture}[scale= 0.75, baseline={([yshift=-2.5pt]current bounding box.center)}]
%
            \coordinate (A) at (0,0,0);
            \coordinate (B) at (0,2,0);
            \coordinate (C) at (2,2,0);
            \coordinate (D) at (2,0,0);
%
            \coordinate (U) at (0,0,-2);
            \coordinate (O) at (2,2,2);
%
%
  %
            \draw (A) -- (B) --(C) -- (D) -- cycle;
            \draw [dotted] (A) -- (U);
            \draw (C) -- (O);
            \foreach \X/\Y in { A/B, B/C, C/D, D/A}
            \node[inner sep=0.5pt, fill=white] at ($( \X )!0.5!( \Y )$) {\footnotesize $X_2$}; 
            \node[inner sep=0.5pt, fill=white] at (0, 0, -1.4) {\footnotesize $Z_1Z_2$};
            \node[inner sep=0.5pt, fill=white] at (2, 2, 1.4) {\footnotesize $Z_2$};
    \end{tikzpicture},
    \begin{tikzpicture}[scale= 0.75, baseline={([yshift=-2.5pt]current bounding box.center)}]
        \coordinate (A) at (0,0,0);
        \coordinate (B) at (2,0,0);
        \coordinate (C) at (2,0,2);
        \coordinate (D) at (0,0,2);
%
        \coordinate (O) at (2,2,2);
        \coordinate (U) at (0,-2,0);
%
%
        \draw  (A) -- (B) -- (C) -- (D) -- cycle;
        \draw [dotted] (A) -- (U);
        \draw (C) -- (O);
%
 %
        \foreach \X/\Y in { A/B, B/C, C/D, D/A}
        \node[inner sep=0.5pt, fill=white] at ($( \X )!0.5!( \Y )$) {\footnotesize $X_2$};
        \node[inner sep=0.5pt, fill=white] at (2,1.4,2) {\footnotesize $Z_2$};
        \node[inner sep=0.5pt, fill=white] at (0,-1.4,0) {\footnotesize $Z_1Z_2$};        
    \end{tikzpicture}.
\end{align*}
The ground state of the 3-fermion Walker-Wang model is a weighted superposition of fermion loops. Since the three fermions braid nontrivially with one another, all of them are confined in the bulk. We choose the smooth boundary condition for the 3-fermion Walker-Wang model, which hosts the 3-fermion anyon theory. The boundary Hamiltonian stabilizers are truncations of the bulk vertex stabilizers and the plaquette stabilizers in the $x-y$ plane~\cite{vonkeyserlingk2013three}. On the smooth boundary, the vertical and horizontal hopping operators for $\ff_1$ are given by
\begin{equation}
\begin{aligned}
\begin{tikzpicture}[scale= 0.8, baseline={([yshift=-2.5pt]current bounding box.center)}]
\def\L{2}
\draw (0,0) -- (\L,0);
\draw (\L, -\L) -- (\L, 0);
\draw (\L*0.5,0) node[inner sep=0.5pt, fill=white] {\footnotesize $Z_1$};
\draw (\L, -\L*0.5) node[inner sep=0.5pt, fill=white] {\footnotesize $X_1$};
\end{tikzpicture}, \quad
\begin{tikzpicture}[scale= 0.8, baseline={([yshift=-2.5pt]current bounding box.center)}]
\def\L{2}
\draw (0,0) -- (\L,0);
\draw (0, 0) -- (0, \L);
\draw (\L*0.5,0) node[inner sep=0.5pt, fill=white] {\footnotesize $X_1$};
\draw (0, \L*0.5) node[inner sep=0.5pt, fill=white] {\footnotesize $Z_1$};
\end{tikzpicture},
\end{aligned}
\end{equation}
Similarly, the vertical and horizontal hopping operators for $\ff_2$ are given by
\begin{equation}
\begin{aligned}
\begin{tikzpicture}[scale= 0.8, baseline={([yshift=-2.5pt]current bounding box.center)}]
\def\L{2}
\draw (0,0) -- (\L,0);
\draw (\L, -\L) -- (\L, 0);
\draw (\L*0.5,0) node[inner sep=0.5pt, fill=white] {\footnotesize $Z_1Z_2$};
\draw (\L, -\L*0.5) node[inner sep=0.5pt, fill=white] {\footnotesize $X_2$};
\end{tikzpicture}, \quad
\begin{tikzpicture}[scale= 0.8, baseline={([yshift=-2.5pt]current bounding box.center)}]
\def\L{2}
\draw (0,0) -- (\L,0);
\draw (0, 0) -- (0, \L);
\draw (\L*0.5,0) node[inner sep=0.5pt, fill=white] {\footnotesize $X_2$};
\draw (0, \L*0.5) node[inner sep=0.5pt, fill=white] {\footnotesize $Z_1 Z_2$};
\end{tikzpicture},
\end{aligned}
\end{equation}
and their products in each direction give the hopping operators for $f_3$. The anyonic ribbon operators on the bottom boundary are defined analogously.

As in Sec.~\ref{sec:gc}, the first qubit corresponds to the anyon $\ff_1$ and the second qubit corresponds to $\ff_2$. Hence, the four states $\ket{00}$, $\ket{01}$, $\ket{10}$ and $\ket{11}$ respectively correspond to the anyon labels $1$, $\ff_2$, $\ff_1$ and $\ff_3$. Since we are decohering the anyon $\ff_1$, we have $S = \{1, \ff_1\}$. Since the only nontrivial anyon which braids trivially with $\ff_1$ is itself, we have $\mathcal{D} = S$. As a result, we project out all the anyon labels in $\mathcal{C}\backslash \mathcal{D} = \{\ff_2, \ff_3\}$. This can be carried out simply by projecting onto $Z_2 = 1$. We see that $B_{v,2}$ becomes the identity operator, leaving only $B_{v,1}$ as a nontrivial vertex stabilizer. The stabilizers $K_{p,1}$ become the usual plaquette terms of the fermionic toric code (see Eq.~\eqref{eq:ferm_tc_plaq}):
\begin{align*}
    \begin{tikzpicture}[scale= 0.8, baseline={([yshift=-2.5pt]current bounding box.center)}]
            \coordinate (A) at (0,0,0);
            \coordinate (B) at (0,0,2);
            \coordinate (C) at (0,2,2);
            \coordinate (D) at (0,2,0);
%
            \coordinate (U) at (-2,0,0);
            \coordinate (O) at (2,2,2);
%
%
            %
            \draw (A) -- (B) --(C) -- (D) -- cycle;
            \draw [dotted] (A) -- (U);
            \draw (C) -- (O);
            \foreach \X/\Y in { A/B, B/C, C/D, D/A}
            \node[inner sep=0.5pt, fill=white] at ($( \X )!0.5!( \Y )$) {\footnotesize $X_1$}; 
            \node[inner sep=0.5pt, fill=white] at (-1.1, 0, 0) {\footnotesize $Z_1$};
            \node[inner sep=0.5pt, fill=white] at (1.1, 2, 2) {\footnotesize $Z_1$};
    \end{tikzpicture}
    \hspace{15pt}
    \begin{tikzpicture}[scale= 0.8, baseline={([yshift=-2.5pt]current bounding box.center)}]
%
            \coordinate (A) at (0,0,0);
            \coordinate (B) at (0,2,0);
            \coordinate (C) at (2,2,0);
            \coordinate (D) at (2,0,0);
%
            \coordinate (U) at (0,0,-2);
            \coordinate (O) at (2,2,2);
%
%
  %
            \draw (A) -- (B) --(C) -- (D) -- cycle;
            \draw [dotted] (A) -- (U);
            \draw (C) -- (O);
            \foreach \X/\Y in { A/B, B/C, C/D, D/A}
            \node[inner sep=0.5pt, fill=white] at ($( \X )!0.5!( \Y )$) {\footnotesize $X_1$}; 
            \node[inner sep=0.5pt, fill=white] at (0, 0, -1.1) {\footnotesize $Z_1$};
            \node[inner sep=0.5pt, fill=white] at (2, 2, 1.1) {\footnotesize $Z_1$};
    \end{tikzpicture}
    \hspace{15pt}
    \begin{tikzpicture}[scale= 0.8, baseline={([yshift=-2.5pt]current bounding box.center)}]
        \coordinate (A) at (0,0,0);
        \coordinate (B) at (2,0,0);
        \coordinate (C) at (2,0,2);
        \coordinate (D) at (0,0,2);
%
        \coordinate (O) at (2,2,2);
        \coordinate (U) at (0,-2,0);
%
%
        \draw  (A) -- (B) -- (C) -- (D) -- cycle;
        \draw [dotted] (A) -- (U);
        \draw (C) -- (O);
%
 %
        \foreach \X/\Y in { A/B, B/C, C/D, D/A}
        \node[inner sep=0.5pt, fill=white] at ($( \X )!0.5!( \Y )$) {\footnotesize $X_1$};
        \node[inner sep=0.5pt, fill=white] at (2,1.1,2) {\footnotesize $Z_1$};
        \node[inner sep=0.5pt, fill=white] at (0,-1.1,0) {\footnotesize $Z_1$};        
    \end{tikzpicture}
\end{align*}
The stabilizers $K_{p,2}$ are the only ones that do not commute with the projection. To cancel the non-commuting $X_2$ factors, we take their product over the faces of a cube, which gives the vertex stabilizer $B_{v,1}$ after projection. Indeed, we recover the stabilizers of the fermionic toric code. At the level of the topological action, this corresponds to simply setting $A_2 = 0$, giving
\begin{align*}
    \mathcal{L}[A_1, A_2 = 0] = \frac{1}{2} A_1\cup A_1,
\end{align*}
which is precisely the topological action of the fermionic toric code~\cite{tata2023anomalies,chen2023higher}. Since we obtain the fermionic toric code with a smooth boundary, the same partial-trace procedure as in Sec.~\ref{sec:wu} recovers $\rho_{TC}^{\dec,\ff}$.

Since the 3-fermion anyon theory is invariant under permutations of the three fermions, the same result can be obtained by decohering $\ff_2$ or $\ff_3$. If we choose to decohere $\ff_2$, then $\mathcal{D} = \{1, \ff_2\}$, and we need to project out $\mathcal{C}\backslash\mathcal{D} = \{\ff_1, \ff_3\}$. This can be carried out by setting $Z_1 = 1$ at the Hamiltonian level, or $A_1 = 0$ at the Lagrangian level. If we choose to decohere $\ff_3$, then $\mathcal{D} = \{1, \ff_3\}$, and we need to project out $\mathcal{C}\backslash\mathcal{D} = \{\ff_1, \ff_2\}$. This can be carried out by setting $Z_1Z_2 = 1$ at the Hamiltonian level, or $A_1 = A_2$ at the Lagrangian level.

This example further demonstrates that the holographic perspective provides a powerful framework for studying mixed-state topological phases. In particular, even for chiral phases that admit no commuting projector Hamiltonian in 2D, this method yields the decohered density matrix and enables a systematic study of its topological properties.

\subsubsection{Decohering Semion-Fermion Topological Order}\label{sec:examples_sf}

The semion-fermion anyon theory, $\nu=4$ in Table~\ref{tab:sixteenfold}, has $\mathbb{Z}_2\times \mathbb{Z}_2$ fusion rule and is generated by a semion $\s$ and a fermion $\ff$, whose mutual statistics is $-1$. In particular, their fusion produces another semion $\s'$ which braids with $\s$ trivially. Together, they give the semion-fermion anyon theory $\mathcal{C} = \{1, \s, \s', \ff\}$. We consider the decoherence of the fermion $\ff$ in $\mathcal{C}$, which again gives $S = \{1, \ff\}$ and $\mathcal{D} = S$.

The first step is again to lift the theory to the 3D Walker-Wang model. If the input anyon theory is the semion theory $\{1,\s\}$, the topological action of its Walker-Wang model is given by the Pontryagin square~\cite{kapustin2017fermionic, sun2026non}:
\begin{align*}
    \frac{1}{4}P(A) = \frac{1}{4}(A\cup A + A \cup_1 \delta A)
\end{align*}
where $\cup_1$ is the higher cup product measuring the failure of graded-commutativity of $\cup$ on cochains. The Hamiltonian corresponding to $\frac{1}{4}P(A)$ was constructed explicitly in Ref.~\cite{vonkeyserlingk2013three} on the trivalently-resolved cubic lattice. Instead, we derive a commuting projector Hamiltonian on the cubic lattice without resolving its vertices into trivalent junctions. For details of this calculation, we refer the reader to Appendix~\ref{app:ta_s_ww}, and we present the key results here. The Hilbert space of the semion Walker-Wang model has one qubit per edge, and the Hamiltonian consists of one term per vertex and plaquette:
\begin{align}
    H_\s = -\sum_v B_v - \sum_p K_p,
\end{align}
where
\begin{equation}
\begin{aligned}
    B_{v} = 
    \begin{tikzpicture}[scale= 0.8, baseline={([yshift=-2.5pt]current bounding box.center)}]
    %
        \coordinate (O) at (0,0,0);
        \coordinate (A) at (-1.5,0,0);
        \coordinate (B) at (1.5,0,0);
        \coordinate (C) at (0,-1.5,0);
        \coordinate (D) at (0,1.5,0);
        \coordinate (E) at (0,0,1.5);
        \coordinate (F) at (0,0,-1.5);
        %
        \draw (A) -- (B);
        \draw (C) -- (D); 
        \draw (E) -- (F);
        \foreach \X/\Y in { O/E, O/F}
        \node[inner sep=0.5pt, fill=white] at ($( \X )!0.5!( \Y )$) {\footnotesize $Z$};
        \foreach \X/\Y in {O/C, O/D}
        \node[inner sep=0.5pt, fill=white] at ($( \X )!0.5!( \Y )$) {\footnotesize $Z$};
        \foreach \X/\Y in {O/A, O/B}
        \node[inner sep=0.5pt, fill=white] at ($( \X )!0.5!( \Y )$) {\footnotesize $Z$}; 
    \end{tikzpicture}.
\end{aligned}
\end{equation}
The plaquette term $K_p$ is given by the following expressions, depending on the plaquette orientation:
\begin{equation}
\begin{aligned}
    \begin{tikzpicture}[scale = 0.7, baseline = 0.8cm]
%
        \coordinate (A) at (0,0,0);
        \coordinate (B) at (0,2,0);
        \coordinate (C) at (2,2,0);
        \coordinate (D) at (2,0,0);
%
        \coordinate (U) at (0,0,-2);
        \coordinate (O) at (2,2,2);
%
%
%
       %
        \draw [dotted] (A) -- (U);
        \draw (C) -- (O);
        \draw (0,0,0) -- (3.5,0,0);
        \draw (2,-1.5,0) -- (2,0,0);
        \draw (2,0,-1.5) -- (2,0,1.5);
        \draw (0,2,0) -- (-1.5,2,0);
        \draw (0,3.5,0) -- (0,2,0);
        \draw (0,2,-1.5) -- (0,2,1.5);
        \draw [blue] (A) -- (B) --(C) -- (D) -- cycle;
        \node [inner sep=0.5pt, fill=white] at (1, 2, 0) {\footnotesize$X\textcolor{red}{S}$};
        \node [inner sep=0.5pt, fill=white] at (0, 1, 0) {\footnotesize$X\textcolor{red}{S^\dagger}$};
        \node [inner sep=0.5pt, fill=white] at (1, 0, 0) {\footnotesize$X\textcolor{red}{S}$};
        \node [inner sep=0.5pt, fill=white] at (2, 1, 0) {\footnotesize$X\textcolor{red}{S^\dagger}$};
        \node [inner sep=0.5pt, fill=white] at (0, 0, -1) {\footnotesize$\textcolor{red}{S}$};
        \node [inner sep=0.5pt, fill=white] at (2, 2, 1) {\footnotesize$\textcolor{red}{S}$};
        \node [inner sep=0.5pt, fill=white] at (0, 3, 0) {\footnotesize$\textcolor{red}{S^\dagger}$};
        \node [inner sep=0.5pt, fill=white] at (-1, 2, 0) {\footnotesize$\textcolor{red}{S}$};
        \node [inner sep=0.5pt, fill=white] at (0, 2, -1) {\footnotesize$\textcolor{red}{S^\dagger}$};
        \node [inner sep=0.5pt, fill=white] at (0, 2, 1) {\footnotesize$\textcolor{red}{S}$};
        \node [inner sep=0.5pt, fill=white] at (3, 0, 0) {\footnotesize$\textcolor{red}{S}$};
        \node [inner sep=0.5pt, fill=white] at (2, -1, 0) {\footnotesize$\textcolor{red}{S^\dagger}$};
        \node [inner sep=0.5pt, fill=white] at (2, 0, -1) {\footnotesize$\textcolor{red}{S}$};
        \node [inner sep=0.5pt, fill=white] at (2, 0, 1) {\footnotesize$\textcolor{red}{S^\dagger}$};
    \end{tikzpicture}
    &\cdot
    \begin{tikzpicture}[scale = 0.7, baseline = 0.8cm]
%
        \coordinate (A) at (0,0,0);
        \coordinate (B) at (0,2,0);
        \coordinate (C) at (2,2,0);
        \coordinate (D) at (2,0,0);
%
        \coordinate (U) at (0,0,-2);
        \coordinate (O) at (2,2,2);
%
%
%
        %
        \draw [dotted] (A) -- (U);
        \draw (C) -- (O);
        \draw (0,0,0) -- (3.5,0,0);
        \draw (2,-1.5,0) -- (2,0,0);
        \draw (2,0,-1.5) -- (2,0,1.5);
        \draw (0,2,0) -- (-1.5,2,0);
        \draw (0,3.5,0) -- (0,2,0);
        \draw (0,2,-1.5) -- (0,2,1.5);
        \draw [blue] (A) -- (B) --(C) -- (D) -- cycle;
        \node [fill, circle, red, inner sep = 1pt] at (2,2,1) {};
        \node [fill, circle, red, inner sep = 1pt] at (2,1,0) {};
        \node [fill, circle, red, inner sep = 1pt] at (1,2,0) {};
        \node [fill, circle, red, inner sep = 1pt] at (0,0,-1) {};
        \node [fill, circle, red, inner sep = 1pt] at (1,0,0) {};
        \node [fill, circle, red, inner sep = 1pt] at (0,1,0) {};
        \node [fill, circle, red, inner sep = 1pt] at (0,2,-1) {};
        \node [fill, circle, red, inner sep = 1pt] at (0,3,0) {};
        \node [fill, circle, red, inner sep = 1pt] at (0,2,1) {};
        \node [fill, circle, red, inner sep = 1pt] at (-1,2,0) {};
        \draw [red] (2,2,1) -- (2,1,0);
        \draw [red] (1,2,0) -- (2,1,0);
        \draw [red] (0,0,-1) -- (1,0,0);
        \draw [red] (0,1,0) -- (1,0,0);
        \draw [red] (0,2,-1) -- (0,3,0);
        \draw [red] (0,2,1) -- (-1,2,0);
    \end{tikzpicture}\\
    \begin{tikzpicture}[scale = 0.7, baseline = -0.6cm]
        \draw (-1.5, 0, 0) -- (1.5, 0, 0);
        \draw[dotted] (0, -1.5, 0) -- (0, 0, 0);
        \draw (0, 1.5, 0) -- (0, 0, 0);
        \draw (0, 0, -1.5) -- (0, 0, 1.5);

        \draw (-1.5, 0, 2) -- (1.5, 0, 2);
        \draw (0, -1.5, 2) -- (0, 1.5, 2);
        \draw (0, 0, 0.5) -- (0, 0, 3.5);

        \draw (0.5, 0, 2) -- (3.5, 0, 2);
        \draw (2, -1.5, 2) -- (2, 1.5, 2);
        \draw (2, 0, 0.5) -- (2, 0, 3.5);

        \draw [blue] (0,0,0) -- (2,0,0) -- (2,0,2) -- (0,0,2) -- cycle;

        \node [inner sep=0.5pt, fill=white] at (1,0,0) {\footnotesize$X\textcolor{red}{Z}$};
        \node [inner sep=0.5pt, fill=white] at (0,0,1) {\footnotesize$X\textcolor{red}{S}$};
        \node [inner sep=0.5pt, fill=white] at (1,0,2) {\footnotesize$X$};
        \node [inner sep=0.5pt, fill=white] at (2,0,1) {\footnotesize$X\textcolor{red}{S^\dagger}$};

        \node [inner sep=0.5pt, fill=white] at (0,-1,0) {\footnotesize$\textcolor{red}{S}$};
        \node [inner sep=0.5pt, fill=white] at (-1,0,2) {\footnotesize$\textcolor{red}{S}$};
        \node [inner sep=0.5pt, fill=white] at (0,0,3) {\footnotesize$\textcolor{red}{S}$};
        \node [inner sep=0.5pt, fill=white] at (0,1,2) {\footnotesize$\textcolor{red}{S}$};
        \node [inner sep=0.5pt, fill=white] at (0,-1,2) {\footnotesize$\textcolor{red}{S}$};

        \node [inner sep=0.5pt, fill=white] at (2,0,3) {\footnotesize$\textcolor{red}{S^\dagger}$};
        \node [inner sep=0.5pt, fill=white] at (3,0,2) {\footnotesize$\textcolor{red}{S}$};
        \node [inner sep=0.5pt, fill=white] at (2,-1,2) {\footnotesize$\textcolor{red}{S^\dagger}$};
    \end{tikzpicture}
    \hspace{5pt}
    &\cdot 
    \hspace{5pt}
    \begin{tikzpicture}[scale = 0.7, baseline = -0.6cm]
        \draw (-1.5, 0, 0) -- (1.5, 0, 0);
        \draw[dotted] (0, -1.5, 0) -- (0, 0, 0);
        \draw (0, 1.5, 0) -- (0, 0, 0);
        \draw (0, 0, -1.5) -- (0, 0, 1.5);

        \draw (-1.5, 0, 2) -- (1.5, 0, 2);
        \draw (0, -1.5, 2) -- (0, 1.5, 2);
        \draw (0, 0, 0.5) -- (0, 0, 3.5);

        \draw (0.5, 0, 2) -- (3.5, 0, 2);
        \draw (2, -1.5, 2) -- (2, 1.5, 2);
        \draw (2, 0, 0.5) -- (2, 0, 3.5);

        \draw [blue] (0,0,0) -- (2,0,0) -- (2,0,2) -- (0,0,2) -- cycle;

        \node [fill, circle, red, inner sep = 1pt] at (0,1,0) {};
        \node [fill, circle, red, inner sep = 1pt] at (1,0,0) {};
        \node [fill, circle, red, inner sep = 1pt] at (0,0,-1) {};
        \node [fill, circle, red, inner sep = 1pt] at (-1,0,0) {};
        \node [fill, circle, red, inner sep = 1pt] at (0,0,1) {};
        \node [fill, circle, red, inner sep = 1pt] at (-1,0,2) {};
        \node [fill, circle, red, inner sep = 1pt] at (0,0,3) {};
        \node [fill, circle, red, inner sep = 1pt] at (0,-1,2) {};
        \node [fill, circle, red, inner sep = 1pt] at (2,0,1) {};
        \node [fill, circle, red, inner sep = 1pt] at (2,1,2) {};

        \draw [red] (0,0,-1) -- (1,0,0);
        \draw [red] (0,1,0) -- (1,0,0);
        \draw [red] (2,0,1) -- (1,0,0);

        \draw [red] (-1,0,0) -- (0,0,1);
        
        \draw [red] (-1,0,2) -- (0,-1,2);
        \draw [red] (-1,0,2) -- (0,0,3);
        \draw [red] (0,-1,2) -- (0,0,3);

        \draw [red] (2,1,2) -- (2,0,1);
    \end{tikzpicture}\\
    \begin{tikzpicture}[scale = 0.7, baseline = 0.6cm]
        \draw [dotted] (-1.5, 0, 0) -- (0, 0, 0);
        \draw (1.5, 0, 0) -- (0, 0, 0);
        \draw (0, -1.5, 0) -- (0, 0, 0);
        \draw (0, 1.5, 0) -- (0, 0, 0);
        \draw (0, 0, -1.5) -- (0, 0, 1.5);

        \draw (-1.5, 2, 0) -- (1.5, 2, 0);
        \draw (0, 0.5, 0) -- (0, 3.5, 0);
        \draw (0, 2, -1.5) -- (0, 2, 1.5);

        \draw (-1.5, 2, 2) -- (1.5, 2, 2);
        \draw (0, 0.5, 2) -- (0, 3.5, 2);
        \draw (0, 2, 0.5) -- (0, 2, 3.5);

        \draw [blue] (0,0,0) -- (0,2,0) -- (0,2,2) -- (0,0,2) -- cycle;
        
        \node [inner sep=0.5pt, fill=white] at (0,1,0) {\footnotesize$X$};
        \node [inner sep=0.5pt, fill=white] at (0,0,1) {\footnotesize$X\textcolor{red}{S^\dagger}$};
        \node [inner sep=0.5pt, fill=white] at (0,1,2) {\footnotesize$X\textcolor{red}{S^\dagger}$};
        \node [inner sep=0.5pt, fill=white] at (0,2,1) {\footnotesize$X$};

        \node [inner sep=0.5pt, fill=white] at (0,0,-1) {\footnotesize$\textcolor{red}{S^\dagger}$};
        \node [inner sep=0.5pt, fill=white] at (1,0,0) {\footnotesize$\textcolor{red}{S^\dagger}$};
        \node [inner sep=0.5pt, fill=white] at (0,-1,0) {\footnotesize$\textcolor{red}{S}$};
        
        \node [inner sep=0.5pt, fill=white] at (0,3,0) {\footnotesize$\textcolor{red}{S}$};
        \node [inner sep=0.5pt, fill=white] at (0,2,-1) {\footnotesize$\textcolor{red}{S}$};
        \node [inner sep=0.5pt, fill=white] at (1,2,0) {\footnotesize$\textcolor{red}{S}$};
        \node [inner sep=0.5pt, fill=white] at (-1,2,0) {\footnotesize$\textcolor{red}{S}$};
        
        \node [inner sep=0.5pt, fill=white] at (0,3,2) {\footnotesize$\textcolor{red}{S^\dagger}$};
        \node [inner sep=0.5pt, fill=white] at (-1,2,2) {\footnotesize$\textcolor{red}{S}$};
        \node [inner sep=0.5pt, fill=white] at (0,2,3) {\footnotesize$\textcolor{red}{S}$};
    \end{tikzpicture}
    \hspace{5pt}
    &\cdot
    \hspace{5pt}
    \begin{tikzpicture}[scale = 0.7, baseline = 0.6cm]
        \draw [dotted] (-1.5, 0, 0) -- (0, 0, 0);
        \draw (1.5, 0, 0) -- (0, 0, 0);
        \draw (0, -1.5, 0) -- (0, 0, 0);
        \draw (0, 1.5, 0) -- (0, 0, 0);
        \draw (0, 0, -1.5) -- (0, 0, 1.5);

        \draw (-1.5, 2, 0) -- (1.5, 2, 0);
        \draw (0, 0.5, 0) -- (0, 3.5, 0);
        \draw (0, 2, -1.5) -- (0, 2, 1.5);

        \draw (-1.5, 2, 2) -- (1.5, 2, 2);
        \draw (0, 0.5, 2) -- (0, 3.5, 2);
        \draw (0, 2, 0.5) -- (0, 2, 3.5);

        \draw [blue] (0,0,0) -- (0,2,0) -- (0,2,2) -- (0,0,2) -- cycle;

        \node [fill, circle, red, inner sep = 1pt] at (0,3,0) {};
        \node [fill, circle, red, inner sep = 1pt] at (0,2,-1) {};
        \node [fill, circle, red, inner sep = 1pt] at (1,2,0) {};
        \node [fill, circle, red, inner sep = 1pt] at (-1,0,0) {};
        \node [fill, circle, red, inner sep = 1pt] at (0,0,1) {};
        \node [fill, circle, red, inner sep = 1pt] at (0,1,2) {};
        \node [fill, circle, red, inner sep = 1pt] at (0,2,1) {};
        \node [fill, circle, red, inner sep = 1pt] at (1,2,2) {};
        \node [fill, circle, red, inner sep = 1pt] at (0,3,2) {};
        \node [fill, circle, red, inner sep = 1pt] at (-1,2,2) {};
        \node [fill, circle, red, inner sep = 1pt] at (0,2,3) {};
        
        \draw [red] (0,3,0) -- (0,2,-1);
        \draw [red] (1,2,0) -- (0,2,-1);
        \draw [red] (1,2,0) -- (0,3,0);

        \draw [red] (-1,0,0) -- (0,0,1);
        \draw [red] (0,1,2) -- (0,0,1);

        \draw [red] (0,2,1) -- (1,2,2);
        \draw [red] (0,3,2) -- (1,2,2);

        \draw [red] (-1,2,2) -- (0,2,3);
    \end{tikzpicture} \label{eq:semion_plaquette}
    \end{aligned}
\end{equation}
where the plaquette $p$ is highlighted in blue, and
$\begin{tikzpicture}[scale = 0.5, baseline = -0.1cm]
    \node [fill, circle, red, inner sep = 1pt] at (0,0) {};
    \node [fill, circle, red, inner sep = 1pt] at (1,0) {};
    \draw [red] (0,0) -- (1,0);
\end{tikzpicture}$
is the CZ gate acting on the two qubits indicated by the pair of red dots. A similar expression is obtained in Ref.~\cite{sun2026non} by condensation in the $\mathbb{Z}_4$ theory $\frac{1}{4} B\cup B$, with $B$ a $\mathbb{Z}_4$ gauge field. Note that these operators themselves are not hermitian. However, their hermitian conjugate are equivalent to the original operators up to multiplication by the vertex star terms.

Now we consider the Walker-Wang model of $\mathcal{C}$, which we present using two choices of generators. First, since $\mathcal{C}$ can be taken to be generated by $\s$ and $\s'$ with trivial mutual braiding, the topological action of the Walker-Wang model of $\mathcal{C}$ is
\begin{align}
    \mathcal{L}_{\s\ff}[A_1, A_2] = \frac{1}{4}(P(A_1) + P(A_2)).
\end{align}
In this case, the Hilbert space is given by two qubits per edge, and the Hamiltonian is simply given by two identical copies of the $B_v$ and $K_p$ terms
\begin{align}
    H_{\s\ff} = -\sum_v (B_{v,1} + B_{v,2}) - \sum_p (K_{p,1} + K_{p,2}),
\end{align}
where $B_{v,i}$ and $K_{p,i}$ act only on the $i$-th qubit on each edge for $i = 1,2$. The 3D transition is given by keeping only the anyons in $\mathcal{D} = \{1,\ff\}$, which in this case can be carried out by projecting onto the $Z_1Z_2 = 1$ subspace. As in the transition of the RBH cluster state in Sec.~\ref{sec:wu}, this identifies
\begin{gather*}
    X_1X_2 \mapsto X,
    \hspace{15pt}
    Z_1 = Z_2 \mapsto Z,
    \hspace{15pt}
    S_1 = S_2 \mapsto S,\\
    \\
    \begin{tikzpicture}[scale = 0.5, baseline = -0.1cm]
    \node [fill, circle, red, inner sep = 1pt] at (0,0) {};
    \node [fill, circle, red, inner sep = 1pt] at (1,0) {};
    \draw [red] (0,0) -- (1,0);
    \end{tikzpicture}_1
    =
    \begin{tikzpicture}[scale = 0.5, baseline = -0.1cm]
    \node [fill, circle, red, inner sep = 1pt] at (0,0) {};
    \node [fill, circle, red, inner sep = 1pt] at (1,0) {};
    \draw [red] (0,0) -- (1,0);
    \end{tikzpicture}_2
    \mapsto
    \begin{tikzpicture}[scale = 0.5, baseline = -0.1cm]
    \node [fill, circle, red, inner sep = 1pt] at (0,0) {};
    \node [fill, circle, red, inner sep = 1pt] at (1,0) {};
    \draw [red] (0,0) -- (1,0);
\end{tikzpicture},
\end{gather*}
where 
$\begin{tikzpicture}[scale = 0.5, baseline = -0.1cm]
    \node [fill, circle, red, inner sep = 1pt] at (0,0) {};
    \node [fill, circle, red, inner sep = 1pt] at (1,0) {};
    \draw [red] (0,0) -- (1,0);
    \end{tikzpicture}_i$
is the CZ gate acting on the $i$-th qubits on the corresponding edges. Following the above identifications, the two vertex terms become:
\begin{align*}
    \begin{tikzpicture}[scale= 0.8, baseline={([yshift=-2.5pt]current bounding box.center)}]
    %
        \coordinate (O) at (0,0,0);
        \coordinate (A) at (-1.5,0,0);
        \coordinate (B) at (1.5,0,0);
        \coordinate (C) at (0,-1.5,0);
        \coordinate (D) at (0,1.5,0);
        \coordinate (E) at (0,0,1.5);
        \coordinate (F) at (0,0,-1.5);
        %
        \draw (A) -- (B);
        \draw (C) -- (D); 
        \draw (E) -- (F);
        \foreach \X/\Y in { O/E, O/F}
        \node[inner sep=0.5pt, fill=white] at ($( \X )!0.5!( \Y )$) {\footnotesize $Z_i$};
        \foreach \X/\Y in {O/C, O/D}
        \node[inner sep=0.5pt, fill=white] at ($( \X )!0.5!( \Y )$) {\footnotesize $Z_i$};
        \foreach \X/\Y in {O/A, O/B}
        \node[inner sep=0.5pt, fill=white] at ($( \X )!0.5!( \Y )$) {\footnotesize $Z_i$}; 
    \end{tikzpicture}
    \hspace{10pt}
    \mapsto
    \hspace{10pt}
    \begin{tikzpicture}[scale= 0.8, baseline={([yshift=-2.5pt]current bounding box.center)}]
    %
        \coordinate (O) at (0,0,0);
        \coordinate (A) at (-1.5,0,0);
        \coordinate (B) at (1.5,0,0);
        \coordinate (C) at (0,-1.5,0);
        \coordinate (D) at (0,1.5,0);
        \coordinate (E) at (0,0,1.5);
        \coordinate (F) at (0,0,-1.5);
        %
        \draw (A) -- (B);
        \draw (C) -- (D); 
        \draw (E) -- (F);
        \foreach \X/\Y in { O/E, O/F}
        \node[inner sep=0.5pt, fill=white] at ($( \X )!0.5!( \Y )$) {\footnotesize $Z$};
        \foreach \X/\Y in {O/C, O/D}
        \node[inner sep=0.5pt, fill=white] at ($( \X )!0.5!( \Y )$) {\footnotesize $Z$};
        \foreach \X/\Y in {O/A, O/B}
        \node[inner sep=0.5pt, fill=white] at ($( \X )!0.5!( \Y )$) {\footnotesize $Z$}; 
    \end{tikzpicture}, \quad
    i = 1,2
\end{align*}
The plaquette terms again do not commute with the projection. As in Sec.~\ref{sec:wu}, we can either multiply the plaquette terms within each copy around a cube, or we can multiply the two plaquette terms together $K_{p,1}K_{p,2}$. Both products commute with the projection. As shown in Appendix~\ref{app:ta}, a product of six plaquette terms around a cube is simply a product of vertex terms and therefore does not give new stabilizers. On the other hand,
since CZ squares to $1$, while $S$ and $S^\dagger$ each square to $Z$, the product $K_{p,1}K_{p,2}$ becomes (depending on the orientation of $p$)
\begin{gather*}
    \begin{tikzpicture}[scale = 0.8, baseline = 0.8cm]
%
        \coordinate (A) at (0,0,0);
        \coordinate (B) at (0,2,0);
        \coordinate (C) at (2,2,0);
        \coordinate (D) at (2,0,0);
%
        \coordinate (U) at (0,0,-2);
        \coordinate (O) at (2,2,2);
%
%
%
       %
        \draw [dotted] (A) -- (U);
        \draw (C) -- (O);
        \draw (0,0,0) -- (3.5,0,0);
        \draw (2,-1.5,0) -- (2,0,0);
        \draw (2,0,-1.5) -- (2,0,1.5);
        \draw (0,2,0) -- (-1.5,2,0);
        \draw (0,3.5,0) -- (0,2,0);
        \draw (0,2,-1.5) -- (0,2,1.5);
        \draw [blue] (A) -- (B) --(C) -- (D) -- cycle;
        \node [inner sep=0.5pt, fill=white] at (1, 2, 0) {\footnotesize$X\textcolor{red}{Z}$};
        \node [inner sep=0.5pt, fill=white] at (0, 1, 0) {\footnotesize$X\textcolor{red}{Z}$};
        \node [inner sep=0.5pt, fill=white] at (1, 0, 0) {\footnotesize$X\textcolor{red}{Z}$};
        \node [inner sep=0.5pt, fill=white] at (2, 1, 0) {\footnotesize$X\textcolor{red}{Z}$};
        \node [inner sep=0.5pt, fill=white] at (0, 0, -1) {\footnotesize$\textcolor{red}{Z}$};
        \node [inner sep=0.5pt, fill=white] at (2, 2, 1) {\footnotesize$\textcolor{red}{Z}$};
        \node [inner sep=0.5pt, fill=white] at (0, 3, 0) {\footnotesize$\textcolor{red}{Z}$};
        \node [inner sep=0.5pt, fill=white] at (-1, 2, 0) {\footnotesize$\textcolor{red}{Z}$};
        \node [inner sep=0.5pt, fill=white] at (0, 2, -1) {\footnotesize$\textcolor{red}{Z}$};
        \node [inner sep=0.5pt, fill=white] at (0, 2, 1) {\footnotesize$\textcolor{red}{Z}$};
        \node [inner sep=0.5pt, fill=white] at (3, 0, 0) {\footnotesize$\textcolor{red}{Z}$};
        \node [inner sep=0.5pt, fill=white] at (2, -1, 0) {\footnotesize$\textcolor{red}{Z}$};
        \node [inner sep=0.5pt, fill=white] at (2, 0, -1) {\footnotesize$\textcolor{red}{Z}$};
        \node [inner sep=0.5pt, fill=white] at (2, 0, 1) {\footnotesize$\textcolor{red}{Z}$};
    \end{tikzpicture}
    \hspace{5pt}
    \begin{tikzpicture}[scale = 0.8, baseline = -0.6cm]
        \draw (-1.5, 0, 0) -- (1.5, 0, 0);
        \draw[dotted] (0, -1.5, 0) -- (0, 0, 0);
        \draw (0, 1.5, 0) -- (0, 0, 0);
        \draw (0, 0, -1.5) -- (0, 0, 1.5);
%
        \draw (-1.5, 0, 2) -- (1.5, 0, 2);
        \draw (0, -1.5, 2) -- (0, 1.5, 2);
        \draw (0, 0, 0.5) -- (0, 0, 3.5);
%
        \draw (0.5, 0, 2) -- (3.5, 0, 2);
        \draw (2, -1.5, 2) -- (2, 1.5, 2);
        \draw (2, 0, 0.5) -- (2, 0, 3.5);
%
        \draw [blue] (0,0,0) -- (2,0,0) -- (2,0,2) -- (0,0,2) -- cycle;
        \node [inner sep=0.5pt, fill=white] at (1,0,0) {\footnotesize$X$};
        \node [inner sep=0.5pt, fill=white] at (0,0,1) {\footnotesize$X\textcolor{red}{Z}$};
        \node [inner sep=0.5pt, fill=white] at (1,0,2) {\footnotesize$X$};
        \node [inner sep=0.5pt, fill=white] at (2,0,1) {\footnotesize$X\textcolor{red}{Z}$};
        \node [inner sep=0.5pt, fill=white] at (0,-1,0) {\footnotesize$\textcolor{red}{Z}$};
        \node [inner sep=0.5pt, fill=white] at (-1,0,2) {\footnotesize$\textcolor{red}{Z}$};
        \node [inner sep=0.5pt, fill=white] at (0,0,3) {\footnotesize$\textcolor{red}{Z}$};
        \node [inner sep=0.5pt, fill=white] at (0,1,2) {\footnotesize$\textcolor{red}{Z}$};
        \node [inner sep=0.5pt, fill=white] at (0,-1,2) {\footnotesize$\textcolor{red}{Z}$};
        \node [inner sep=0.5pt, fill=white] at (2,0,3) {\footnotesize$\textcolor{red}{Z}$};
        \node [inner sep=0.5pt, fill=white] at (3,0,2) {\footnotesize$\textcolor{red}{Z}$};
        \node [inner sep=0.5pt, fill=white] at (2,-1,2) {\footnotesize$\textcolor{red}{Z}$};
    \end{tikzpicture}\\
    \begin{tikzpicture}[scale = 0.8, baseline = 0.6cm]
        \draw [dotted] (-1.5, 0, 0) -- (0, 0, 0);
        \draw (1.5, 0, 0) -- (0, 0, 0);
        \draw (0, -1.5, 0) -- (0, 0, 0);
        \draw (0, 1.5, 0) -- (0, 0, 0);
        \draw (0, 0, -1.5) -- (0, 0, 1.5);
%
        \draw (-1.5, 2, 0) -- (1.5, 2, 0);
        \draw (0, 0.5, 0) -- (0, 3.5, 0);
        \draw (0, 2, -1.5) -- (0, 2, 1.5);
%
        \draw (-1.5, 2, 2) -- (1.5, 2, 2);
        \draw (0, 0.5, 2) -- (0, 3.5, 2);
        \draw (0, 2, 0.5) -- (0, 2, 3.5);
%
        \draw [blue] (0,0,0) -- (0,2,0) -- (0,2,2) -- (0,0,2) -- cycle;
        \node [inner sep=0.5pt, fill=white] at (0,1,0) {\footnotesize$X$};
        \node [inner sep=0.5pt, fill=white] at (0,0,1) {\footnotesize$X\textcolor{red}{Z}$};
        \node [inner sep=0.5pt, fill=white] at (0,1,2) {\footnotesize$X\textcolor{red}{Z}$};
        \node [inner sep=0.5pt, fill=white] at (0,2,1) {\footnotesize$X$};
        \node [inner sep=0.5pt, fill=white] at (0,0,-1) {\footnotesize$\textcolor{red}{Z}$};
        \node [inner sep=0.5pt, fill=white] at (1,0,0) {\footnotesize$\textcolor{red}{Z}$};
        \node [inner sep=0.5pt, fill=white] at (0,-1,0) {\footnotesize$\textcolor{red}{Z}$};
        \node [inner sep=0.5pt, fill=white] at (0,3,0) {\footnotesize$\textcolor{red}{Z}$};
        \node [inner sep=0.5pt, fill=white] at (0,2,-1) {\footnotesize$\textcolor{red}{Z}$};
        \node [inner sep=0.5pt, fill=white] at (1,2,0) {\footnotesize$\textcolor{red}{Z}$};
        \node [inner sep=0.5pt, fill=white] at (-1,2,0) {\footnotesize$\textcolor{red}{Z}$};
        \node [inner sep=0.5pt, fill=white] at (0,3,2) {\footnotesize$\textcolor{red}{Z}$};
        \node [inner sep=0.5pt, fill=white] at (-1,2,2) {\footnotesize$\textcolor{red}{Z}$};
        \node [inner sep=0.5pt, fill=white] at (0,2,3) {\footnotesize$\textcolor{red}{Z}$};
    \end{tikzpicture}
\end{gather*}
Up to multiplication by the vertex terms, these recover the plaquette terms of the fermionic toric code in Eq.~\eqref{eq:ferm_tc_plaq}. At the level of the topological action, this corresponds to setting $A_1 = A_2 = A$ as discussed in Appendix~\ref{app:ta_lm}, giving
\begin{align}
    \mathcal{L}_{\s\ff}[A_1 = A_2 = A] = \frac{1}{4} (2 P(A)) = \frac{1}{2} A\cup A,
\end{align}
which is the topological action of the fermionic toric code.

Alternatively, we can take $\mathcal{C}$ to be generated by $\s$ and $\ff$. In this case, after the basis transformation $A_1 \to A_1 + A_2$, the topological action takes the form
\begin{align}
    \mathcal{L}_{\s\ff}'[A_1, A_2] = \frac{1}{4}P(A_1) + \frac{1}{2}(A_1\cup A_2 + A_2 \cup A_2).
\end{align}
The corresponding Hamiltonian is
\begin{align}
    H_{\s\ff}' = -\sum_{v} (B_{v,1} + B_{v,2}) - \sum_{p}(K_{p,\s} + K_{p, \ff}), 
\end{align}
where the vertex terms $B_{v,i}$ are the usual six-body terms for $i = 1,2$. The plaquette terms $K_{p,\s}$ are given by
\begin{align*}
\hspace{-0.3cm}
K_{\textcolor{blue}{p},1}
\cdot
    \begin{tikzpicture}[scale= 0.7, baseline={([yshift=-2.5pt]current bounding box.center)}]
            \coordinate (A) at (0,0,0);
            \coordinate (B) at (0,0,2);
            \coordinate (C) at (0,2,2);
            \coordinate (D) at (0,2,0);
%
            \coordinate (U) at (-2,0,0);
            \coordinate (O) at (2,2,2);
%
%
            %
            \draw [blue] (A) -- (B) --(C) -- (D) -- cycle;
            \draw (C) -- (O);
            \node[inner sep=0.5pt, fill=white] at (1.3, 2, 2) {\footnotesize $Z_2$};
    \end{tikzpicture}\
    ,
    K_{\textcolor{blue}{p},1}
\cdot\begin{tikzpicture}[scale= 0.7, baseline={([yshift=-2.5pt]current bounding box.center)}]
%
            \coordinate (A) at (0,0,0);
            \coordinate (B) at (0,2,0);
            \coordinate (C) at (2,2,0);
            \coordinate (D) at (2,0,0);
%
            \coordinate (U) at (0,0,-2);
            \coordinate (O) at (2,2,2);
%
%
  %
            \draw [blue](A) -- (B) --(C) -- (D) -- cycle;
            \draw (C) -- (O);
            \node[inner sep=0.5pt, fill=white] at (2, 2, 1.1) {\footnotesize $Z_2$};
    \end{tikzpicture}\
    ,
    K_{\textcolor{blue}{p},1}
\cdot\begin{tikzpicture}[scale= 0.7, baseline={([yshift=-2.5pt]current bounding box.center)}]
        \coordinate (A) at (0,0,0);
        \coordinate (B) at (2,0,0);
        \coordinate (C) at (2,0,2);
        \coordinate (D) at (0,0,2);
%
        \coordinate (O) at (2,2,2);
        \coordinate (U) at (0,-2,0);
%
%
        \draw [blue] (A) -- (B) -- (C) -- (D) -- cycle;
        \draw (C) -- (O);
        \node[inner sep=0.5pt, fill=white] at (2,1.3,2) {\footnotesize $Z_2$};        
    \end{tikzpicture},
\end{align*}
where the blue plaquette $p$ is highlighted in the figure, and $K_{p,1}$ is given by Eq.~\eqref{eq:semion_plaquette}. The plaquette terms $K_{p, \ff}$ are given by
\begin{align*}
    \begin{tikzpicture}[scale= 0.8, baseline={([yshift=-2.5pt]current bounding box.center)}]
            \coordinate (A) at (0,0,0);
            \coordinate (B) at (0,0,2);
            \coordinate (C) at (0,2,2);
            \coordinate (D) at (0,2,0);
%
            \coordinate (U) at (-2,0,0);
            \coordinate (O) at (2,2,2);
%
%
            %
            \draw (A) -- (B) --(C) -- (D) -- cycle;
            \draw [dotted] (A) -- (U);
            \draw (C) -- (O);
            \foreach \X/\Y in { A/B, B/C, C/D, D/A}
            \node[inner sep=0.5pt, fill=white] at ($( \X )!0.5!( \Y )$) {\footnotesize $X_2$}; 
            \node[inner sep=0.5pt, fill=white] at (-1.4, 0, 0) {\footnotesize $Z_1Z_2$};
            \node[inner sep=0.5pt, fill=white] at (1.4, 2, 2) {\footnotesize $Z_2$};
    \end{tikzpicture}
    \hspace{15pt}
    \begin{tikzpicture}[scale= 0.8, baseline={([yshift=-2.5pt]current bounding box.center)}]
%
            \coordinate (A) at (0,0,0);
            \coordinate (B) at (0,2,0);
            \coordinate (C) at (2,2,0);
            \coordinate (D) at (2,0,0);
%
            \coordinate (U) at (0,0,-2);
            \coordinate (O) at (2,2,2);
%
%
  %
            \draw (A) -- (B) --(C) -- (D) -- cycle;
            \draw [dotted] (A) -- (U);
            \draw (C) -- (O);
            \foreach \X/\Y in { A/B, B/C, C/D, D/A}
            \node[inner sep=0.5pt, fill=white] at ($( \X )!0.5!( \Y )$) {\footnotesize $X_2$}; 
            \node[inner sep=0.5pt, fill=white] at (0, 0, -1.4) {\footnotesize $Z_1Z_2$};
            \node[inner sep=0.5pt, fill=white] at (2, 2, 1.4) {\footnotesize $Z_2$};
    \end{tikzpicture}
    \hspace{15pt}
    \begin{tikzpicture}[scale= 0.8, baseline={([yshift=-2.5pt]current bounding box.center)}]
        \coordinate (A) at (0,0,0);
        \coordinate (B) at (2,0,0);
        \coordinate (C) at (2,0,2);
        \coordinate (D) at (0,0,2);
%
        \coordinate (O) at (2,2,2);
        \coordinate (U) at (0,-2,0);
%
%
        \draw  (A) -- (B) -- (C) -- (D) -- cycle;
        \draw [dotted] (A) -- (U);
        \draw (C) -- (O);
%
 %
        \foreach \X/\Y in { A/B, B/C, C/D, D/A}
        \node[inner sep=0.5pt, fill=white] at ($( \X )!0.5!( \Y )$) {\footnotesize $X_2$};
        \node[inner sep=0.5pt, fill=white] at (2,1.4,2) {\footnotesize $Z_2$};
        \node[inner sep=0.5pt, fill=white] at (0,-1.4,0) {\footnotesize $Z_1Z_2$};        
    \end{tikzpicture}
\end{align*}
In this case, the transition given by projecting onto $\mathcal{D} = \{1, \ff\}$ is carried out simply by projecting onto the $Z_1 = 1$ subspace. Under the projection, $B_{v,1}$ becomes the identity, leaving only $B_{v,2}$ as a nontrivial vertex term. The plaquette terms $K_{p,\s}$ must be multiplied around a cube in order to commute with the projection. The resulting operator is the product of one $B_{v,2}$ with several $B_{v,1}$ terms, which in turn reduces to $B_{v,2}$ after the projection. Finally, $B_{p,\ff}$ becomes the fermionic toric code plaquette term in Eq.~\eqref{eq:ferm_tc_plaq}. At the Lagrangian level, this projection is given by $A_1 = 0$, which results in
\begin{align}
    \mathcal{L}_{\s\ff}'[A_1 = 0, A_2] = \frac{1}{2}A_2 \cup A_2,
\end{align}
which indeed gives the fermionic toric code.

In both cases, we see explicitly at both the Hamiltonian and Lagrangian levels that the bulk of the Walker-Wang model of $\mathcal{C}$ transitions to the fermionic toric code. Since the transition preserves smooth boundaries, the same partial-trace procedure as in Sec.~\ref{sec:wu} yields the density matrix $\rho_{TC}^{\dec, \ff}$.

We note that in addition to the semion-fermion anyon theory, $\nu=12$ in Table~\ref{tab:sixteenfold}, there is an anti-semion-fermion anyon theory with anyons $\{1, \bar{\s}, \bar{\s}', \ff\}$. We can obtain the explicit Hamiltonian of this theory by exchanging $S\leftrightarrow S^\dagger$ throughout the semion-fermion Hamiltonian. It follows that if we decohere $\mathcal{D} = \{1,\ff\}$, the same calculations apply, and we obtain the density matrix $\rho_{TC}^{\dec, \ff}$ as well.

\subsubsection{Decohering $\mathbb{Z}_4$-Type Topological Order}

After describing all the $\mathbb{Z}_2\times \mathbb{Z}_2$ cases in the sixteenfold way, in this section we consider the remaining four abelian anyon theories, $\nu=2,6,10,14$ in Table~\ref{tab:sixteenfold}. These theories have $\mathbb{Z}_4$ fusion rules and are generated by an anyon $a$ with topological spin $e^{k\pi i/4}$ for $k$ an odd integer mod $8$. The anyon $a^2$ is a fermion, so the anyon content is $\mathcal{C} = \{ 1, a, a^2 = \ff, a^3\}$. The topological action is given by
\begin{align}
\mathcal{L}_k[A] = \frac{k}{8} P(A),
\hspace{25pt}
k = 1,3,5,7,
\end{align}
where $A$ is a $\mathbb{Z}_4$ 2-form gauge field. We consider the decoherence of $S = \{1,\ff\}$, leading to $\mathcal{D} = S$.

The derivation of the Hamiltonian of the Walker-Wang model is very similar to the semionic case in Appendix~\ref{app:ta_s_ww}. However, the calculation quickly becomes unwieldy, since the entangler lies one level higher in the Clifford hierarchy. An easier case-specific method is to condense the boson in the $\mathbb{Z}_8$ theory $\frac{k}{8} B\cup B$, with $B$ a $\mathbb{Z}_8$ 2-form gauge field, similar to the construction in~\cite{sun2026non}. Here, instead of presenting the explicit Hamiltonian, we only work with the Lagrangian.

To carry out the restriction to $\mathcal{D}$, we first decompose $A$ into two $\mathbb{Z}_2$ 2-cochains $A_1$ and $A_2$ by $A = A_1 + 2A_2$, subject to the original flatness constraint. The projection sets $A_1 = 0$, so $A_2$ must be closed in $\mathbb{Z}_2$. Substituting this into the Lagrangian, the following equality holds for all four cases:
\begin{align}
\mathcal{L}[A = 2A_2] = \frac{1}{2} A_2 \cup A_2,
\end{align}
with $A_2$ a $\mathbb{Z}_2$ 2-form gauge field, giving the fermionic toric code. Applying the same partial-trace procedure as in Sec.~\ref{sec:wu}, we again obtain the density matrix $\rho_{TC}^{\dec, \ff}$.

\subsubsection{Decohering Ising-Type Topological Order}\label{sec:examples_ising}

In this section, we consider the Ising-type anyon theories, $\nu=\text{odd}$ in Table~\ref{tab:sixteenfold}. These are nonabelian anyon theories with three anyon types $\mathcal{C} = \{1, \ff, \sigma\}$, where $\ff$ is a fermion whose mutual braiding with $\sigma$ has phase $-1$, and $\sigma$ has topological spin $e^{k\pi i/8}$ for $k$ being an odd integer mod $16$. The fusion rules are given by
\begin{equation}
    \begin{aligned}
        &\ff \times \ff = 1,\\
        &\ff \times \sigma = \sigma,\\
        & \sigma \times \sigma = 1+\ff.
    \end{aligned}
\end{equation}
We consider decohering the fermion $\ff$ in $\mathcal{C}$, which corresponds to the choice $S = \{1,\ff\}$ and $\mathcal{D} = S$.

\begin{figure}
\centering
\includegraphics[width=0.5\linewidth]{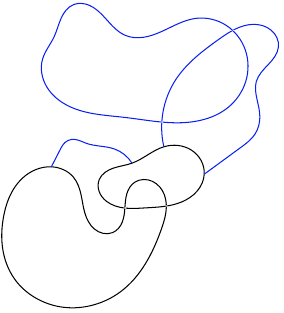}
\caption{A generic string-net state of the Ising Walker-Wang model. The black lines are labeled by $\sigma$, which must form loops. The blue lines are labeled by $\ff$, which can either form loops or end on $\sigma$-loops.}
\label{fig:ising_WW}
\end{figure}

First, we lift the anyon theory $\mathcal{C}$ to 3D using the Walker-Wang construction. Formally, the Hamiltonian is given by
\begin{align}
H = -\sum_{v}B_v - \frac{1}{4}\sum_p(1 + K_p^\ff + \sqrt{2}K_p^\sigma),
\end{align}
where $B_v$ enforces the fusion rules at each vertex, whereas $K_p^\ff$ and $K_p^\sigma$ respectively fuse an $\ff$ loop and a $\sigma$ loop into the plaquette $p$. However, it is easy to describe the ground state wavefunction. Due to the fusion rules, a string-net state in $\mathcal{C}$ is given by a collection of $\ff$-loops and $\sigma$-loops, where points on the $\sigma$-loops may be connected by $\ff$-lines~\cite{ellison2025toward}. A generic string-net state is given in Fig.~\ref{fig:ising_WW}. A ground state wavefunction is given by a superposition of admissible string-nets. Within a chosen ground-state sector, their relative coefficients are determined by the $R$-matrix and $F$-symbols together with the loop evaluation rules~\cite{walker20123+,vonkeyserlingk2013three}.

Next, the transition of the Ising Walker-Wang model is given by keeping only the anyon labels in $\mathcal{D}$ and annihilating all string-net states with labels in $S\backslash\mathcal{D} = \{\sigma\}$. This projection can be carried out using the $\mathbb{Z}_2$ grading of the Ising-type anyon theories~\cite{ellison2025toward}. Specifically, we can define the operator $\mu^Z$ acting on the local basis $\{\ket{1}, \ket{\ff}, \ket{\sigma}\}$ by
\begin{align}
\mu^Z\ket{1} = \ket{1},
\hspace{15pt}
\mu^Z\ket{\ff} = \ket{\ff},
\hspace{15pt}
\mu^Z\ket{\sigma} = -\ket{\sigma}.
\end{align}
It follows that the projection is carried out by $\mu^Z = 1$ on every edge. After applying the projection to the ground state, the wavefunction contains only $\ff$-loops, with exchange phase $R_1^{\ff\ff} = -1$. This is precisely the ground state of the fermionic toric code. Since the transition preserves smooth boundaries, the same partial-trace procedure as in Sec.~\ref{sec:wu} again gives the density matrix $\rho_{TC}^{\dec,\ff}$.

\subsection{Decohering Double Semion Topological Order}

In this section, we consider the double semion anyon theory, which has $\mathbb{Z}_2 \times \mathbb{Z}_2$ fusion rules and is generated by a semion $\s$ and an anti-semion $\bar{\s}$. Moreover, $\s$ and $\bar{\s}$ have trivial mutual braiding, so $\s\bar{\s}$ is a boson. The anyon content is thus $\mathcal{C} = \{1, \s, \bar{\s}, \s\bar{\s}\}$. We consider the decoherence of the semion, giving $S = \{1, \s\}$ and $\mathcal{D} = \{1, \bar{\s}\}$. Since $\s$ and $\bar{\s}$ braid trivially, we expect the decoherence of $\s$ to preserve the $\bar{\s}$ sector described by $\{1, \bar{\s}\}$~\cite{sohal2025noisy}.

First, we lift $\mathcal{C}$ to the 3D Walker-Wang model. Taking $\s$ and $\bar{\s}$ as generators, the Lagrangian is
\begin{align*}
\mathcal{L}_{\s\bar{\s}}[A_1, A_2] = \frac{1}{4}P(A_1) + \frac{3}{4}P(A_2),
\end{align*}
and the corresponding Hilbert space is spanned by two qubits per edge, with the Hamiltonian,
\begin{align*}
H_{\s\bar{\s}} = -\sum_{\text{\st{b}} \textcolor{myBlue}{v}} (B_{v,1} + B_{v,2}) - \sum_p (K_{p,1} + \bar{K}_{p,2}),
\end{align*}
where the subscripts indicate which set of qubits the operator acts on, and $\bar{K}_{p}$'s are the plaquette terms of the anti-semion Walker-Wang model, given by exchanging $S$ and $S^\dagger$ in $K_p$. The projection onto $\mathcal{D} = \{1, \bar{\s}\}$ is given by enforcing $Z_1 = 1$ at the Hamiltonian level, or $A_1 = 0$ at the Lagrangian level. In each case, we are left with the anti-semion Walker-Wang model. Since the anti-semion anyon theory is modular, the bulk of the Walker-Wang model is confined, leaving the anyon theory $\{1, \bar{\s}\}$ at the top smooth boundary. As a result, after tracing out the bottom half of the system, we expect to obtain a density matrix that is in the same phase as a properly defined pure anti-semion ground state, as both states share the same strong anti-semionic 1-form symmetry.

\section{Conclusion and Outlook}\label{sec:conclusion}

In this work, we studied 2D decoherence phase transitions from a 3D holographic perspective. Decoherence in a chiral topological order $\mathcal{C}$ is hard to study because such a phase does not admit a commuting projector Hamiltonian in 2D. One way is to consider the Drinfeld center $\mathcal{Z}(\mathcal{C})$ of $\mathcal{C}$, which admits a commuting projector Hamiltonian through the Levin-Wen construction~\cite{lw05}. It is known that decoherence in $\mathcal{C}$ corresponds to condensation in $\mathcal{Z}(\mathcal{C})$ under vectorization of the density matrix~\cite{bao2026mixed,sohal2025noisy}. After the condensation, one needs to undo the vectorization, with $\bar{{\mathcal{C}}}$ representing the bra sector, to obtain a density matrix. Although the process of condensation can be carried out with a commuting projector Hamiltonian, it is unclear how to identify the required microscopic ket and bra factors in a generic Levin-Wen realization of $\mathcal{Z}(\mathcal{C})$.

In the present work, we conjecture a holographic description of decoherence in $\mathcal{C}$ by first lifting the theory to a 3D Walker-Wang model, then performing a phase transition and tracing out the bottom half of the system. Each step can be carried out explicitly within a lattice framework based on commuting projector Hamiltonians. This construction is related to condensation in $\mathcal{Z}(\mathcal{C})$. Indeed, for a modular anyon theory $\mathcal{C}$, the bulk of the Walker-Wang model is confined, and the top smooth boundary and bottom smooth boundary support the anyon theories $\mathcal{C}$ and $\bar{\mathcal{C}}$, respectively. By considering the Walker-Wang model on a thin slab, we recover the $\mathcal{Z}(\mathcal{C})$ topological order. We motivated this procedure using two decoherence channels in the toric code. For the bosonic decoherence, the transition is from the RBH cluster state to the 3D toric code, and tracing out the bulk yields the SWSSB state. For the fermionic decoherence, the transition is from the RBH cluster state to the 3D fermionic toric code, and tracing out the bulk yields the intrinsically mixed-state topological order. We applied the procedure to fermionic decoherence in Kitaev's 16-fold way, and explicitly demonstrated the transitions from various Walker-Wang models to the 3D fermionic toric code, at the Hamiltonian, Lagrangian, or wavefunction level. In all cases, we recover the same intrinsically mixed-state topological order.

There are several avenues for future work. First, although we identified the transition in 3D which corresponds to a 2D decoherence phase transition, it is not clear how the two phase transitions are precisely related. It would be interesting to investigate whether the two critical points are related and, if so, how. For instance, can we infer the universal behavior of one critical point from the other?

Secondly, the phase transition defined in Sec.~\ref{sec:gc} may seem transparent, as we are simply restricting the Levin-Wen or Walker-Wang model from a category $\mathcal{C}$ to a subcategory. However, if the restriction is not onto a subcategory, the resulting phase is not immediately determined by the retained labels. The simplest example is given by taking $\mathcal{C}$ as the Ising anyon theory $\{1, \ff, \sigma\}$ as in Sec.~\ref{sec:examples_ising}, where $\sigma$ has spin $\theta_\sigma = e^{\pi i/8}$ and the nonabelian fusion rule
\begin{align}
\sigma\times \sigma = 1+\ff.
\end{align}
Consider taking $\mathcal{D} = \{1, \sigma\}$, which is not a subcategory, since $\mathcal{D}$ is clearly not closed under fusion. Then the projection on the Ising Levin-Wen or Walker-Wang ground state is given by annihilating all the string-nets with labels in $\mathcal{C}\backslash \mathcal{D} = \{\ff\}$. Under the projection, the fusion channel $\sigma \times \sigma \mapsto \ff$ is no longer allowed at a vertex, and we are only left with the fusion channel $\sigma\times\sigma \mapsto 1$. We see that the presence or absence of $\sigma$ defines a $\mathbb{Z}_2$ variable, and the ground state after the projection becomes a superposition of $\mathbb{Z}_2$ loops. In Figure~\ref{fig:ising_WW} this projection corresponds to deleting all the blue lines, and retaining only the black loops. One may ask whether the phase after the transition is a topological phase built from the Vec$_{\mathbb{Z}_2}$ fusion category. In the 2D Levin-Wen model, this requires an analysis of the projected loop amplitudes. However, in the 3D Walker-Wang model, braiding is involved. Interestingly, the original spin of $\sigma$ is $e^{2\pi i/16}$, which precludes interpreting the surviving label as a $\mathbb{Z}_2$ anyon, whose topological spin can only be a $4$-th root of unity. Nevertheless, the wavefunction after the transition is clearly well-defined as a superspoition of $\mathbb{Z}_2$ loops. However, since this question is not the central topic of the paper, we do not pursue this direction further, and leave the investigation to future work. 

In general, such a failure of the naive anyon interpretation can arise when the anyon labels we project onto are not closed under fusion. An open question is what phase such a wavefunction belongs to, and what local Hamiltonian, if any, it is a ground state of. Moreover, recall that projecting the RBH ground state can be used to perform gauging dualities. It would be interesting to investigate whether the projections on Walker-Wang models defined here can be used to generalize gauging dualities in a nonabelian setting.

\begin{acknowledgments}
    We thank Yu-An Chen, Yabo Li, Ruochen Ma, and Nathanan Tantivasadakarn for insightful discussions. This work was partially supported by the U.S. National Science Foundation under Grant No. PHY 2310614 (T.-C.W. and S.Z.), Grant No.~DRL I-TEST 2148467 (Z.S.), and Grant No.~DMR-2339319 (Z.B.) T.-C.W. and S.Z. acknowledge the support from Stony Brook University's Center for Distributed Quantum Processing. L.H.L. and Z.B. acknowledge a startup fund and a Quantum SuperSEED fund (ICDS\_QS25\_029093) from the Institute for Computational and Data Sciences at the Pennsylvania State University. 
\end{acknowledgments}

\bibliography{biblio}

\onecolumngrid
\clearpage


\appendix

\section{An introduction to the Walker-Wang model}\label{app:ww}

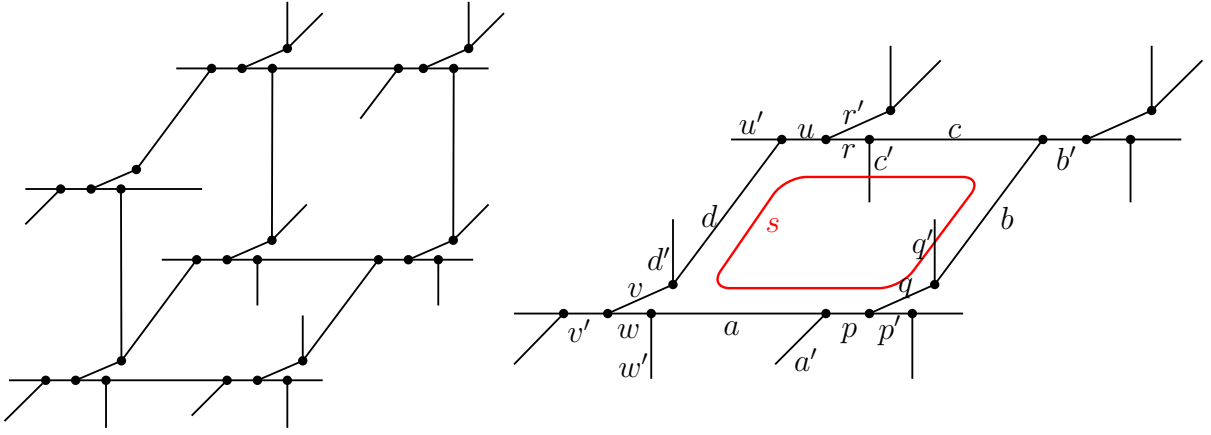
\begin{figure}
    \centering
    \begin{tikzpicture}[scale= 0.6,baseline={([yshift=-2.5pt]current bounding box.center)},
    x=0.01cm,
    y=-0.01cm,
    bond/.style={line width=0.7pt},
    site/.style={circle, fill=black, inner sep=1.25pt}
]

\draw[bond] (0,835) -- (692,835);
\draw[bond] (36,413) -- (426,413);
\draw[bond] (337,569) -- (1024,569);
\draw[bond] (369,147) -- (1058,147);

\coordinate (b1) at (81,835);
\coordinate (b2) at (147,835);
\coordinate (b3) at (214,835);
\coordinate (b4) at (481,835);
\coordinate (b5) at (548,835);
\coordinate (b6) at (614,835);

\coordinate (l1) at (114,413);
\coordinate (l2) at (181,413);
\coordinate (l3) at (247,413);

\coordinate (m1) at (414,569);
\coordinate (m2) at (481,569);
\coordinate (m3) at (548,569);
\coordinate (m4) at (815,569);
\coordinate (m5) at (881,569);
\coordinate (m6) at (947,569);

\coordinate (t1) at (447,147);
\coordinate (t2) at (514,147);
\coordinate (t3) at (581,147);
\coordinate (t4) at (859,147);
\coordinate (t5) at (914,147);
\coordinate (t6) at (981,147);

\coordinate (pL)  at (248,792);
\coordinate (qL)  at (281,370);
\coordinate (pM)  at (580,526);
\coordinate (qM)  at (648,792);
\coordinate (uM)  at (614,103);
\coordinate (pR)  at (980,526);
\coordinate (uR)  at (1014,103);

\draw[bond] (-10,925) -- (b1);
\draw[bond] (b3) -- (214,940);
\draw[bond] (b4) -- (403,913);
\draw[bond] (b6) -- (614,940);

\draw[bond] (b2) -- (pL);
\draw[bond] (pL) -- (l3);

\draw[bond] (b5) -- (qM);
\draw[bond] (qM) -- (648,692);
\draw[bond] (qM) -- (m4);

\draw[bond] (l1) -- (36,491);
\draw[bond] (l2) -- (qL);
\draw[bond] (qL) -- (t1);

\draw[bond] (m2) -- (pM);
\draw[bond] (m3) -- (548,670);
\draw[bond] (t3) -- (pM);
\draw[bond] (pM) -- (657,447);

\draw[bond] (m5) -- (pR);
\draw[bond] (m6) -- (947,670);
\draw[bond] (t6) -- (pR);
\draw[bond] (pR) -- (1058,447);

\draw[bond] (t2) -- (uM);
\draw[bond] (uM) -- (614,0);
\draw[bond] (uM) -- (692,25);

\draw[bond] (t4) -- (775,258);
\draw[bond] (t5) -- (uR);
\draw[bond] (uR) -- (1014,0);
\draw[bond] (uR) -- (1092,25);

\draw[bond] (pL) -- (m1);

\foreach \P in {b1,b2,b3,b4,b5,b6,l1,l2,l3,m1,m2,m3,m4,m5,m6,t1,t2,t3,t4,t5,t6,pL,qL,pM,qM,uM,pR,uR}
{
    \node[site] at (\P) {};
}

\end{tikzpicture}
\begin{tikzpicture}[scale= 0.8,baseline={([yshift=-2.5pt]current bounding box.center)},
    x=0.01cm,
    y=-0.01cm,
    bond/.style={line width=0.7pt},
    site/.style={circle, fill=black, inner sep=1.25pt},
    lab/.style={font=\large}
]


\draw[bond] (4,447) -- (747,447);
\draw[bond] (363,159) -- (1108,159);


\draw[bond] (4,531) -- (86,447);
\draw[bond] (159,447) -- (267,399);
\draw[bond] (267,399) -- (447,159);
\draw[bond] (267,399) -- (267,291);
\draw[bond] (231,447) -- (231,555);


\draw[bond] (520,447) -- (436,531);
\draw[bond] (592,447) -- (700,399);
\draw[bond] (700,399) -- (879,159);
\draw[bond] (663,447) -- (663,555);


\draw[bond] (520,159) -- (627,111);
\draw[bond] (627,111) -- (627,4);
\draw[bond] (627,111) -- (710,28);
\draw[bond] (592,159) -- (592,263);


\draw[bond] (951,159) -- (1059,111);
\draw[bond] (1059,111) -- (1059,4);
\draw[bond] (1059,111) -- (1143,28);
\draw[bond] (1024,159) -- (1024,263);


\draw[red, line width=0.9pt, rounded corners=8pt]
    (326,405) -- (643,405) -- (781,221) -- (453,221) -- cycle;

\draw[bond] (700,399) -- (700,291);

\node[site] at (86,447) {};
\node[site] at (159,447) {};
\node[site] at (231,447) {};
\node[site] at (267,399) {};
\node[site] at (447,159) {};
\node[site] at (520,159) {};
\node[site] at (520,447) {};
\node[site] at (592,159) {};
\node[site] at (592,447) {};
\node[site] at (627,111) {};
\node[site] at (663,447) {};
\node[site] at (700,399) {};
\node[site] at (879,159) {};
\node[site] at (951,159) {};
\node[site] at (1024,159) {};
\node[site] at (1059,111) {};


\node[lab] at (112,475) {$v'$};
\node[lab] at (194,476) {$w$};
\node[lab] at (202,532) {$w'$};
\node[lab] at (205,408) {$v$};
\node[lab] at (244,356) {$d'$};
\node[lab] at (327,285) {$d$};

\node[lab] at (397,131) {$u'$};
\node[lab] at (487,145) {$u$};
\node[lab] at (566,114) {$r'$};
\node[lab] at (558,178) {$r$};

\node[lab] at (364,471) {$a$};
\node[lab] at (487,523) {$a'$};
\node[lab] at (559,480) {$p$};
\node[lab] at (626,475) {$p'$};
\node[lab] at (652,405) {$q$};
\node[lab] at (681,332) {$q'$};

\node[lab] at (616,190) {$c'$};
\node[lab] at (733,146) {$c$};
\node[lab] at (820,288) {$b$};
\node[lab] at (918,190) {$b'$};

\node[lab] at (432,301) {\textcolor{red}{$s$}};

\end{tikzpicture}
    \caption{(Left) The lattice structure of the Walker-Wang model. Each vertex is trivalent, and a local Hilbert space spanned by the simple objects of the input category is associated with each edge. Each plaquette has 10 boundary edges, together with 10 outgoing edges attached to its 10 vertices. (Right) Inserting an $s$ loop on the plaquette.}
    \label{fig:WW_lattice}
\end{figure}

In this section we review the Walker-Wang model~\cite{walker20123+}. Conventionally, the Walker-Wang model is a $(3{+}1)$D commuting projector Hamiltonian, defined on the cubic lattice in which each vertex is resolved into a trivalent junction as shown on the left of Fig.~\ref{fig:WW_lattice}. The Walker-Wang construction takes a braided fusion category $\mathcal{C}$ as input, which is defined by a set of simple objects $\mathcal{C} = \{a,b,c, \cdots\}$, a set of $F$-symbols $F^{abc}_{d,ef}$, and $R$-matrices $R^{ab}_c$ which satisfy the pentagon and hexagon equations. For a detailed review of braided fusion category, we refer the reader to Ref.~\cite{wang2010topological, simon2023topological} for more detailed discussions. For simplicity, we assume all simple objects are self-dual, so we do not need to orient the edges of the lattice. The Hilbert space of the Walker-Wang model is formed by a $|\mathcal{C}|$-dimensional qudit per edge, where $|\mathcal{C}|$ is the cardinality of simple objects, forming a tensor product Hilbert space. Thus a generic state is given via labeling each edge by a simple object in $\mathcal{C}$. After choosing a projection of the 3D lattice onto 2D, the Hamiltonian consists of two types of terms
\begin{equation}
H_{\mathcal{C}} \;=\; -\sum_{v} B_{v} \;-\; \sum_{p} K_{p},
\end{equation}
The vertex term $B_{v}$ is defined by enforcing fusion compatibility at the trivalent vertex $v$. That is, $B_v$ projects onto the subspace where the labels $a$, $b$ and $c$ on the three edges of $v$ satisfy $N^{c}_{ab}=1$, i.e., that $c$ appears as a fusion product in $a\otimes b$. Since the objects are self-dual, reversing an edge does not change its label. Restricting to the subspace with $B_v = 1$ gives the space of string nets in three dimensions, analogous to the string-net space of the Levin-Wen model~\cite{lw05}. Such a state is given by a superposition of anyon loops with three-point junctions respecting the fusion rules. On the other hand, the plaquette term $K_{p}$ is built from loop-insertion operators. First, consider the operator $K_p^s$, which is defined by inserting a loop of anyon $s$ into the plaquette $p$, as shown on the right of Fig.~\ref{fig:WW_lattice}, followed by fusing the $s$-loop into the lattice. Thus, we can express the matrix elements of $K_p^s$ in terms of $F$-symbols and $R$-matrices
\begin{gather}
    K_p^s |\Psi_{a, \ldots, w}\rangle = \sum_{a'', \ldots, w'' \in \mathcal{C}} K_{p, a'', \ldots, w''}^s\, |\Psi_{a'', \ldots, w''}\rangle,\\
    K_{p, a'', \ldots, w''}^s = R^{q'b}_q \overline{R^{c'r}_{c}} \overline{R^{q'b''}_{q''}} R^{c'r''}_{c''} F^{a''sp}_{a';ap''} F^{p''sq}_{p';pq''} F^{q''sb}_{q';qb''} F^{b''sc}_{b';bc''} F^{c''sr}_{c';cr''} F^{r''su}_{r';ru''} F^{u''sd}_{u';ud''} F^{d''sv}_{d';dv''} F^{v''sw}_{v';vw''} F^{w''sa}_{w';wa''}.
\end{gather}
A detailed derivation of this formula can be found in Ref.~\cite{walker20123+}. In terms of the loop insertion operators $K_p^s$, the plaquette terms are given by
\begin{equation}
K_{p} = \frac{1}{\mathcal{D}^{2}}\sum_{a\in\mathcal{C}} d_{s} K_{p}^{s},
\end{equation}
where $d_s$ is the quantum dimension of $s$, and $\mathcal{D}$ is the total qunautm dimension $\mathcal{D}^2 = \sum_{s\in\mathcal{C}} d_s^2$.

\begin{figure}
    \begin{tikzpicture}[scale= 0.8,baseline={([yshift=-2.5pt]current bounding box.center)},
    x=0.01cm,
    y=-0.01cm,
    bond/.style={line width=0.7pt},
    site/.style={circle, fill=black, inner sep=1.25pt},
    lab/.style={font=\large}
]


\draw[bond] (4,447) -- (747,447);
\draw[bond] (363,159) -- (1108,159);


\draw[bond] (4,531) -- (86,447);
\draw[bond] (159,447) -- (447,159);
\draw[bond] (231,447) -- (231,555);


\draw[bond] (520,447) -- (436,531);
\draw[bond] (592,447) -- (879,159);
\draw[bond] (663,447) -- (663,555);


\draw[bond] (520,159) -- (627,111);
\draw[bond] (592,159) -- (592,263);


\draw[bond] (951,159) -- (1059,111);
\draw[bond] (1024,159) -- (1024,263);


\draw[red, line width=0.9pt, rounded corners=8pt]
    (246,405) -- (573,405) -- (791,191) -- (463,191) -- cycle;


\node[site] at (86,447) {};
\node[site] at (159,447) {};
\node[site] at (231,447) {};
\node[site] at (447,159) {};
\node[site] at (520,159) {};
\node[site] at (520,447) {};
\node[site] at (592,159) {};
\node[site] at (592,447) {};
\node[site] at (663,447) {};
\node[site] at (879,159) {};
\node[site] at (951,159) {};
\node[site] at (1024,159) {};






\node[lab] at (402,301) {\textcolor{red}{$s$}};

\end{tikzpicture}
    \caption{The smooth boundary of the WW model. The boundary terminates in the vertical direction, hence all edges pointing up from the boundary are removed. The fusion of the anyon loop $s$ into the lattice defines the boundary plaquette operators.}
    \label{fig:WW_lattice_boundary}
\end{figure}
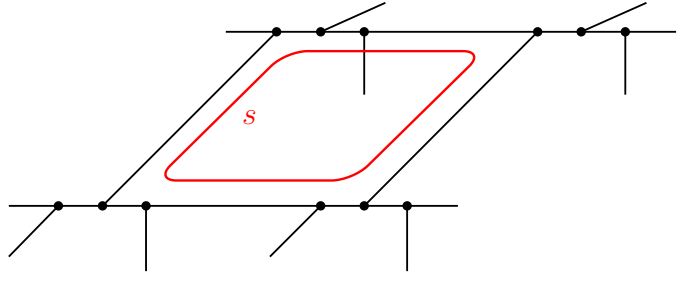

If the input category $\mathcal{C}$ is modular, that is for each anyon $a$ in $ \mathcal{C}$, there exists another anyon which braids non-trivially with $a$, then the Walker-Wang model $H_\mathcal{C}$ is in an invertible phase in 3D, and has a unique ground state on a manifold without boundary. However, on a manifold with boundary, we can choose a boundary condition such that the boundary hosts the 2D topological order $\mathcal{C}$. Such a boundary is the smooth boundary~\cite{vonkeyserlingk2013three}. To define the smooth boundary, we remove all edges sticking out from the boundary of the cubic lattice and resolve the vertices into trivalent junctions as usual. An example is shown in Fig.~\ref{fig:WW_lattice_boundary}, where we choose the boundary to be perpendicular to the vertical direction. The boundary terms in the Hamiltonian are defined in the same way as the bulk terms. That is, the vertex term enforces fusion, while the boundary plaquette terms are given by superposing an anyon loop as in Fig.~\ref{fig:WW_lattice_boundary} and fusing the loop into the lattice. It follows that logical operators supported strictly near the boundary can be defined to obey the fusion rules of $\mathcal{C}$. Thus, the smooth boundary hosts the topological order $\mathcal{C}$. For example, on the solid donut $S^1\times D^2$, the ground-state degeneracy of the Walker-Wang model is precisely $|\mathcal{C}|$.

\section{Topological Action and Hamiltonian}\label{app:ta}

In this appendix we review the topological action formalism~\cite{tata2023anomalies,tantivasadakarn2024long,chen2023higher,li2023measuring,li2024domain} of various topological phases, and how to derive their corresponding Hamiltonians systematically following Ref.~\cite{chen2023higher}. 

Let $\mathcal{L}[A_i]$ be the Lagrangian of an abelian topological phase. Consider a $d+1$-dimensional spacetime manifold $M$ equipped with a simplicial or CW structure on which the theory is defined. Each $A_i$ is an $n_i$-form gauge field valued in an abelian group $G_i$ and is an $n_i$-cocycle~\cite{w17} on $M$. That is, each $A_i$ assigns a group element in $G_i$ to every $n_i$-simplex or $n_i$-cell and satisfies $\delta^{n_i}(A_i) = 0$, where $\delta^{n_i}$ is the simplicial or cellular coboundary map. The partition function on $M$ is
\begin{align}
    Z = e^{2\pi i S[A_i]},
    \hspace{35pt}
    S[A_i] = \int_M \mathcal{L}[A_i],
\end{align}
where the topological action $S$ is given by integration over the spacetime manifold $M$. To find the corresponding Hamiltonian, we first ungauge the theory by replacing $A_i \mapsto \delta a_i$ for all $i$~\cite{chen2023higher}, where $a_i$ is an ($n_i-1$)-cochain on $M$. After ungauging, the Lagrangian becomes a total derivative
\begin{align}
    S[\delta a_i] = \int_M \mathcal{L}[\delta a_i] = \int_M \delta\mathcal{U}[a_i] = \int_{\partial M} \mathcal{U}[a_i].
\end{align}
On the boundary $\partial M$, a Hilbert space is defined, which is spanned by the cochains $a_i$ restricted to the boundary. Specifically, $a_i$ corresponds to a $|G_i|$-dimensional qudit on each $(n_i-1)$-simplex. The ground state is given by~\cite{li2023measuring, li2024domain}
\begin{align}
    \ket{\psi} = \int \mathcal{D}[a_i] e^{2\pi i \mathcal{U}[a_i]} \ket{a_i},
\end{align}
where $\ket{a_i}$ corresponds to the state given by $\ket{a_i(\Delta^{(n_i-1)})}$, with $a_i(\Delta^{(n_i-1)})\in G_i$, for each $(n_i-1)$-simplex $\Delta^{(n_i-1)}$.
The phases $e^{2\pi i \mathcal{U}[a_i]}$ give rise to an entangler $U$. 

To prepare $\ket{\psi}$, we start with a paramagnet state $\ket{\psi_0} = \int \mathcal{D}[a_i] \,\ket{a_i}$. The Hamiltonian $H_0$ of the paramagnet state $\ket{\psi_0}$ is simply given by a sum of shift operators $X$ acting on each qudit. Therefore the Hamiltonian of $\ket{\psi}$ is given by conjugating $H_0$ with the entangler $U$, which gives the ungauged Hamiltonian $UH_0U^\dagger$. We finally gauge the symmetry in $U H_0U^\dagger$ to obtain the Hamiltonian of the Lagrangian $\mathcal{L}[A_i]$. In the remainder of this appendix, we first illustrate the procedure with a few examples, such as the 3D bosonic (and fermionic) toric code and the RBH model, and then present the more complicated semion Walker-Wang model.

\subsection{3D bosonic toric code}

The topological action of the 3D toric code is trivial, $\mathcal{L}[A] = 0$~\cite{chen2023higher}, where $A$ is a $\mathbb{Z}_2$ $2$-form gauge field. In this case, the entangler is trivial $U = 1$, and the ungauged theory is simply the paramagnet with qubits on edges $H_0 = -\sum_e X_e$. We gauge the $1$-form symmetry generated by the vertex term
\begin{align}\label{eq:1_form}
    \begin{tikzpicture}[scale= 1, baseline={([yshift=-2.5pt]current bounding box.center)}]
    %
        \coordinate (O) at (0,0,0);
        \coordinate (A) at (-1.5,0,0);
        \coordinate (B) at (1.5,0,0);
        \coordinate (C) at (0,-1.5,0);
        \coordinate (D) at (0,1.5,0);
        \coordinate (E) at (0,0,1.5);
        \coordinate (F) at (0,0,-1.5);
        %
        \draw (A) -- (B);
        \draw (C) -- (D); 
        \draw (E) -- (F);
        \foreach \X/\Y in { O/E, O/F}
        \node[inner sep=0.5pt, fill=white] at ($( \X )!0.5!( \Y )$) {\footnotesize $X$};
        \foreach \X/\Y in {O/C, O/D}
        \node[inner sep=0.5pt, fill=white] at ($( \X )!0.5!( \Y )$) {\footnotesize $X$};
        \foreach \X/\Y in {O/A, O/B}
        \node[inner sep=0.5pt, fill=white] at ($( \X )!0.5!( \Y )$) {\footnotesize $X$}; 
    \end{tikzpicture}
\end{align}
The gauging map is given by~\cite{y16, ky18}
\begin{align}\label{eq:gauge_1_form}
\begin{tikzpicture}[baseline = 1cm]
    \draw(0,0) -- (0,2);
    \node at (0,1) {$X$};
\end{tikzpicture}
\hspace{15pt}\mapsto
\hspace{15pt}
\begin{tikzpicture}[scale = 1, baseline = 1cm]
    \coordinate (O) at (0,0,0); 
    \coordinate (A) at (2,0,0);
    \coordinate (B) at (-2,0,0);
    \coordinate (C) at (0,0,2);
    \coordinate (D) at (0,0,-2);
    \coordinate (E) at (0,2,0);
    \coordinate (F) at (2,2,0);
    \coordinate (G) at (-2,2,0);
    \coordinate (H) at (0,2,2);
    \coordinate (I) at (0,2,-2);
    \node at (1, 1, 0) {$X$};
    \node at (-1, 1, 0) {$X$};
    \node  at (0, 1, 1) {$X$};
    \node at (0, 1, -1) {$X$};
%
    %
    \draw (A) -- (O);
    \draw [dotted](O) -- (-0.8, 0,0);
    \draw (-0.77,0,0) -- (B);
    \draw (O) -- (C);
    \draw [dotted] (O) -- (D);
    \draw (O) -- (E);
    \draw (F) -- (G);
    \draw (H) -- (E) -- (I);
    \draw (A) -- (F);
    \draw (B) -- (G);
    \draw (C) -- (H);
    \draw [dotted](D) -- (0, 1.3,-2);
    \draw (0,1.24,-2) -- (I);
    \draw [red](-1, 0.95, -1) -- (1,0.95,-1) -- (1,0.95,1) -- (-1,0.95,1) -- cycle;
\end{tikzpicture}
\hspace{35pt}
\begin{tikzpicture}[baseline = 1cm]
    \draw (0,0) -- (2,0) -- (2,2) -- (0,2) -- cycle;
    \node at (1,0) {$Z$};
    \node at (1,2) {$Z$};
    \node at (0,1) {$Z$};
    \node at (2,1) {$Z$};
\end{tikzpicture}
\hspace{15pt} \mapsto
\hspace{15pt}
\begin{tikzpicture}[baseline = 1cm]
    \draw (0,0,0) -- (2,0,0) -- (2,2,0) -- (0,2,0) -- cycle;
    \draw [red](1,1,2) --
    (1,1,0);
    \draw [red, dotted] (1,1,0) -- (1,1,-2);
    \node at (1,1,0) {$Z$};
\end{tikzpicture}
\end{align}
where the red lines are edges of the dual lattice. We see after gauging, $H_0$ becomes a sum of plaquette terms on the dual lattice
\begin{align}
K_{p}
\hspace{5pt}=
\hspace{5pt}
    \begin{tikzpicture}[scale= 1, baseline={([yshift=-2.5pt]current bounding box.center)}]
            \coordinate (A) at (0,0,0);
            \coordinate (B) at (0,0,2);
            \coordinate (C) at (0,2,2);
            \coordinate (D) at (0,2,0);
%
            \coordinate (U) at (-2,0,0);
            \coordinate (O) at (2,2,2);
%
%
            %
            \draw (A) -- (B) --(C) -- (D) -- cycle;
            \foreach \X/\Y in { A/B, B/C, C/D, D/A}
            \node[inner sep=0.5pt, fill=white] at ($( \X )!0.5!( \Y )$) {\footnotesize $X$}; 
    \end{tikzpicture},
    \hspace{25pt}
    \begin{tikzpicture}[scale= 1, baseline={([yshift=-2.5pt]current bounding box.center)}]
%
            \coordinate (A) at (0,0,0);
            \coordinate (B) at (0,2,0);
            \coordinate (C) at (2,2,0);
            \coordinate (D) at (2,0,0);
%
            \coordinate (U) at (0,0,-2);
            \coordinate (O) at (2,2,2);
%
%
  %
            \draw (A) -- (B) --(C) -- (D) -- cycle;
            \foreach \X/\Y in { A/B, B/C, C/D, D/A}
            \node[inner sep=0.5pt, fill=white] at ($( \X )!0.5!( \Y )$) {\footnotesize $X$}; 
    \end{tikzpicture},
    \hspace{25pt}
    \begin{tikzpicture}[scale= 1, baseline={([yshift=-2.5pt]current bounding box.center)}]
        \coordinate (A) at (0,0,0);
        \coordinate (B) at (2,0,0);
        \coordinate (C) at (2,0,2);
        \coordinate (D) at (0,0,2);
%
        \coordinate (O) at (2,2,2);
        \coordinate (U) at (0,-2,0);
%
%
        \draw  (A) -- (B) -- (C) -- (D) -- cycle;
%
 %
        \foreach \X/\Y in { A/B, B/C, C/D, D/A}
        \node[inner sep=0.5pt, fill=white] at ($( \X )!0.5!( \Y )$) {\footnotesize $X$};
    \end{tikzpicture}.
\end{align}
Upon gauging a $\mathbb{Z}_2$ $1$-form symmetry in 3D, we obtain a dual $\mathbb{Z}_2$ $1$-form symmetry, which in this case is given by the following vertex term on the dual lattice
\begin{align}\label{eq:dual_1_form}
    B_v
    \hspace{5pt}=\hspace{5pt}
    \begin{tikzpicture}[scale= 1, baseline={([yshift=-2.5pt]current bounding box.center)}]
    %
        \coordinate (O) at (0,0,0);
        \coordinate (A) at (-1.5,0,0);
        \coordinate (B) at (1.5,0,0);
        \coordinate (C) at (0,-1.5,0);
        \coordinate (D) at (0,1.5,0);
        \coordinate (E) at (0,0,1.5);
        \coordinate (F) at (0,0,-1.5);
        %
        \draw (A) -- (B);
        \draw (C) -- (D); 
        \draw (E) -- (F);
        \foreach \X/\Y in { O/E, O/F}
        \node[inner sep=0.5pt, fill=white] at ($( \X )!0.5!( \Y )$) {\footnotesize $Z$};
        \foreach \X/\Y in {O/C, O/D}
        \node[inner sep=0.5pt, fill=white] at ($( \X )!0.5!( \Y )$) {\footnotesize $Z$};
        \foreach \X/\Y in {O/A, O/B}
        \node[inner sep=0.5pt, fill=white] at ($( \X )!0.5!( \Y )$) {\footnotesize $Z$}; 
    \end{tikzpicture}.
\end{align}
Together, we obtain the Hamiltonian for the 3D bosonic toric code
\begin{align}
    H = -\sum_v B_v - \sum_p K_p.
\end{align}

\subsection{3D fermionic toric code}\label{app:ta_ftc}

The topological action of the 3D fermionic toric code is given by~\cite{tata2023anomalies,chen2023higher}
\begin{align}
    \mathcal{L}[A] = \frac{1}{2} A\cup A,
\end{align}
where $A$ is a 2-form gauge field with $\mathbb{Z}_2$ coefficient. $\cup$ is the cup product on cochains, we are mostly interested in the cup product on the cubic lattice, and we refer to~\cite{chen2023higher} for a detailed definition, as well as a geometric interpretation~\cite{tata2020geometrically}. Here we demonstrate how to obtain the Hamiltonian starting from $\mathcal{L}$. First we ungauge by replacing $A\mapsto \delta a$, where $a$ is a $1$-cochain with $\mathbb{Z}_2$ coefficient. The Lagrangian becomes
\begin{align}
    \mathcal{L}[\delta a] = \frac{1}{2} \delta a\cup \delta a = \delta
    \mathcal{U}[a],
    \hspace{35pt}
    \mathcal{U}[a] = \frac{1}{2} a\cup \delta a.
\end{align}
Upon integrating over the spacetime manifold $M$, the action can be restricting to the boundary $\partial M$ on which the Hilbert space is defined. Since the field $a$ is a $1$-cochain with $\mathbb{Z}_2$ coefficient, the Hilbert space is given by one qubit per edge. The ground state is
\begin{align}
    \ket{\psi} = \sum_{a} (-1)^{a\cup \delta a} \ket{a}
    =
    U \ket{\psi_0},
    \hspace{35pt}
    \ket{\psi_0} = \sum_a \ket{a},
\end{align}
where $U$ is the entangler given by the phase $e^{2\pi i \mathcal{U}[a]} = (-1)^{a\cup \delta a} $. Following the definition of the cup product, it is not difficult to derive the entangler $U$. If we fix one edge, which is colored in blue in the figure below, all the components of $U$ which is supported on this edge is given by
\begin{align}
\begin{tikzpicture}[scale=1, baseline={([yshift=-2.5pt]current bounding box.center)}]
\coordinate (A) at (0,0,0);
\coordinate (B) at (0,0,2);
\coordinate (C) at (0,2,2);
\coordinate (D) at (0,2,0);
\coordinate (E) at (2,2,2);
\coordinate (F) at (2,2,4);
\coordinate (G) at (2,4,4);
\coordinate (H) at (2,4,2);
\coordinate (O) at (2,2,2);
%
\draw (A) -- (B) -- (C) -- (D) -- cycle;
\draw (E) -- (F) -- (G) -- (H) -- cycle;
\draw[blue] (C) -- (1.23,2,2);
\draw[dashed,blue] (1.23,2,2) -- (O);
%
\coordinate (X) at ($(C)!0.5!(O)$);
%
\foreach \P/\Q in {A/B,C/D,D/A,E/F,F/G,G/H}
\draw[white,line width=1.8pt] (X) -- ($(\P)!0.5!(\Q)$);
%
\draw[white,line width=1.8pt]
(X) to[bend right=30] ($(B)!0.5!(C)$);
\draw[white,line width=1.8pt]
(X) to[bend right=30] ($(H)!0.5!(E)$);
%
\foreach \P/\Q in {A/B,C/D,D/A,E/F,F/G,G/H}
\draw[red] (X) -- ($(\P)!0.5!(\Q)$);
%
\draw[red]
(X) to[bend right=30] ($(B)!0.5!(C)$);
\draw[red]
(X) to[bend right=30] ($(H)!0.5!(E)$);
%
\foreach \P/\Q in {A/B,B/C,C/D,D/A,E/F,F/G,G/H,H/E}
\node[fill,circle,red,inner sep=1pt] at ($(\P)!0.5!(\Q)$) {};
\node[fill,circle,red,inner sep=1pt] at (X) {};
\end{tikzpicture}
\hspace{35pt}
\begin{tikzpicture}[scale=1, baseline={([yshift=-2.5pt]current bounding box.center)}]
\coordinate (A) at (0,0,0);
\coordinate (B) at (0,2,0);
\coordinate (C) at (2,2,0);
\coordinate (D) at (2,0,0);
\coordinate (E) at (2,2,2);
\coordinate (F) at (2,4,2);
\coordinate (G) at (4,4,2);
\coordinate (H) at (4,2,2);
\coordinate (O) at (2,2,2);
%
\draw (A) -- (B);
\draw (B) -- (1.23,2,0);
\draw[dotted] (C) -- (1.23,2,0);
\draw (D) -- (2,1.23,0);
\draw[dotted] (C) -- (2,1.23,0);
\draw (D) -- (A);
\draw (E) -- (F) -- (G) -- (H) -- cycle;
\draw[dashed,blue] (C) -- (O);
%
\coordinate (X) at ($(C)!0.5!(O)$);
%
\foreach \P/\Q in {A/B,B/C,C/D,D/A,E/F,F/G,G/H,H/E}
\draw[white,line width=1.8pt] (X) -- ($(\P)!0.5!(\Q)$);
%
\foreach \P/\Q in {A/B,B/C,C/D,D/A,E/F,F/G,G/H,H/E}
\draw[red] (X) -- ($(\P)!0.5!(\Q)$);
%
\foreach \P/\Q in {A/B,B/C,C/D,D/A,E/F,F/G,G/H,H/E}
\node[fill,circle,red,inner sep=1pt] at ($(\P)!0.5!(\Q)$) {};
\node[fill,circle,red,inner sep=1pt] at (X) {};
\end{tikzpicture}
\hspace{25pt}
\begin{tikzpicture}[scale=1, baseline={([yshift=-2.5pt]current bounding box.center)}]
\coordinate (A) at (0,0,0);
\coordinate (B) at (2,0,0);
\coordinate (C) at (2,0,2);
\coordinate (D) at (0,0,2);
\coordinate (E) at (2,2,2);
\coordinate (F) at (4,2,2);
\coordinate (G) at (4,2,4);
\coordinate (H) at (2,2,4);
\coordinate (O) at (2,2,2);
%
\draw (A) -- (B) -- (C) -- (D) -- cycle;
\draw (E) -- (F) -- (G) -- (H) -- cycle;
\draw[blue] (C) -- (2,1.23,2);
\draw[dashed,blue] (O) -- (2,1.23,2);
%
\coordinate (X) at ($(C)!0.5!(O)$);
%
\foreach \P/\Q in {A/B,B/C,D/A,F/G,G/H,H/E}
\draw[white,line width=1.8pt] (X) -- ($(\P)!0.5!(\Q)$);
%
\draw[white,line width=1.8pt]
(X) to[bend left=30] ($(C)!0.5!(D)$);
\draw[white,line width=1.8pt]
(X) to[bend left=30] ($(E)!0.5!(F)$);
%
\foreach \P/\Q in {A/B,B/C,D/A,F/G,G/H,H/E}
\draw[red] (X) -- ($(\P)!0.5!(\Q)$);
%
\draw[red]
(X) to[bend left=30] ($(C)!0.5!(D)$);
\draw[red]
(X) to[bend left=30] ($(E)!0.5!(F)$);
%
\foreach \P/\Q in {A/B,B/C,C/D,D/A,E/F,F/G,G/H,H/E}
\node[fill,circle,red,inner sep=1pt] at ($(\P)!0.5!(\Q)$) {};
\node[fill,circle,red,inner sep=1pt] at (X) {};
\end{tikzpicture}
\end{align}
where $\begin{tikzpicture}[scale = 0.5, baseline = -0.1cm]
    \node [fill, circle, red, inner sep = 1pt] at (0,0) {};
    \node [fill, circle, red, inner sep = 1pt] at (1,0) {};
    \draw [red] (0,0) -- (1,0);
\end{tikzpicture}$
is the CZ gate. The state $\ket{\psi_0}$ is the paramagnet state, with Hamiltonian
\begin{align}
    \ket{\psi_0} = \bigotimes_e \ket{+}_e,
    \hspace{35pt}
    H_0 = -\sum_e X_e,
\end{align}
where the tensor and sum is over all edges $e$, $X_e$ is the Pauli $X$ matrix, and $\ket{+} = \frac{1}{\sqrt{2}}( \ket{0} + \ket{1})$ is the $+1$ eigenstate of $X$. Conjugating $H_0$ by $U$, we obtain stabilizers
\begin{align}
   \begin{tikzpicture}[scale=1, baseline={([yshift=-2.5pt]current bounding box.center)}]
\coordinate (A) at (0,0,0);
\coordinate (B) at (0,0,2);
\coordinate (C) at (0,2,2);
\coordinate (D) at (0,2,0);
\coordinate (E) at (2,2,2);
\coordinate (F) at (2,2,4);
\coordinate (G) at (2,4,4);
\coordinate (H) at (2,4,2);
\coordinate (O) at (2,2,2);
\draw (A) -- (B) -- (C) -- (D) -- cycle;
\draw (E) -- (F) -- (G) -- (H) -- cycle;
\draw[blue] (C) -- (1.23,2,2);
\draw[dashed,blue] (1.23,2,2) -- (O);
\DrawGateConnections{A/B/0,B/C/-30,C/D/0,D/A/0,E/F/0,F/G/0,G/H/0,H/E/-30}
\end{tikzpicture}
\hspace{35pt}
\begin{tikzpicture}[scale=1, baseline={([yshift=-2.5pt]current bounding box.center)}]
\coordinate (A) at (0,0,0);
\coordinate (B) at (0,2,0);
\coordinate (C) at (2,2,0);
\coordinate (D) at (2,0,0);
\coordinate (E) at (2,2,2);
\coordinate (F) at (2,4,2);
\coordinate (G) at (4,4,2);
\coordinate (H) at (4,2,2);
\coordinate (O) at (2,2,2);
\draw (A) -- (B);
\draw (B) -- (1.23,2,0);
\draw[dotted] (C) -- (1.23,2,0);
\draw (D) -- (2,1.23,0);
\draw[dotted] (C) -- (2,1.23,0);
\draw (D) -- (A);
\draw (E) -- (F) -- (G) -- (H) -- cycle;
\draw[dashed,blue] (C) -- (O);
\DrawGateConnections{A/B/0,B/C/0,C/D/0,D/A/0,E/F/0,F/G/0,G/H/0,H/E/0}
\end{tikzpicture}
\hspace{25pt}
\begin{tikzpicture}[scale=1, baseline={([yshift=-2.5pt]current bounding box.center)}]
\coordinate (A) at (0,0,0);
\coordinate (B) at (2,0,0);
\coordinate (C) at (2,0,2);
\coordinate (D) at (0,0,2);
\coordinate (E) at (2,2,2);
\coordinate (F) at (4,2,2);
\coordinate (G) at (4,2,4);
\coordinate (H) at (2,2,4);
\coordinate (O) at (2,2,2);
\draw (A) -- (B) -- (C) -- (D) -- cycle;
\draw (E) -- (F) -- (G) -- (H) -- cycle;
\draw[blue] (C) -- (2,1.23,2);
\draw[dashed,blue] (O) -- (2,1.23,2);
\DrawGateConnections{A/B/0,B/C/0,C/D/30,D/A/0,E/F/30,F/G/0,G/H/0,H/E/0}
\end{tikzpicture}
\end{align}
Finally, in order to obtain the Hamiltonian of $\mathcal{L}[A]$, we gauge the $1$-form symmetry~\eqref{eq:1_form}. Note one can check that the $1$-form symmetry generators indeed commute with the Hamiltonian $UH_0U^\dagger$, or more generally with the entangler $U$. Applying the gauging map~\eqref{eq:gauge_1_form} and translating to the dual lattice, we see $U H_0U^\dagger$ becomes
\begin{align}
    K_p
    \hspace{5pt}
    =
    \hspace{5pt}
    \begin{tikzpicture}[scale= 1, baseline={([yshift=-2.5pt]current bounding box.center)}]
            \coordinate (A) at (0,0,0);
            \coordinate (B) at (0,0,2);
            \coordinate (C) at (0,2,2);
            \coordinate (D) at (0,2,0);
%
            \coordinate (U) at (-2,0,0);
            \coordinate (O) at (2,2,2);
%
%
            %
            \draw (A) -- (B) --(C) -- (D) -- cycle;
            \draw [dotted] (A) -- (U);
            \draw (C) -- (O);
            \foreach \X/\Y in { A/B, B/C, C/D, D/A}
            \node[inner sep=0.5pt, fill=white] at ($( \X )!0.5!( \Y )$) {\footnotesize $X$}; 
            \node[inner sep=0.5pt, fill=white] at (-1.1, 0, 0) {\footnotesize $Z$};
            \node[inner sep=0.5pt, fill=white] at (1.1, 2, 2) {\footnotesize $Z$};
    \end{tikzpicture},
    \hspace{15pt}
    \begin{tikzpicture}[scale= 1, baseline={([yshift=-2.5pt]current bounding box.center)}]
%
            \coordinate (A) at (0,0,0);
            \coordinate (B) at (0,2,0);
            \coordinate (C) at (2,2,0);
            \coordinate (D) at (2,0,0);
%
            \coordinate (U) at (0,0,-2);
            \coordinate (O) at (2,2,2);
%
%
  %
            \draw (A) -- (B) --(C) -- (D) -- cycle;
            \draw [dotted] (A) -- (U);
            \draw (C) -- (O);
            \foreach \X/\Y in { A/B, B/C, C/D, D/A}
            \node[inner sep=0.5pt, fill=white] at ($( \X )!0.5!( \Y )$) {\footnotesize $X$}; 
            \node[inner sep=0.5pt, fill=white] at (0, 0, -1.1) {\footnotesize $Z$};
            \node[inner sep=0.5pt, fill=white] at (2, 2, 1.1) {\footnotesize $Z$};
    \end{tikzpicture},
    \hspace{15pt}
    \begin{tikzpicture}[scale= 1, baseline={([yshift=-2.5pt]current bounding box.center)}]
        \coordinate (A) at (0,0,0);
        \coordinate (B) at (2,0,0);
        \coordinate (C) at (2,0,2);
        \coordinate (D) at (0,0,2);
%
        \coordinate (O) at (2,2,2);
        \coordinate (U) at (0,-2,0);
%
%
        \draw  (A) -- (B) -- (C) -- (D) -- cycle;
        \draw [dotted] (A) -- (U);
        \draw (C) -- (O);
%
 %
        \foreach \X/\Y in { A/B, B/C, C/D, D/A}
        \node[inner sep=0.5pt, fill=white] at ($( \X )!0.5!( \Y )$) {\footnotesize $X$};
        \node[inner sep=0.5pt, fill=white] at (2,1.1,2) {\footnotesize $Z$};
        \node[inner sep=0.5pt, fill=white] at (0,-1.1,0) {\footnotesize $Z$};        
    \end{tikzpicture},
\end{align}
which are the plaquette terms of the fermionic toric code. Together with the dual symmetry in~\eqref{eq:dual_1_form}, we see that the Lagrangian $\mathcal{L}[A] = \frac{1}{2} A\cup A$ indeed recovers the Hamiltonian of the fermionic toric code
\begin{align}
    H = -\sum_v B_v - \sum_p K_p.
\end{align}

\subsection{RBH Cluster state}

The topological action of the RBH cluster state is given by 
\begin{align}
    \mathcal{L}[A_1, A_2] = \frac{1}{2} A_1\cup A_2,
\end{align}
where $A_2$ and $A_2$ are both $\mathbb{Z}_2$ $2$-form gauge fields. To derive the Hamiltonian, first we ungauge by replacing $A_i\mapsto \delta a_i$, where both $a_1$ and $a_2$ are $1$-cochains with $\mathbb{Z}_2$ coefficients, and obtain a total derivative
\begin{align}
    \mathcal{L}[\delta a_1, \delta a_2] = \delta\mathcal{U}[a_1,a_2],
    \hspace{35pt}
    \mathcal{U}[a_1,a_2] = \frac{1}{2} (a_1\cup \delta a_2).
\end{align}
On the boundary of the manifold, the Hilbert space is defined. Since $a_1$ and $a_2$ are $1$-cochains with $\mathbb{Z}_2$ coefficients, the Hilbert space consists of two qubits per edge. The ground state is given by
\begin{align}
    \ket{\psi} = \sum_{a} (-1)^{a_1\cup \delta a_2} \ket{a_1, a_2}
    =
    U \ket{\psi_0},
    \hspace{35pt}
    \ket{\psi_0} = \sum_a \ket{a_1, a_2},
\end{align}
where $U$ is the entangler given by the phase $(-1)^{a_1\cup \delta a_2}$. Similar as in Appendix~\ref{app:ta_ftc}, if we fix the blue edge in the figure below, all the components of $U$ which is supported on a single qubit on this edge is given by
\begin{equation}
\begin{aligned}
    \begin{tikzpicture}[scale=1, baseline={([yshift=-2.5pt]current bounding box.center)}]
        \coordinate (A) at (0,0,0);
        \coordinate (B) at (0,0,2);
        \coordinate (C) at (0,2,2);
        \coordinate (D) at (0,2,0);
        \coordinate (E) at (2,2,2);
        \coordinate (F) at (2,2,4);
        \coordinate (G) at (2,4,4);
        \coordinate (H) at (2,4,2);
        \coordinate (O) at (2,2,2);
        \coordinate (X) at ($(C)!0.5!(O)$);
%
        \draw (E) -- (F) -- (G) -- (H) -- cycle;
        \draw[blue] (C) -- (1.23,2,2);
        \draw[dashed,blue] (1.23,2,2) -- (O);
%
        \foreach \P/\Q in {E/F,F/G,G/H}
        \draw[white,line width=1.8pt] (X) -- ($(\P)!0.5!(\Q)$);
%
        \draw[white,line width=1.8pt]
        (X) to[bend right=30] ($(H)!0.5!(E)$);
%
        \foreach \P/\Q in {E/F,F/G,G/H}
        \draw[red] (X) -- ($(\P)!0.5!(\Q)$);
        \draw[red]
        (X) to[bend right=30] ($(H)!0.5!(E)$);
%
        \foreach \P/\Q in {E/F,F/G,G/H,H/E}
        \node[fill,circle,myGreen,inner sep=1pt] at ($(\P)!0.5!(\Q)$) {};
        \node[fill,circle,myBlue,inner sep=1pt] at (X) {};
    \end{tikzpicture}
    \hspace{35pt}
    \begin{tikzpicture}[scale=1, baseline={([yshift=-2.5pt]current bounding box.center)}]
        \coordinate (A) at (0,0,0);
        \coordinate (B) at (0,2,0);
        \coordinate (C) at (2,2,0);
        \coordinate (D) at (2,0,0);
        \coordinate (E) at (2,2,2);
        \coordinate (F) at (2,4,2);
        \coordinate (G) at (4,4,2);
        \coordinate (H) at (4,2,2);
        \coordinate (O) at (2,2,2);
        \coordinate (X) at ($(C)!0.5!(O)$);
%
        
        \draw (E) -- (F) -- (G) -- (H) -- cycle;
        \draw[dashed,blue] (C) -- (O);
%
        \foreach \P/\Q in {E/F,F/G,G/H,H/E}
        \draw[white,line width=1.8pt] (X) -- ($(\P)!0.5!(\Q)$);
%
        \foreach \P/\Q in {E/F,F/G,G/H,H/E}
        \draw[red] (X) -- ($(\P)!0.5!(\Q)$);
%
        \foreach \P/\Q in {E/F,F/G,G/H,H/E}
        \node[fill,circle,myGreen,inner sep=1pt] at ($(\P)!0.5!(\Q)$) {};
        \node[fill,circle,myBlue,inner sep=1pt] at (X) {};
    \end{tikzpicture}
    \hspace{25pt}
    \begin{tikzpicture}[scale=1, baseline={([yshift=-2.5pt]current bounding box.center)}]
        \coordinate (A) at (0,0,0);
        \coordinate (B) at (2,0,0);
        \coordinate (C) at (2,0,2);
        \coordinate (D) at (0,0,2);
        \coordinate (E) at (2,2,2);
        \coordinate (F) at (4,2,2);
        \coordinate (G) at (4,2,4);
        \coordinate (H) at (2,2,4);
        \coordinate (O) at (2,2,2);
        \coordinate (X) at ($(C)!0.5!(O)$);
%
        \draw (E) -- (F) -- (G) -- (H) -- cycle;
        \draw[blue] (C) -- (2,1.23,2);
        \draw[dashed,blue] (O) -- (2,1.23,2);
%
        \foreach \P/\Q in {F/G,G/H,H/E}
        \draw[white,line width=1.8pt] (X) -- ($(\P)!0.5!(\Q)$);
%
        \draw[white,line width=1.8pt]
        (X) to[bend left=30] ($(E)!0.5!(F)$);
%
        \foreach \P/\Q in {F/G,G/H,H/E}
        \draw[red] (X) -- ($(\P)!0.5!(\Q)$);
        \draw[red]
        (X) to[bend left=30] ($(E)!0.5!(F)$);
%
        \foreach \P/\Q in {E/F,F/G,G/H,H/E}
        \node[fill,circle,myGreen,inner sep=1pt] at ($(\P)!0.5!(\Q)$) {};
        \node[fill,circle,myBlue,inner sep=1pt] at (X) {};
    \end{tikzpicture}
\\
    \begin{tikzpicture}[scale=1, baseline={([yshift=-2.5pt]current bounding box.center)}]
        \coordinate (A) at (0,0,0);
        \coordinate (B) at (0,0,2);
        \coordinate (C) at (0,2,2);
        \coordinate (D) at (0,2,0);
        \coordinate (E) at (2,2,2);
        \coordinate (F) at (2,2,4);
        \coordinate (G) at (2,4,4);
        \coordinate (H) at (2,4,2);
        \coordinate (O) at (2,2,2);
        \coordinate (X) at ($(C)!0.5!(O)$);
%
        \draw (A) -- (B) -- (C) -- (D) -- cycle;
        \draw[blue] (C) -- (1.23,2,2);
        \draw[blue] (1.23,2,2) -- (O);
%
        \foreach \P/\Q in {A/B,C/D,D/A}
        \draw[white,line width=1.8pt] (X) -- ($(\P)!0.5!(\Q)$);
%
        \draw[white,line width=1.8pt]
        (X) to[bend right=30] ($(B)!0.5!(C)$);
%
        \foreach \P/\Q in {A/B,C/D,D/A}
        \draw[red] (X) -- ($(\P)!0.5!(\Q)$);
        \draw[red]
        (X) to[bend right=30] ($(B)!0.5!(C)$);
%
        \foreach \P/\Q in {A/B,B/C,C/D,D/A}
        \node[fill,circle,myBlue,inner sep=1pt] at ($(\P)!0.5!(\Q)$) {};
        \node[fill,circle,myGreen,inner sep=1pt] at (X) {};
    \end{tikzpicture}
    \hspace{35pt}
    \begin{tikzpicture}[scale=1, baseline={([yshift=-2.5pt]current bounding box.center)}]
        \coordinate (A) at (0,0,0);
        \coordinate (B) at (0,2,0);
        \coordinate (C) at (2,2,0);
        \coordinate (D) at (2,0,0);
        \coordinate (E) at (2,2,2);
        \coordinate (F) at (2,4,2);
        \coordinate (G) at (4,4,2);
        \coordinate (H) at (4,2,2);
        \coordinate (O) at (2,2,2);
        \coordinate (X) at ($(C)!0.5!(O)$);
%
        \draw (A) -- (B);
        \draw (B) -- (1.23,2,0);
        \draw (C) -- (1.23,2,0);
        \draw (D) -- (2,1.23,0);
        \draw (C) -- (2,1.23,0);
        \draw (D) -- (A);
        \draw[blue] (C) -- (O);
%
        \foreach \P/\Q in {A/B,B/C,C/D,D/A}
        \draw[white,line width=1.8pt] (X) -- ($(\P)!0.5!(\Q)$);
%
        \foreach \P/\Q in {A/B,B/C,C/D,D/A}
        \draw[red] (X) -- ($(\P)!0.5!(\Q)$);
%
        \foreach \P/\Q in {A/B,B/C,C/D,D/A}
        \node[fill,circle,myBlue,inner sep=1pt] at ($(\P)!0.5!(\Q)$) {};
        \node[fill,circle,myGreen,inner sep=1pt] at (X) {};
    \end{tikzpicture}
    \hspace{25pt}
    \begin{tikzpicture}[scale=1, baseline={([yshift=-2.5pt]current bounding box.center)}]
        \coordinate (A) at (0,0,0);
        \coordinate (B) at (2,0,0);
        \coordinate (C) at (2,0,2);
        \coordinate (D) at (0,0,2);
        \coordinate (E) at (2,2,2);
        \coordinate (F) at (4,2,2);
        \coordinate (G) at (4,2,4);
        \coordinate (H) at (2,2,4);
        \coordinate (O) at (2,2,2);
        \coordinate (X) at ($(C)!0.5!(O)$);
%
        \draw (A) -- (B) -- (C) -- (D) -- cycle;
        \draw[blue] (C) -- (2,1.23,2);
        \draw[blue] (O) -- (2,1.23,2);
%
        \foreach \P/\Q in {A/B,B/C,D/A}
        \draw[white,line width=1.8pt] (X) -- ($(\P)!0.5!(\Q)$);
%
        \draw[white,line width=1.8pt]
        (X) to[bend left=30] ($(C)!0.5!(D)$);
%
        \foreach \P/\Q in {A/B,B/C,D/A}
        \draw[red] (X) -- ($(\P)!0.5!(\Q)$);
        \draw[red]
        (X) to[bend left=30] ($(C)!0.5!(D)$);
%
        \foreach \P/\Q in {A/B,B/C,C/D,D/A}
        \node[fill,circle,myBlue,inner sep=1pt] at ($(\P)!0.5!(\Q)$) {};
        \node[fill,circle,myGreen,inner sep=1pt] at (X) {};
    \end{tikzpicture}
\end{aligned}
\end{equation}
where 
$\begin{tikzpicture}[baseline = -0.1cm]
    \node [fill, circle, myBlue, inner sep = 1pt] at (0,0) {}; 
\end{tikzpicture}$\,
is the first qubit, 
$\begin{tikzpicture}[baseline = -0.1cm]
    \node [fill, circle, myGreen, inner sep = 1pt] at (0,0) {}; 
\end{tikzpicture}$\,
is the second qubit, and the red line is the CZ gate. The state $\ket{\psi_0}$ is the paramagnet state, with Hamiltonian
\begin{align}
    H_0 = -\sum_e X_{e,1} - \sum_e X_{e,2}.
\end{align}
Conjugating $H_0$ by the entangler $U$, we obtain the stabilizers
\begin{equation}
\begin{aligned}
\hspace{-20pt}
    \begin{tikzpicture}[scale= 1, baseline={([yshift=-2.5pt]current bounding box.center)}]
            \coordinate (A) at (0,0,0);
            \coordinate (B) at (0,0,2);
            \coordinate (C) at (0,2,2);
            \coordinate (D) at (0,2,0);
%
            \coordinate (U) at (-2,0,0);
            \coordinate (O) at (2,2,2);
%
%
            %
            \draw (A) -- (B) --(C) -- (D) -- cycle;
            \draw [dotted] (A) -- (U);
            \foreach \X/\Y in { A/B, B/C, C/D, D/A}
            \node[inner sep=0.5pt, fill=white] at ($( \X )!0.5!( \Y )$) {\footnotesize $Z_2$}; 
            \node[inner sep=0.5pt, fill=white] at (-1.3, 0, 0) {\footnotesize $X_1$};
    \end{tikzpicture}
    \hspace{40pt}
    \begin{tikzpicture}[scale= 1, baseline={([yshift=-2.5pt]current bounding box.center)}]
%
            \coordinate (A) at (0,0,0);
            \coordinate (B) at (0,2,0);
            \coordinate (C) at (2,2,0);
            \coordinate (D) at (2,0,0);
%
            \coordinate (U) at (0,0,-2);
            \coordinate (O) at (2,2,2);
%
%
  %
            \draw (A) -- (B) --(C) -- (D) -- cycle;
            \draw [dotted] (A) -- (U);
            \foreach \X/\Y in { A/B, B/C, C/D, D/A}
            \node[inner sep=0.5pt, fill=white] at ($( \X )!0.5!( \Y )$) {\footnotesize $Z_2$}; 
            \node[inner sep=0.5pt, fill=white] at (0, 0, -1.1) {\footnotesize $X_1$};
    \end{tikzpicture}
    \hspace{25pt}
    \begin{tikzpicture}[scale= 1, baseline={([yshift=-2.5pt]current bounding box.center)}]
        \coordinate (A) at (0,0,0);
        \coordinate (B) at (2,0,0);
        \coordinate (C) at (2,0,2);
        \coordinate (D) at (0,0,2);
%
        \coordinate (O) at (2,2,2);
        \coordinate (U) at (0,-2,0);
%
%
        \draw  (A) -- (B) -- (C) -- (D) -- cycle;
        \draw [dotted] (A) -- (U);
%
 %
        \foreach \X/\Y in { A/B, B/C, C/D, D/A}
        \node[inner sep=0.5pt, fill=white] at ($( \X )!0.5!( \Y )$) {\footnotesize $Z_2$};
        \node[inner sep=0.5pt, fill=white] at (0,-1.3,0) {\footnotesize $X_1$};        
    \end{tikzpicture}\\
    \begin{tikzpicture}[scale= 1, baseline={([yshift=-2.5pt]current bounding box.center)}]
            \coordinate (A) at (0,0,0);
            \coordinate (B) at (0,0,2);
            \coordinate (C) at (0,2,2);
            \coordinate (D) at (0,2,0);
%
            \coordinate (U) at (-2,0,0);
            \coordinate (O) at (2,2,2);
%
%
            %
            \draw (A) -- (B) --(C) -- (D) -- cycle;
            \draw (C) -- (O);
            \foreach \X/\Y in { A/B, B/C, C/D, D/A}
            \node[inner sep=0.5pt, fill=white] at ($( \X )!0.5!( \Y )$) {\footnotesize $Z_1$}; 
            \node[inner sep=0.5pt, fill=white] at (1.3, 2, 2) {\footnotesize $X_2$};
    \end{tikzpicture}
    \hspace{25pt}
    \begin{tikzpicture}[scale= 1, baseline={([yshift=-2.5pt]current bounding box.center)}]
%
            \coordinate (A) at (0,0,0);
            \coordinate (B) at (0,2,0);
            \coordinate (C) at (2,2,0);
            \coordinate (D) at (2,0,0);
%
            \coordinate (U) at (0,0,-2);
            \coordinate (O) at (2,2,2);
%
%
  %
            \draw (A) -- (B) --(C) -- (D) -- cycle;
            \draw (C) -- (O);
            \foreach \X/\Y in { A/B, B/C, C/D, D/A}
            \node[inner sep=0.5pt, fill=white] at ($( \X )!0.5!( \Y )$) {\footnotesize $Z_1$}; 
            \node[inner sep=0.5pt, fill=white] at (2, 2, 1.1) {\footnotesize $X_2$};
    \end{tikzpicture}
    \hspace{25pt}
    \begin{tikzpicture}[scale= 1, baseline={([yshift=-2.5pt]current bounding box.center)}]
        \coordinate (A) at (0,0,0);
        \coordinate (B) at (2,0,0);
        \coordinate (C) at (2,0,2);
        \coordinate (D) at (0,0,2);
%
        \coordinate (O) at (2,2,2);
        \coordinate (U) at (0,-2,0);
%
%
        \draw  (A) -- (B) -- (C) -- (D) -- cycle;
        \draw (C) -- (O);
%
 %
        \foreach \X/\Y in { A/B, B/C, C/D, D/A}
        \node[inner sep=0.5pt, fill=white] at ($( \X )!0.5!( \Y )$) {\footnotesize $Z_1$};
        \node[inner sep=0.5pt, fill=white] at (2,1.3,2) {\footnotesize $X_2$};        
    \end{tikzpicture},
\end{aligned}
\end{equation}

Finally, to obtain the Hamiltonian of the theory $\mathcal{L}[A_1, A_2] = \frac{1}{2} A_1\cup A_2$, we now need to gauge two $1$-form symmetries generated by two copies of~\eqref{eq:1_form}
\begin{align}
    \begin{tikzpicture}[scale= 1, baseline={([yshift=-2.5pt]current bounding box.center)}]
    %
        \coordinate (O) at (0,0,0);
        \coordinate (A) at (-1.5,0,0);
        \coordinate (B) at (1.5,0,0);
        \coordinate (C) at (0,-1.5,0);
        \coordinate (D) at (0,1.5,0);
        \coordinate (E) at (0,0,1.5);
        \coordinate (F) at (0,0,-1.5);
        %
        \draw (A) -- (B);
        \draw (C) -- (D); 
        \draw (E) -- (F);
        \foreach \X/\Y in { O/E, O/F}
        \node[inner sep=0.5pt, fill=white] at ($( \X )!0.5!( \Y )$) {\footnotesize $X_1$};
        \foreach \X/\Y in {O/C, O/D}
        \node[inner sep=0.5pt, fill=white] at ($( \X )!0.5!( \Y )$) {\footnotesize $X_1$};
        \foreach \X/\Y in {O/A, O/B}
        \node[inner sep=0.5pt, fill=white] at ($( \X )!0.5!( \Y )$) {\footnotesize $X_1$}; 
    \end{tikzpicture}
    \hspace{35pt}
    \begin{tikzpicture}[scale= 1, baseline={([yshift=-2.5pt]current bounding box.center)}]
    %
        \coordinate (O) at (0,0,0);
        \coordinate (A) at (-1.5,0,0);
        \coordinate (B) at (1.5,0,0);
        \coordinate (C) at (0,-1.5,0);
        \coordinate (D) at (0,1.5,0);
        \coordinate (E) at (0,0,1.5);
        \coordinate (F) at (0,0,-1.5);
        %
        \draw (A) -- (B);
        \draw (C) -- (D); 
        \draw (E) -- (F);
        \foreach \X/\Y in { O/E, O/F}
        \node[inner sep=0.5pt, fill=white] at ($( \X )!0.5!( \Y )$) {\footnotesize $X_2$};
        \foreach \X/\Y in {O/C, O/D}
        \node[inner sep=0.5pt, fill=white] at ($( \X )!0.5!( \Y )$) {\footnotesize $X_2$};
        \foreach \X/\Y in {O/A, O/B}
        \node[inner sep=0.5pt, fill=white] at ($( \X )!0.5!( \Y )$) {\footnotesize $X_2$}; 
    \end{tikzpicture}
\end{align}
The gauging map is given by~\eqref{eq:gauge_1_form} acting on the two sets of qubits respectively. After gauging $UH_0U^\dagger$, on the dual lattice we obtain plaquette terms
\begin{align}
\hspace{-5pt}
K_{p,1}
\hspace{5pt}=\hspace{15pt}
    \begin{tikzpicture}[scale= 1, baseline={([yshift=-2.5pt]current bounding box.center)}]
            \coordinate (A) at (0,0,0);
            \coordinate (B) at (0,0,2);
            \coordinate (C) at (0,2,2);
            \coordinate (D) at (0,2,0);
%
            \coordinate (U) at (-2,0,0);
            \coordinate (O) at (2,2,2);
%
%
            %
            \draw (A) -- (B) --(C) -- (D) -- cycle;
            \draw (C) -- (O);
            \foreach \X/\Y in { A/B, B/C, C/D, D/A}
            \node[inner sep=0.5pt, fill=white] at ($( \X )!0.5!( \Y )$) {\footnotesize $X_1$}; 
            \node[inner sep=0.5pt, fill=white] at (1.3, 2, 2) {\footnotesize $Z_2$};
    \end{tikzpicture},
    \hspace{15pt}
    \begin{tikzpicture}[scale= 1, baseline={([yshift=-2.5pt]current bounding box.center)}]
%
            \coordinate (A) at (0,0,0);
            \coordinate (B) at (0,2,0);
            \coordinate (C) at (2,2,0);
            \coordinate (D) at (2,0,0);
%
            \coordinate (U) at (0,0,-2);
            \coordinate (O) at (2,2,2);
%
%
  %
            \draw (A) -- (B) --(C) -- (D) -- cycle;
            \draw (C) -- (O);
            \foreach \X/\Y in { A/B, B/C, C/D, D/A}
            \node[inner sep=0.5pt, fill=white] at ($( \X )!0.5!( \Y )$) {\footnotesize $X_1$}; 
            \node[inner sep=0.5pt, fill=white] at (2, 2, 1.1) {\footnotesize $Z_2$};
    \end{tikzpicture},
    \hspace{25pt}
    \begin{tikzpicture}[scale= 1, baseline={([yshift=-2.5pt]current bounding box.center)}]
        \coordinate (A) at (0,0,0);
        \coordinate (B) at (2,0,0);
        \coordinate (C) at (2,0,2);
        \coordinate (D) at (0,0,2);
%
        \coordinate (O) at (2,2,2);
        \coordinate (U) at (0,-2,0);
%
%
        \draw  (A) -- (B) -- (C) -- (D) -- cycle;
        \draw (C) -- (O);
%
 %
        \foreach \X/\Y in { A/B, B/C, C/D, D/A}
        \node[inner sep=0.5pt, fill=white] at ($( \X )!0.5!( \Y )$) {\footnotesize $X_1$};
        \node[inner sep=0.5pt, fill=white] at (2,1.3,2) {\footnotesize $Z_2$};        
    \end{tikzpicture},
\end{align}
\begin{align}
K_{p,2}
\hspace{5pt}=\hspace{5pt}
    \begin{tikzpicture}[scale= 1, baseline={([yshift=-2.5pt]current bounding box.center)}]
            \coordinate (A) at (0,0,0);
            \coordinate (B) at (0,0,2);
            \coordinate (C) at (0,2,2);
            \coordinate (D) at (0,2,0);
%
            \coordinate (U) at (-2,0,0);
            \coordinate (O) at (2,2,2);
%
%
            %
            \draw (A) -- (B) --(C) -- (D) -- cycle;
            \draw [dotted] (A) -- (U);
            \foreach \X/\Y in { A/B, B/C, C/D, D/A}
            \node[inner sep=0.5pt, fill=white] at ($( \X )!0.5!( \Y )$) {\footnotesize $X_2$}; 
            \node[inner sep=0.5pt, fill=white] at (-1.3, 0, 0) {\footnotesize $Z_1$};
    \end{tikzpicture},
    \hspace{25pt}
    \begin{tikzpicture}[scale= 1, baseline={([yshift=-2.5pt]current bounding box.center)}]
%
            \coordinate (A) at (0,0,0);
            \coordinate (B) at (0,2,0);
            \coordinate (C) at (2,2,0);
            \coordinate (D) at (2,0,0);
%
            \coordinate (U) at (0,0,-2);
            \coordinate (O) at (2,2,2);
%
%
  %
            \draw (A) -- (B) --(C) -- (D) -- cycle;
            \draw [dotted] (A) -- (U);
            \foreach \X/\Y in { A/B, B/C, C/D, D/A}
            \node[inner sep=0.5pt, fill=white] at ($( \X )!0.5!( \Y )$) {\footnotesize $X_2$}; 
            \node[inner sep=0.5pt, fill=white] at (0, 0, -1.1) {\footnotesize $Z_1$};
    \end{tikzpicture},
    \hspace{25pt}
    \begin{tikzpicture}[scale= 1, baseline={([yshift=-2.5pt]current bounding box.center)}]
        \coordinate (A) at (0,0,0);
        \coordinate (B) at (2,0,0);
        \coordinate (C) at (2,0,2);
        \coordinate (D) at (0,0,2);
%
        \coordinate (O) at (2,2,2);
        \coordinate (U) at (0,-2,0);
%
%
        \draw  (A) -- (B) -- (C) -- (D) -- cycle;
        \draw [dotted] (A) -- (U);
%
 %
        \foreach \X/\Y in { A/B, B/C, C/D, D/A}
        \node[inner sep=0.5pt, fill=white] at ($( \X )!0.5!( \Y )$) {\footnotesize $X_2$};
        \node[inner sep=0.5pt, fill=white] at (0,-1.3,0) {\footnotesize $Z_1$};        
    \end{tikzpicture}.
\end{align}
After gauging, there are two dual $1$-form symmetries given by two copies of~\eqref{eq:dual_1_form}
\begin{align}
    B_{v,1}
    \hspace{5pt}=\hspace{5pt}
    \begin{tikzpicture}[scale= 1, baseline={([yshift=-2.5pt]current bounding box.center)}]
    %
        \coordinate (O) at (0,0,0);
        \coordinate (A) at (-1.5,0,0);
        \coordinate (B) at (1.5,0,0);
        \coordinate (C) at (0,-1.5,0);
        \coordinate (D) at (0,1.5,0);
        \coordinate (E) at (0,0,1.5);
        \coordinate (F) at (0,0,-1.5);
        %
        \draw (A) -- (B);
        \draw (C) -- (D); 
        \draw (E) -- (F);
        \foreach \X/\Y in { O/E, O/F}
        \node[inner sep=0.5pt, fill=white] at ($( \X )!0.5!( \Y )$) {\footnotesize $Z_1$};
        \foreach \X/\Y in {O/C, O/D}
        \node[inner sep=0.5pt, fill=white] at ($( \X )!0.5!( \Y )$) {\footnotesize $Z_1$};
        \foreach \X/\Y in {O/A, O/B}
        \node[inner sep=0.5pt, fill=white] at ($( \X )!0.5!( \Y )$) {\footnotesize $Z_1$}; 
    \end{tikzpicture}
    \hspace{35pt}
    B_{v,1}
    \hspace{5pt}=\hspace{5pt}
    \begin{tikzpicture}[scale= 1, baseline={([yshift=-2.5pt]current bounding box.center)}]
    %
        \coordinate (O) at (0,0,0);
        \coordinate (A) at (-1.5,0,0);
        \coordinate (B) at (1.5,0,0);
        \coordinate (C) at (0,-1.5,0);
        \coordinate (D) at (0,1.5,0);
        \coordinate (E) at (0,0,1.5);
        \coordinate (F) at (0,0,-1.5);
        %
        \draw (A) -- (B);
        \draw (C) -- (D); 
        \draw (E) -- (F);
        \foreach \X/\Y in { O/E, O/F}
        \node[inner sep=0.5pt, fill=white] at ($( \X )!0.5!( \Y )$) {\footnotesize $Z_2$};
        \foreach \X/\Y in {O/C, O/D}
        \node[inner sep=0.5pt, fill=white] at ($( \X )!0.5!( \Y )$) {\footnotesize $Z_2$};
        \foreach \X/\Y in {O/A, O/B}
        \node[inner sep=0.5pt, fill=white] at ($( \X )!0.5!( \Y )$) {\footnotesize $Z_2$}; 
    \end{tikzpicture}
\end{align}
However, it is clear that $B_{v,i}$ are just products of the plaquette terms around a cube $c$
\begin{align}
    B_{v,1} = \prod_{p\in c} K_{p,2},
    \hspace{25pt}
    B_{v,2} = \prod_{p\in c} K_{p,1}.
\end{align}
Therefore, the vertex terms are redundant. The Hamiltonian of the RBH cluster state is
\begin{align}
    H = -\sum_p K_{p,1} -\sum_p K_{p,2}.
\end{align}

\subsection{Semionic Walker-Wang model}\label{app:ta_s_ww}

The topological action of the semionic Walker-Wang model is given by
\begin{align}
    \mathcal{L}[A] = \frac{1}{4} P(A)
    \hspace{35pt}
    P(A) = A \cup A + A\cup_1 \delta A
\end{align}
where $A$ is a $2$-form gauge field with $\mathbb{Z}_2$  coefficient, and $P(A)$ is the Pontryagin square~\cite{kapustin2017fermionic, sun2026non}, and $\cup_1$ is the higher cup product measuring the failure of graded-commutivity of $\cup$ on cochains. We again refer to Ref.~\cite{chen2023higher} for detailed definitions on a cubic lattice. The semionic Walker-Wang model has been formulated on a cubic lattice whose vertices are resolved into trivalent junctions~\cite{vonkeyserlingk2013three}. Here, we derive a formulation directly on the cubic lattice, without resolving its vertices into trivalent junctions. A recent paper, Ref.~\cite{sun2026non}, presents a related formulation on the unresolved cubic lattice using a condensation-based approach.

We first ungauge the semionic Walker-Wang model. However, because the Pontryagin square lifts the coefficient from $\mathbb{Z}_2$ to $\mathbb{Z}_4$, this step requires care. Instead of substituting $A \mapsto \delta a$, we use the replacement described in Ref.~\cite{chen2021exactly}:
\begin{align}
    A \mapsto \delta a + 2(a \cup a + a\cup_1 \delta a)
\end{align}
where $a$ is a $1$-cochain with $\mathbb{Z}_2$ coefficient and $\delta$ is now the coboundary map in $\mathbb{Z}$ coefficient. We have
\begin{equation}
\begin{aligned}
    P(\delta a) &=
    \delta a \cup \delta a + 2 \delta a \cup (a\cup a + a\cup_1 \delta a)
    + 2 (a\cup a + a\cup_1 \delta a) \cup \delta a
    + 2
    \delta a \cup_1 \delta (a\cup a + a\cup_1 \delta a)\\
    & = \delta(a\cup \delta a + 2 \delta a \cup_1 (a\cup a + a\cup_1 \delta a))
\end{aligned}
\end{equation}
where we have used
\begin{align}
    \delta(a \cup_1 b) = \delta a\cup_1 b + a\cup_1 \delta b + a\cup b + b\cup a
\end{align}
As a result, we have
\begin{align}
    \mathcal{L}[\delta a] = \delta\mathcal{U}[a]
    \hspace{35pt}
    \mathcal{U}[a] = \frac{1}{4}a\cup \delta a + \frac{1}{2}( \delta a \cup_1 (a\cup a) +\delta a \cup_1( a\cup_1 \delta a))
\end{align}
Note the ordering of the cup products matters. After ungauging, we can again restrict to the boundary where the Hilbert space is defined. Since the field $a$ is a $1$-cochain with $\mathbb{Z}_2$ coefficient, the Hilbert space is again given by one qubit per edge. The ground state is
\begin{align}
    \ket{\psi} = \sum_{a} e^{2\pi i\mathcal{U}[a]} \ket{a}
    =
    U \ket{\psi_0}
    \hspace{35pt}
    \ket{\psi_0} = \sum_e \ket{+}_e
\end{align}
where $U$ is the entangler given by the phases $e^{2\pi i\mathcal{U}[a]}$. Thus the ungauged Hamiltonian is given by $UH_0U^\dagger$, and $H_0 = -\sum_e X_e$. The entangler $U$ is much more complicated than before. Since the coefficient of $\mathcal{U}[a]$ is $\mathbb{Z}_4$, the entangler $U$ is in the third Clifford Hierarchy given by a product of C$S$, C$S^\dagger$, CCZ, CZ and $Z$ gates, where C$S$ is the controlled-phase gate, and CCZ is the controlled-controlled-$Z$ gate.

Similar to Appendix~\ref{app:ta_ftc}, we fix an edge, and focus on all the gates in $U$ which have support on this edge. Due to the complexity of the presentation, we only exhibit the gates acting on the edge $e$ that points into the page. The rest of the two orientations can be worked out following the same procedure. First we consider the quadratic term $\frac{1}{4} a\cup \delta a$, which simply gives C$S$ and C$S^\dagger$ gates (recall $\delta$ is the cobondary in $\mathbb{Z}$ coefficient). Pictorially, it gives the entangler
\begin{align}
    \begin{tikzpicture}[scale= 1, baseline={([yshift=-2.5pt]current bounding box.center)}]
%
            \coordinate (A) at (0,0,0);
            \coordinate (B) at (0,2,0);
            \coordinate (C) at (2,2,0);
            \coordinate (D) at (2,0,0);
            \coordinate (E) at (2,2,2);
            \coordinate (F) at (2,4,2);
            \coordinate (G) at (4,4,2);
            \coordinate (H) at (4,2,2);
%
            \coordinate (O) at (2,2,2);
%
%
            %
            \draw (A) -- (B);
            \draw (B) -- (1.23, 2, 0);
            \draw [dotted](C) -- (1.23, 2, 0);
            \draw (D) -- (2,1.23,  0);
            \draw [dotted](C) -- (2,1.23, 0);
            \draw (D) -- (A);
            \draw (E) -- (F) --(G) -- (H) -- cycle;
            \draw [dashed, blue](C)-- (O);
            \foreach \X/\Y in {A/B, B/C, E/F, F/G}
            \draw [pink] ($( \X )!0.5!( \Y )$) -- ($(C)!0.5!(O)$);
            \foreach \X/\Y in {  C/D, D/A, G/H, H/E}
            \draw [myGreen] ($( \X )!0.5!( \Y )$) -- ($(C)!0.5!(O)$);
            \foreach \X/\Y in { A/B, B/C, C/D, D/A, E/F, F/G, G/H, H/E, C/O}
            \node [fill, circle, red, inner sep = 1pt] at ($( \X )!0.5!( \Y )$) {}; 
    \end{tikzpicture}
    \hspace{45pt}
    \begin{tikzpicture}[scale = 0.5, baseline = -0.1cm]
    \draw [pink] (0,0) -- (1,0);
    \node [fill, circle, red, inner sep = 1pt] at (0,0) {};
    \node [fill, circle, red, inner sep = 1pt] at (1,0) {};
\end{tikzpicture} = CS^\dagger
\hspace{15pt}
\begin{tikzpicture}[scale = 0.5, baseline = -0.1cm]
    \draw [myGreen] (0,0) -- (1,0);
    \node [fill, circle, red, inner sep = 1pt] at (0,0) {};
    \node [fill, circle, red, inner sep = 1pt] at (1,0) {};
\end{tikzpicture} = CS,
\end{align}
where the blue edge is the edge $e$. The two cubic terms are more complicated. When none of the three factors are the same, it gives a CCZ gate. When two of the three factors coincide and the third is distinct, it gives a CZ gate. When all three factors coincide, it gives a $Z$ gate. First look at the cubic term $\frac{1}{2}\delta a \cup_1( a\cup_1 \delta a)$, the CCZ part of this term acting on the edge $e$ is given by

\begin{minipage}{0.3\textwidth}
\begin{gather*}
    \frac{1}{2}\delta \textcolor{blue}{a} \cup_1 (a \cup_1 \delta a)\\
    \includegraphics[width=1\linewidth]{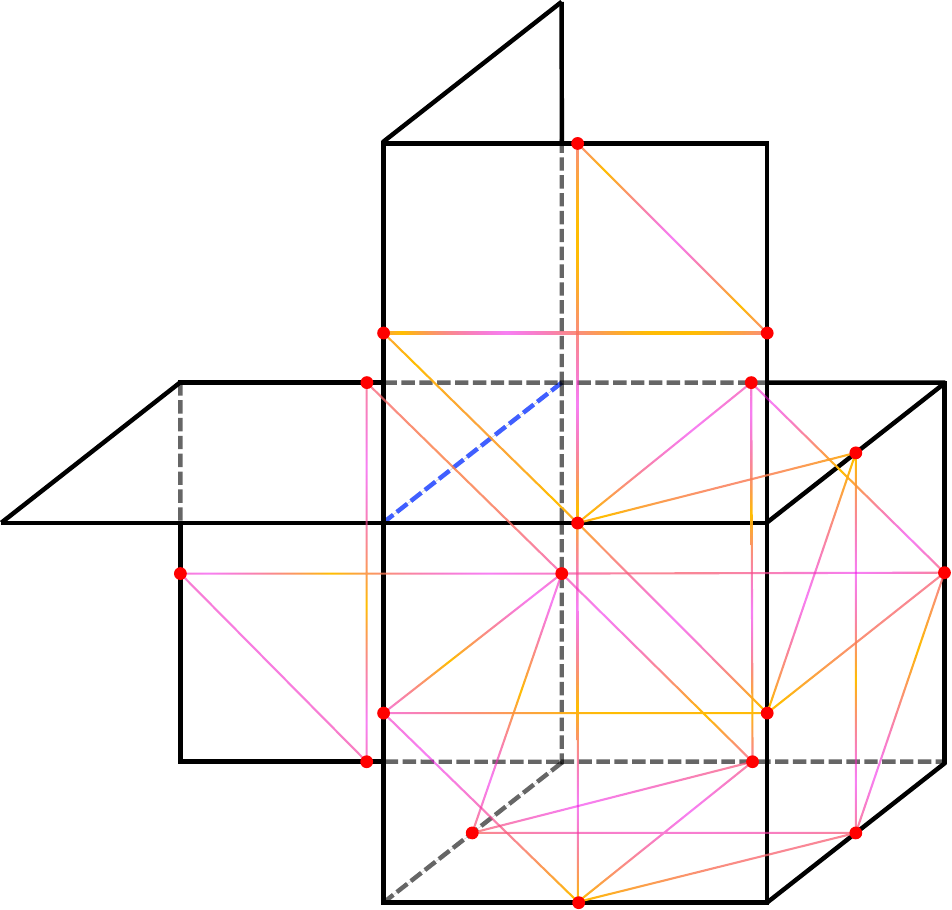}
\end{gather*}
\end{minipage}
\hspace{25pt}
\begin{minipage}{0.3\textwidth}
\begin{gather*}
\frac{1}{2}\delta a \cup_1 (\textcolor{blue}{a} \cup_1 \delta a)\\\includegraphics[width=1\linewidth]{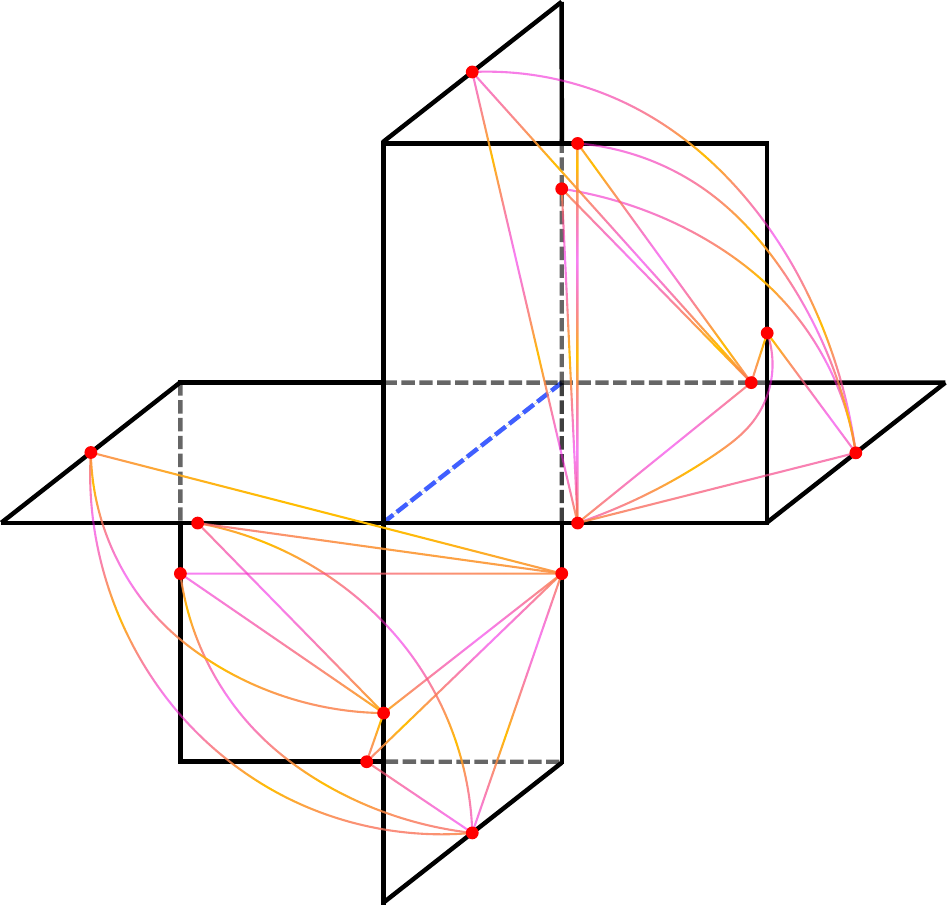}
\end{gather*}
\end{minipage}
\hspace{25pt}
\begin{minipage}{0.3\textwidth}
\begin{gather*}
    \frac{1}{2}\delta a \cup_1 (a \cup_1 \delta \textcolor{blue}{a})\\\includegraphics[width=1\linewidth]{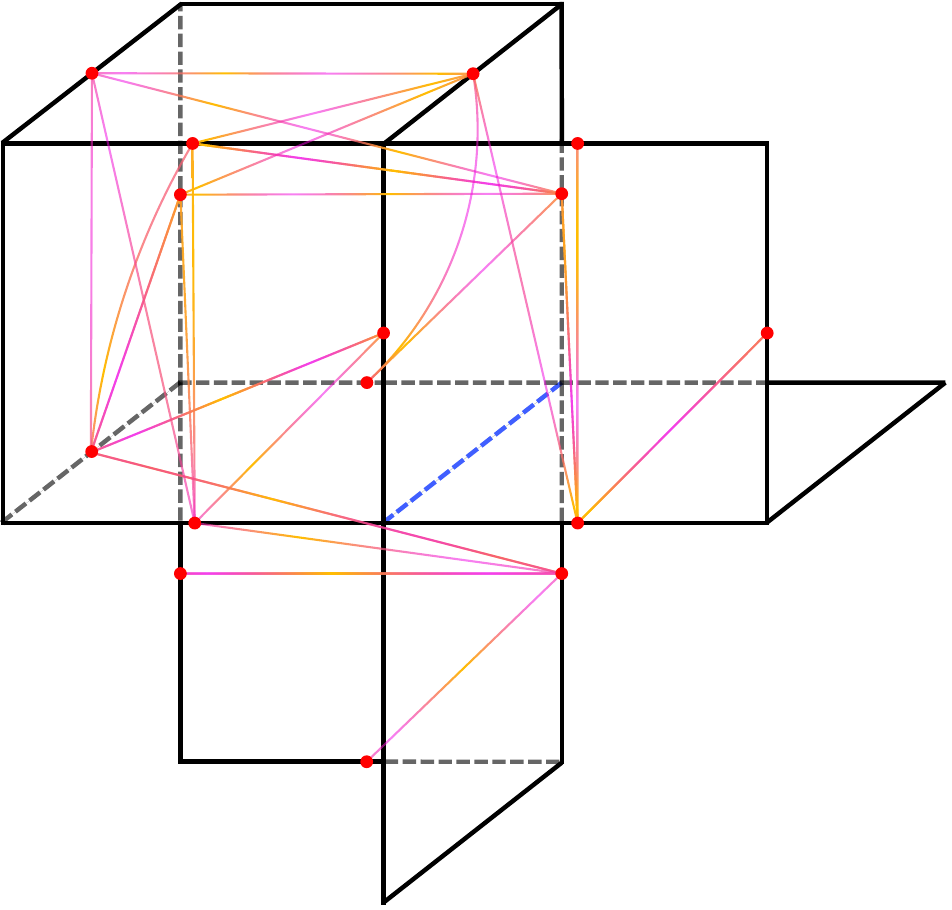}
\end{gather*}
\end{minipage}

\noindent Note in the figures here, the rainbow line is \textit{not} a CZ gate, it is a CCZ gate whose support is the two qubits at the end points of the rainbow line, as well as the qubit on the blue edge. The blue edge is again the edge $e$, and the blue factors \textcolor{blue}{$a$} in the expressions above the figures indicate which factor of $a$ correspond the the blue edge $e$. The CZ part of $\frac{1}{2}\delta a \cup_1( a\cup_1 \delta a)$ arise when two of the three factors coincide, which gives entanglers

\begin{minipage}{0.3\textwidth}
\begin{gather*}
\frac{1}{2}\delta a \cup_1 (\textcolor{blue}{a} \cup_1 \delta \textcolor{blue}{a} )\\
\includegraphics[width=1\linewidth]{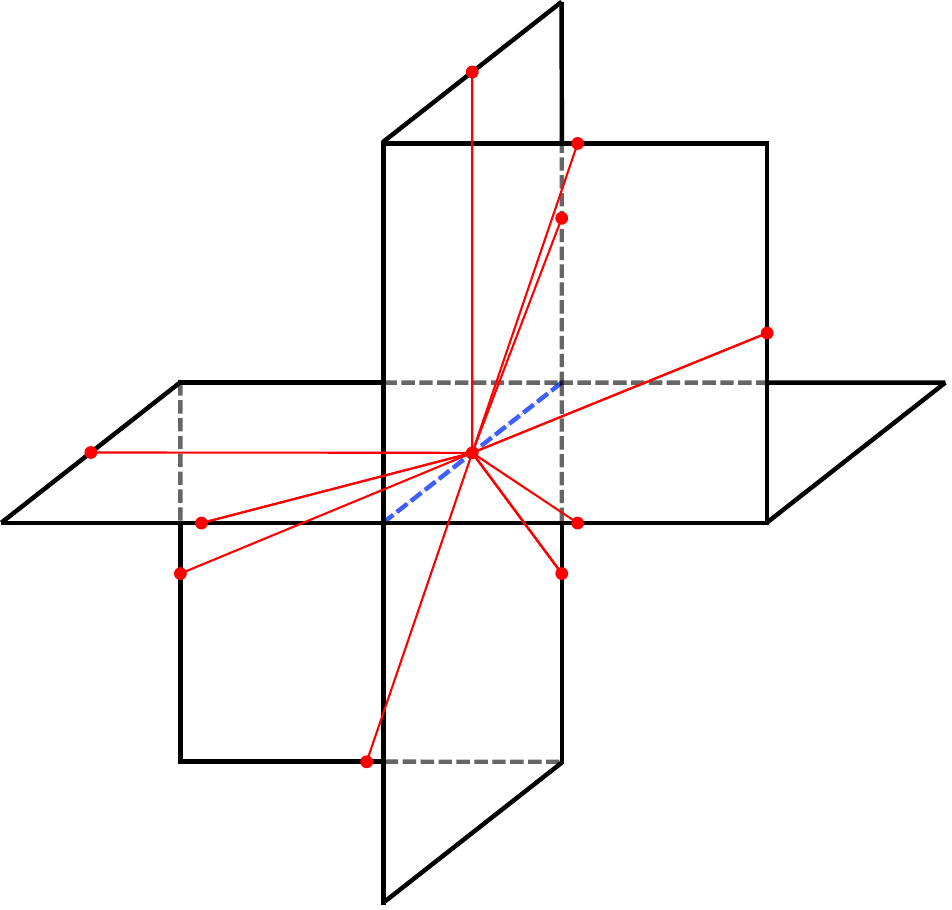}
\end{gather*}
\end{minipage}
\begin{minipage}{0.3\textwidth}
\begin{gather*}
\frac{1}{2}\delta \textcolor{blue}{a} \cup_1 (a \cup_1 \delta \textcolor{blue}{a} )\\
\includegraphics[width=1\linewidth]{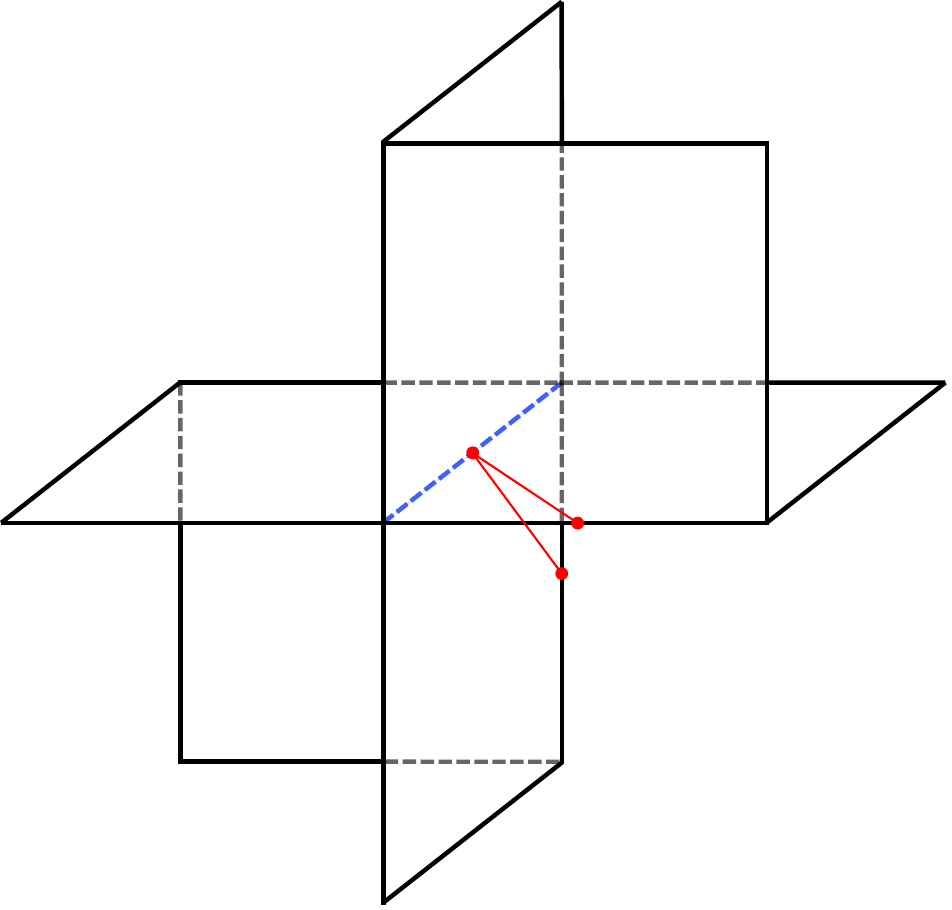}
\end{gather*}
\end{minipage}
\begin{minipage}{0.3\textwidth}
\begin{gather*}
\frac{1}{2}\delta \textcolor{blue}{a} \cup_1 (\textcolor{blue}{a} \cup_1 \delta a )\\
\includegraphics[width=1\linewidth]{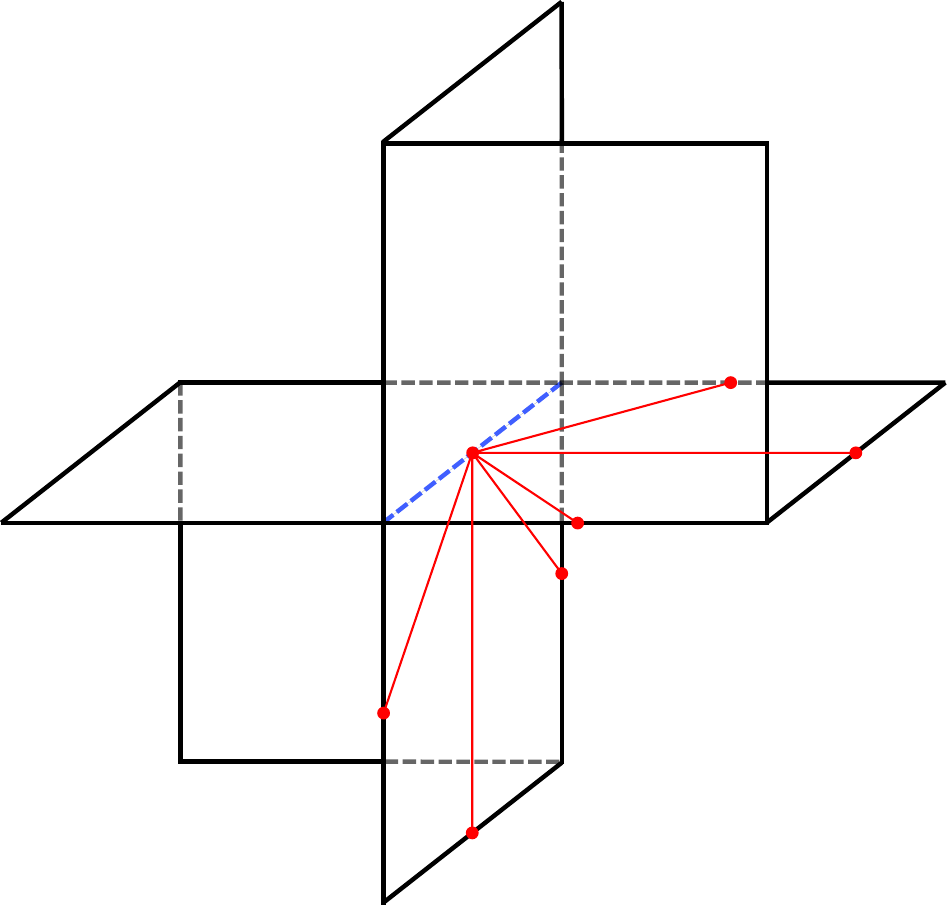}
\end{gather*}
\end{minipage}

\begin{minipage}{0.3\textwidth}
\begin{gather*}
\frac{1}{2}\delta \textcolor{blue}{a} \cup_1 (\tikzmarknode{term1}{a}\cup_1 \delta \tikzmarknode{term2}{a} ) \\[0.5cm]
\includegraphics[width=1\linewidth]{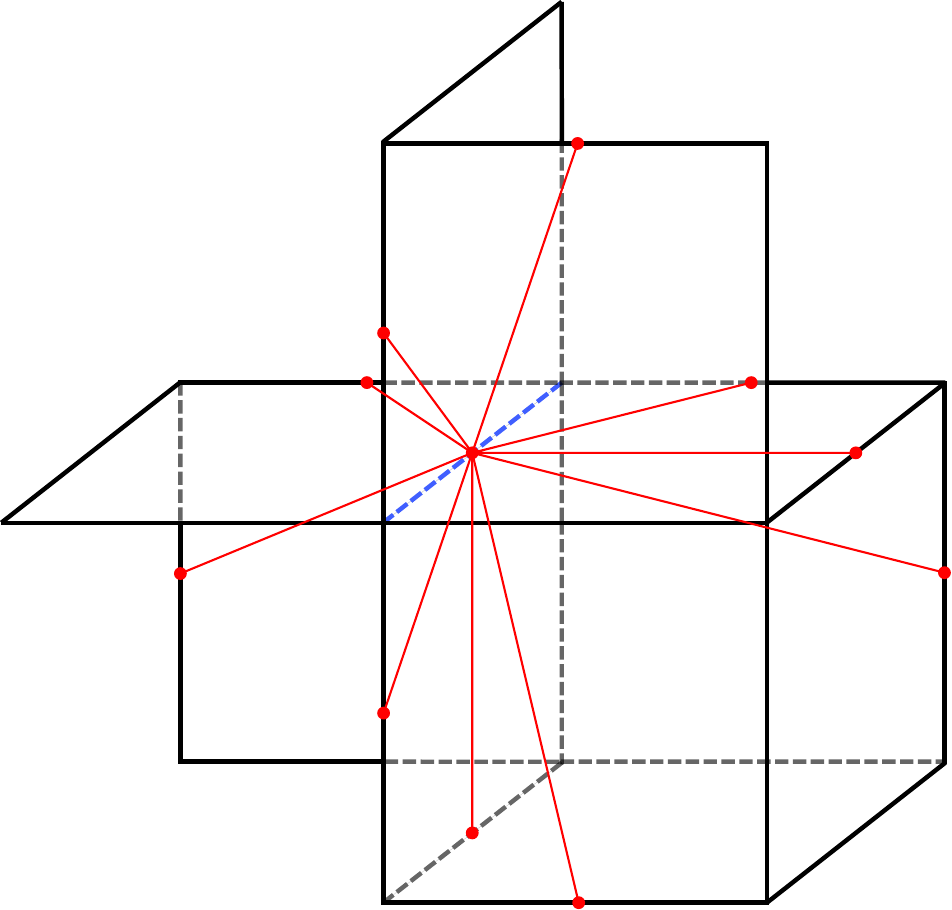}
\end{gather*}
\begin{tikzpicture}[remember picture, overlay]
  \draw[thick, black, <->, >=stealth] ([yshift=-2pt]term1.south) to[out=-60, in=-120] 
  node[below, font=\small] {same}([yshift=-2pt]term2.south);
\end{tikzpicture}
\end{minipage}
\begin{minipage}{0.3\textwidth}
\begin{gather*}
\frac{1}{2}\delta \tikzmarknode{term1}{a} \cup_1 (\textcolor{blue}{a} \cup_1 \delta \tikzmarknode{term2}{a} )\\[0.5cm]
\includegraphics[width=1\linewidth]{figures/daada_CZ_2.pdf}
\end{gather*}
\begin{tikzpicture}[remember picture, overlay]
  \draw[thick, black, <->, >=stealth] ([yshift=-2pt]term1.south) to[out=-30, in=-150] 
  node[below, font=\small] {same}([yshift=-2pt]term2.south);
\end{tikzpicture}
\end{minipage}
\begin{minipage}{0.3\textwidth}
\begin{gather*}
\frac{1}{2}\delta \tikzmarknode{term1}{a} \cup_1 ( \tikzmarknode{term2}{a}\cup_1 \delta  \textcolor{blue}{a})\\[0.5cm]\includegraphics[width=1\linewidth]{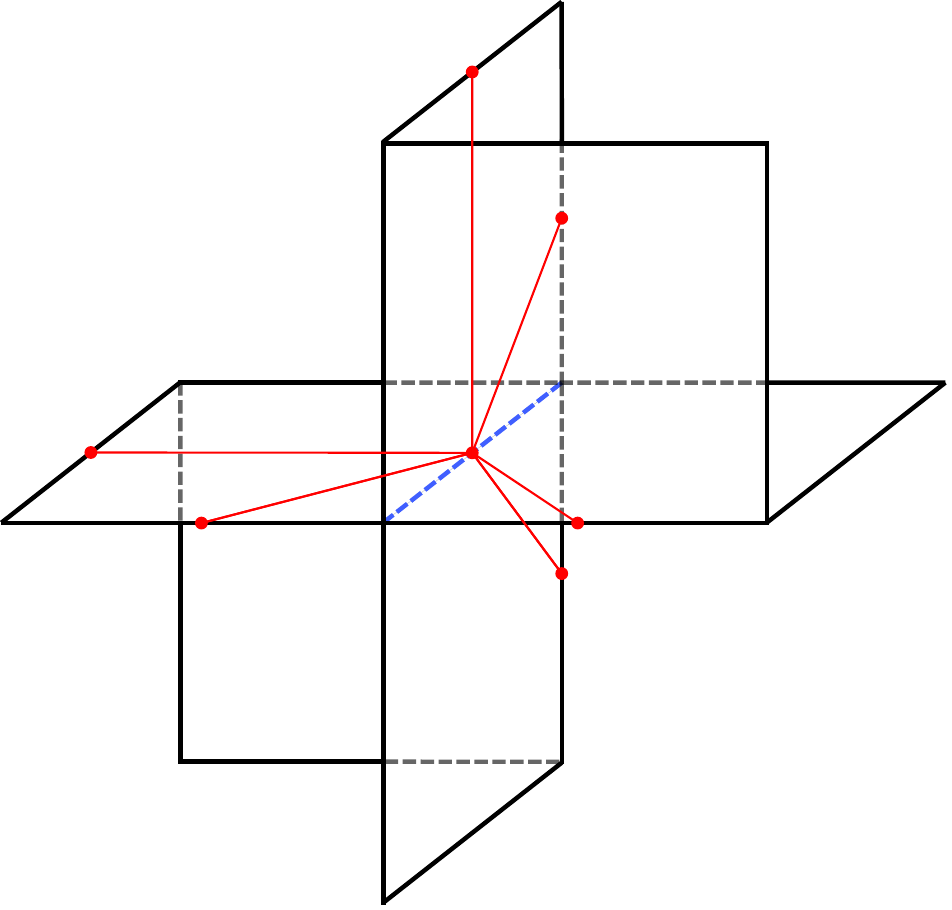}
\end{gather*}

\begin{tikzpicture}[remember picture, overlay]
  \draw[thick, black, <->, >=stealth] ([yshift=-2pt]term1.south) to[out=-60, in=-120]
  node[below, font=\small] {same}([yshift=-2pt]term2.south);
\end{tikzpicture}
\end{minipage}

\noindent In the six figures above, the red line $\begin{tikzpicture}[scale = 0.5, baseline = -0.1cm]
    \draw [red] (0,0) -- (1,0);
    \node [fill, circle, red, inner sep = 1pt] at (0,0) {};
    \node [fill, circle, red, inner sep = 1pt] at (1,0) {};
\end{tikzpicture}$ is the CZ gate. The three terms on the first line is the CZ gate arise due to the blue edge $e$ appears twice in the cubit term, and the double occurrence is indicated by the blue $a$ in the expressions immediately above the figures. On the other hand, the three terms one the second line is the CZ gate arise due the blue edge $a$ appears once, but the rest of the two factors coincide, and the factor which appears twice is indicated by the expressions immediately above the figures. Finally, when all three factors coincide, that is $\frac{1}{2}\delta \textcolor{blue}{a}\cup_1 ( \textcolor{blue}{a}\cup_1 \delta  \textcolor{blue}{a})$, a $Z$ gate can arise. However, the $Z$ gates cancel in this case.

Next we look at the cubic term $\frac{1}{2} \delta a \cup_1 (a\cup a)$, the CCZ part of this term acting on the edge $e$ is given by

\hspace{50pt}
\begin{minipage}{0.3\textwidth}
\begin{gather*}
\frac{1}{2}\delta \textcolor{blue}{a} \cup_1 (a \cup  a)\\
\includegraphics[width=1\linewidth]{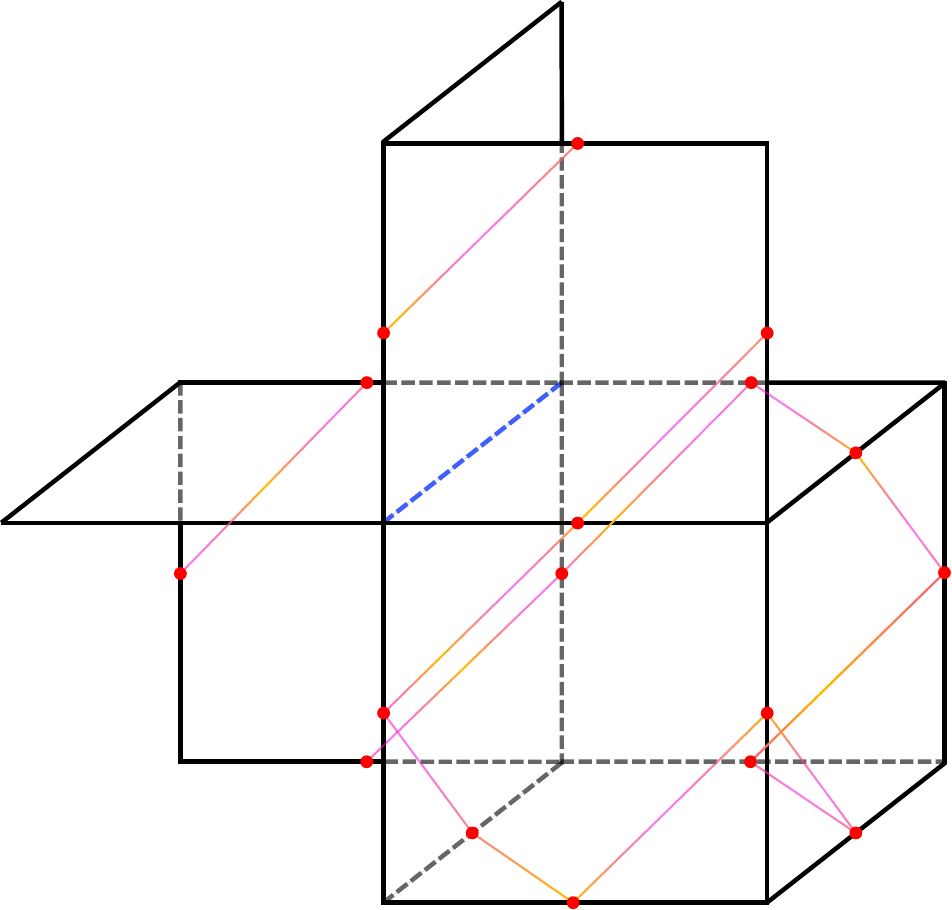}
\end{gather*}
\end{minipage}
\begin{minipage}{0.3\textwidth}
\hspace{35pt}
\end{minipage}
\hspace{-100pt}
\begin{minipage}{0.3\textwidth}
\begin{gather*}
\frac{1}{2}\delta a \cup_1 (\textcolor{blue}{a} \cup  a)
\hspace{5pt}\&\hspace{5pt}
\frac{1}{2}\delta a \cup_1 (a \cup  \textcolor{blue}{a})\\
\includegraphics[width=1\linewidth]{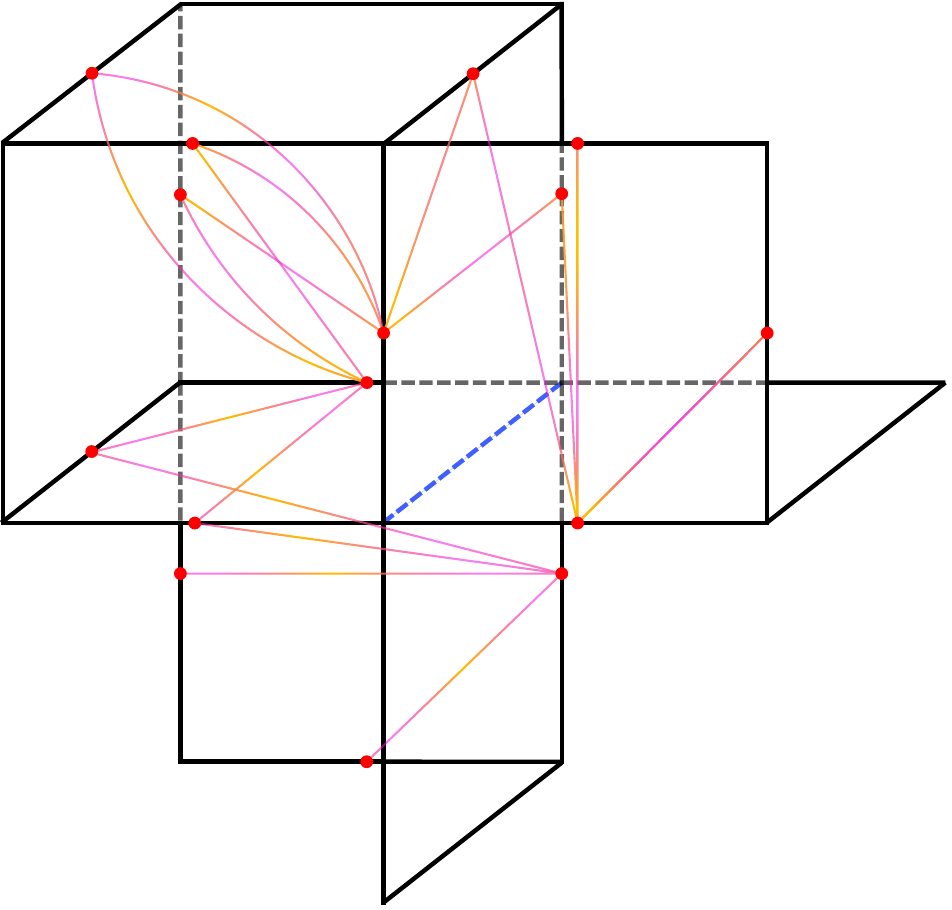}
\end{gather*}
\end{minipage}

\noindent where the rainbow lines are again the CCZ gates whose support is the two qubits at the end points of the rainbow line, as well as the qubit on the blue edge. The blue \textcolor{blue}{$a$} in the expressions indicate which factor the blue edge $e$ corresponds to. Note the second terms is two contributions combined. Next consider the CZ part of $\frac{1}{2}\delta a \cup_1 (a \cup a )$. The only nonzero contributions are

\hspace{50pt}
\begin{minipage}{0.3\textwidth}
\vspace{-0.63cm}
\begin{gather*}
\frac{1}{2}\delta \textcolor{blue}{a} \cup_1 (\textcolor{blue}{a} \cup a )
\hspace{5pt}\& \hspace{5pt}
\frac{1}{2}\delta \textcolor{blue}{a} \cup_1 ( a\cup \textcolor{blue}{a})\\[0.5cm]
\includegraphics[width=1\linewidth]{figures/daada_CZ_2.pdf}
\end{gather*}
\end{minipage}
\begin{minipage}{0.3\textwidth}
\hspace{35pt}
\end{minipage}
\hspace{-100pt}
\begin{minipage}{0.3\textwidth}
\begin{gather*}
\frac{1}{2}\delta \tikzmarknode{term1}{a} \cup_1 (\textcolor{blue}{a} \cup \tikzmarknode{term2}{a} )
\hspace{5pt}\& \hspace{5pt}
\frac{1}{2}\delta \tikzmarknode{term3}{a} \cup_1 ( \tikzmarknode{term4}{a}\cup \textcolor{blue}{a})\\[0.4cm]
\includegraphics[width=1\linewidth]{figures/daada_CZ_2.pdf}
\end{gather*}
\begin{tikzpicture}[remember picture, overlay]
  \draw[thick, black, <->, >=stealth] ([yshift=-2pt]term1.south) to[out=-30, in=-150] 
  node[below, font=\small] {same}([yshift=-2pt]term2.south);
\end{tikzpicture}
\begin{tikzpicture}[remember picture, overlay]
  \draw[thick, black, <->, >=stealth] ([yshift=-2pt]term3.south) to[out=-60, in=-120]
  node[below, font=\small] {same}([yshift=-2pt]term4.south);
\end{tikzpicture}
\end{minipage}

\noindent where the red lines $\begin{tikzpicture}[scale = 0.5, baseline = -0.1cm]
    \draw [red] (0,0) -- (1,0);
    \node [fill, circle, red, inner sep = 1pt] at (0,0) {};
    \node [fill, circle, red, inner sep = 1pt] at (1,0) {};
\end{tikzpicture}$ is the CZ gates. Finally, when all three factors coincide, $\frac{1}{2}\delta \textcolor{blue}{a}\cup_1 ( \textcolor{blue}{a}\cup \textcolor{blue}{a})$ gives a $Z$ gate, which is again canceled in this case.

The product of all the figures above gives the component of the entangler $U$ supported on the blue edge $e$ that points into the page. The support of $U$ on edges in the other two orientations can be derived similarly. After writing down all the terms in $U$, one can check that this entangler is symmetric under the usual $1$-form symmetry in~\eqref{eq:1_form}.

Recall that the ungauged Hamiltonian is given by $UH_0U^\dagger$, where $H_0 = -\sum_e X_e $. We again focus on an edge $e$ that points into the page. The term $U X_e U^\dagger$ expands into dozens of terms, making it difficult to work with. Instead, we can break $UX_eU^\dagger$ into the following simpler building blocks:
\hspace{50pt}
\begin{align}\label{eq:building_block}
G_S 
\hspace{5pt} := \hspace{5pt}
\raisebox{-2cm}{\includegraphics[width=0.25\linewidth]{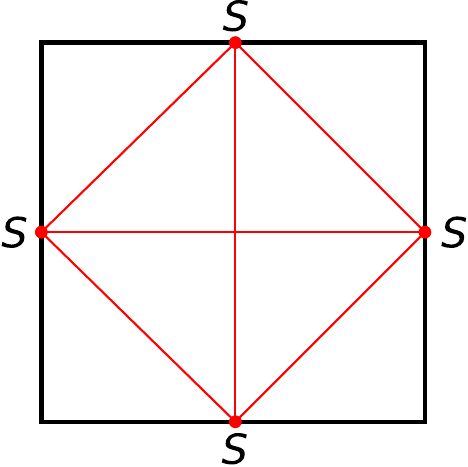}}
\hspace{50pt}
\vspace{-0.5cm}
G_{CZ} 
\hspace{5pt} := \hspace{5pt}
\raisebox{-1.5cm}{\includegraphics[width=0.25\linewidth]{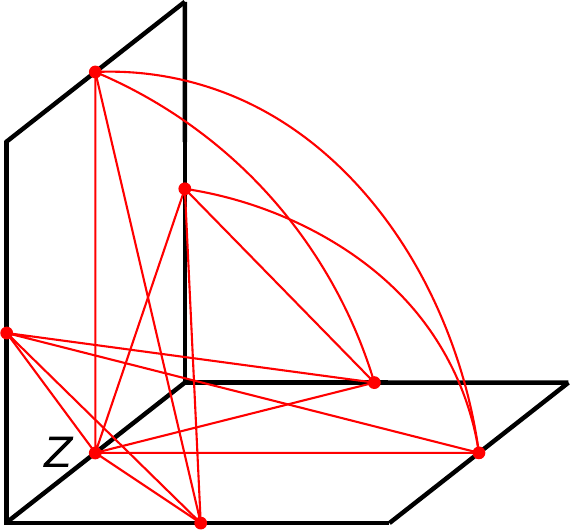}}
\end{align}
\noindent The operators $G_S$ and $G_{CZ}$ should be thought of as a plaquette operator and a two-plaquette operator respectively. Similarly we define $G_{S^\dagger}$, which is given by replacing the $S$ in $G_S$ by $S^\dagger$, and we do not draw it out here. A nice thing about the building blocks $G_S$ and $G_{CZ}$ is that they are symmetric under the $1$-form symmetry Eq.~\eqref{eq:1_form}. In terms of which we have
\begin{gather*}
    UX_eU^\dagger 
    = X_e
    G_{S^\dagger}^{(1)}G_{S^\dagger}^{(2)}G_S^{(3)}G_S^{(4)}G_{S^\dagger}^{(5)}G_S^{(6)}G_S^{(13)}G_S^{(14)}
    G_{CZ}^{(1,10)}
    G_{CZ}^{(1,13)}
    G_{CZ}^{(3,8)}
    G_{CZ}^{(3,14)}
    G_{CZ}^{(9,11)}
    G_{CZ}^{(7,12)}
    G_{S^\dagger}^{(7)}
    G_{S^\dagger}^{(8)}
    G_S^{(9)}
    G_S^{(10)}
    G_S^{(11)}
    G_{S^\dagger}^{(12)}\\[0.3cm]
    \includegraphics[width=0.35\linewidth]{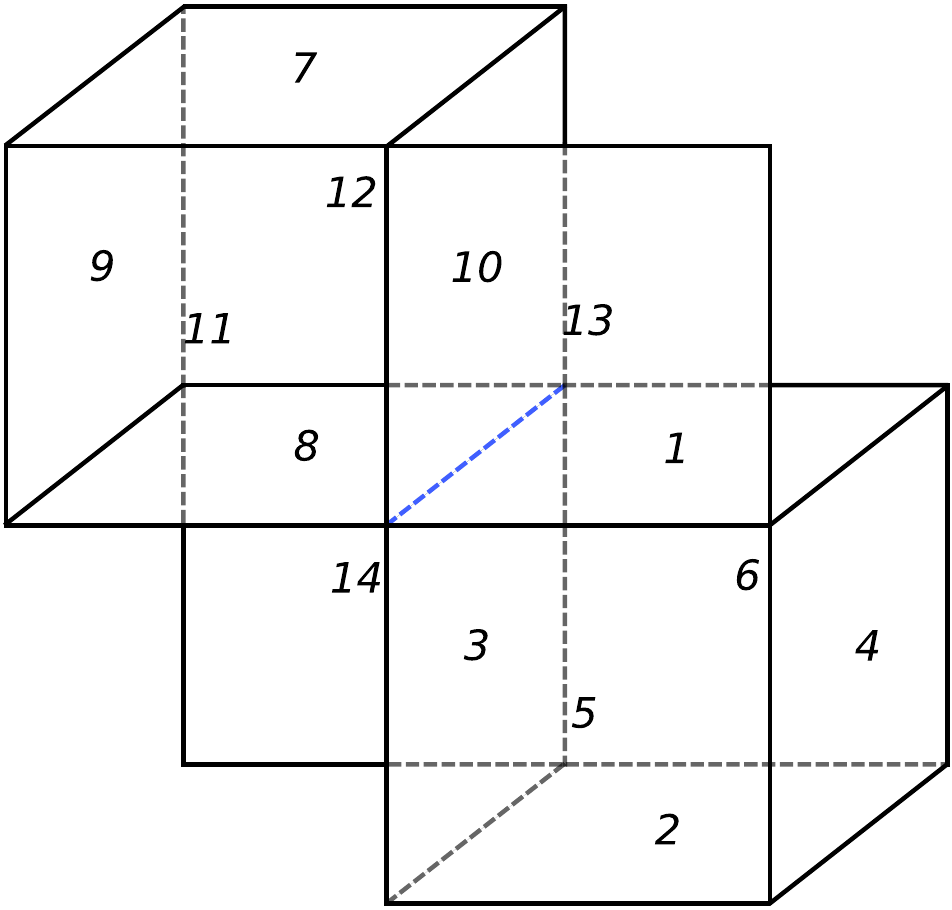}
\end{gather*}
where the superscripts $G_S^{(i)}$ and $G_{CZ}^{(i,j)}$ indicates which plaquette or pair of plaquettes the operators act on, and the numbering of the plaquettes in given by the figure above, the blue edge is the edge $e$. In this form, it is obvious the $UX_eU^\dagger$ is symmetric under the $1$-form symmetry Eq.~\eqref{eq:1_form}.

Finally, in order to obtain the Hamiltonian of $\mathcal{L}[A]$, we gauge the $1$-form symmetry Eq.~\eqref{eq:1_form}. In addition to the gauging map in~\eqref{eq:gauge_1_form}, gauging also gives
\begin{align*}
    G_S^{(i)} \mapsto S_i
    \hspace{35pt}
    G_{S^\dagger}^{(i)} \mapsto S^\dagger_i
    \hspace{35pt}
    G_{CZ}^{(i,j)} \mapsto CZ_{ij}
\end{align*}
After gauging $UX_eU^\dagger$ and going to the dual lattice, we obtain

\begin{align*}
    UX_eU^\dagger 
    \hspace{5pt}\mapsto \hspace{5pt}
    K_p
    \hspace{5pt}:=\hspace{5pt}
    \begin{tikzpicture}[scale = 1, baseline = 0.8cm]
%
        \coordinate (A) at (0,0,0);
        \coordinate (B) at (0,2,0);
        \coordinate (C) at (2,2,0);
        \coordinate (D) at (2,0,0);
%
        \coordinate (U) at (0,0,-2);
        \coordinate (O) at (2,2,2);
%
%
%
       %
        \draw [dotted] (A) -- (U);
        \draw (C) -- (O);
        \draw (0,0,0) -- (3.5,0,0);
        \draw (2,-1.5,0) -- (2,0,0);
        \draw (2,0,-1.5) -- (2,0,1.5);
        \draw (0,2,0) -- (-1.5,2,0);
        \draw (0,3.5,0) -- (0,2,0);
        \draw (0,2,-1.5) -- (0,2,1.5);
        \draw [blue] (A) -- (B) --(C) -- (D) -- cycle;
        \node [inner sep=0.5pt, fill=white] at (1, 2, 0) {\footnotesize$X\textcolor{red}{S}$};
        \node [inner sep=0.5pt, fill=white] at (0, 1, 0) {\footnotesize$X\textcolor{red}{S^\dagger}$};
        \node [inner sep=0.5pt, fill=white] at (1, 0, 0) {\footnotesize$X\textcolor{red}{S}$};
        \node [inner sep=0.5pt, fill=white] at (2, 1, 0) {\footnotesize$X\textcolor{red}{S^\dagger}$};
        \node [inner sep=0.5pt, fill=white] at (0, 0, -1) {\footnotesize$\textcolor{red}{S}$};
        \node [inner sep=0.5pt, fill=white] at (2, 2, 1) {\footnotesize$\textcolor{red}{S}$};
        \node [inner sep=0.5pt, fill=white] at (0, 3, 0) {\footnotesize$\textcolor{red}{S^\dagger}$};
        \node [inner sep=0.5pt, fill=white] at (-1, 2, 0) {\footnotesize$\textcolor{red}{S}$};
        \node [inner sep=0.5pt, fill=white] at (0, 2, -1) {\footnotesize$\textcolor{red}{S^\dagger}$};
        \node [inner sep=0.5pt, fill=white] at (0, 2, 1) {\footnotesize$\textcolor{red}{S}$};
        \node [inner sep=0.5pt, fill=white] at (3, 0, 0) {\footnotesize$\textcolor{red}{S}$};
        \node [inner sep=0.5pt, fill=white] at (2, -1, 0) {\footnotesize$\textcolor{red}{S^\dagger}$};
        \node [inner sep=0.5pt, fill=white] at (2, 0, -1) {\footnotesize$\textcolor{red}{S}$};
        \node [inner sep=0.5pt, fill=white] at (2, 0, 1) {\footnotesize$\textcolor{red}{S^\dagger}$};
    \end{tikzpicture}
    \cdot
    \begin{tikzpicture}[scale = 1, baseline = 0.8cm]
%
        \coordinate (A) at (0,0,0);
        \coordinate (B) at (0,2,0);
        \coordinate (C) at (2,2,0);
        \coordinate (D) at (2,0,0);
%
        \coordinate (U) at (0,0,-2);
        \coordinate (O) at (2,2,2);
%
%
%
        %
        \draw [dotted] (A) -- (U);
        \draw (C) -- (O);
        \draw (0,0,0) -- (3.5,0,0);
        \draw (2,-1.5,0) -- (2,0,0);
        \draw (2,0,-1.5) -- (2,0,1.5);
        \draw (0,2,0) -- (-1.5,2,0);
        \draw (0,3.5,0) -- (0,2,0);
        \draw (0,2,-1.5) -- (0,2,1.5);
        \draw [blue] (A) -- (B) --(C) -- (D) -- cycle;
        \node [fill, circle, red, inner sep = 1pt] at (2,2,1) {};
        \node [fill, circle, red, inner sep = 1pt] at (2,1,0) {};
        \node [fill, circle, red, inner sep = 1pt] at (1,2,0) {};
        \node [fill, circle, red, inner sep = 1pt] at (0,0,-1) {};
        \node [fill, circle, red, inner sep = 1pt] at (1,0,0) {};
        \node [fill, circle, red, inner sep = 1pt] at (0,1,0) {};
        \node [fill, circle, red, inner sep = 1pt] at (0,2,-1) {};
        \node [fill, circle, red, inner sep = 1pt] at (0,3,0) {};
        \node [fill, circle, red, inner sep = 1pt] at (0,2,1) {};
        \node [fill, circle, red, inner sep = 1pt] at (-1,2,0) {};
        \draw [red] (2,2,1) -- (2,1,0);
        \draw [red] (1,2,0) -- (2,1,0);
        \draw [red] (0,0,-1) -- (1,0,0);
        \draw [red] (0,1,0) -- (1,0,0);
        \draw [red] (0,2,-1) -- (0,3,0);
        \draw [red] (0,2,1) -- (-1,2,0);
    \end{tikzpicture}
\end{align*}
where the edge $e$ becomes the blue plaquette $p$ on the dual lattice, and the red lines $\begin{tikzpicture}[scale = 0.5, baseline = -0.1cm]
    \draw [red] (0,0) -- (1,0);
    \node [fill, circle, red, inner sep = 1pt] at (0,0) {};
    \node [fill, circle, red, inner sep = 1pt] at (1,0) {};
\end{tikzpicture}$ are again the CZ gates. Similarly, the terms after gauging the edge terms in the rest of the two orientations give the following two plaquette terms $K_p$
\begin{equation*}
    \begin{tikzpicture}[scale = 1, baseline = -0.6cm]
        \draw (-1.5, 0, 0) -- (1.5, 0, 0);
        \draw[dotted] (0, -1.5, 0) -- (0, 0, 0);
        \draw (0, 1.5, 0) -- (0, 0, 0);
        \draw (0, 0, -1.5) -- (0, 0, 1.5);

        \draw (-1.5, 0, 2) -- (1.5, 0, 2);
        \draw (0, -1.5, 2) -- (0, 1.5, 2);
        \draw (0, 0, 0.5) -- (0, 0, 3.5);

        \draw (0.5, 0, 2) -- (3.5, 0, 2);
        \draw (2, -1.5, 2) -- (2, 1.5, 2);
        \draw (2, 0, 0.5) -- (2, 0, 3.5);

        \draw [blue] (0,0,0) -- (2,0,0) -- (2,0,2) -- (0,0,2) -- cycle;

        \node [inner sep=0.5pt, fill=white] at (1,0,0) {\footnotesize$X\textcolor{red}{Z}$};
        \node [inner sep=0.5pt, fill=white] at (0,0,1) {\footnotesize$X\textcolor{red}{S}$};
        \node [inner sep=0.5pt, fill=white] at (1,0,2) {\footnotesize$X$};
        \node [inner sep=0.5pt, fill=white] at (2,0,1) {\footnotesize$X\textcolor{red}{S^\dagger}$};

        \node [inner sep=0.5pt, fill=white] at (0,-1,0) {\footnotesize$\textcolor{red}{S}$};
        \node [inner sep=0.5pt, fill=white] at (-1,0,2) {\footnotesize$\textcolor{red}{S}$};
        \node [inner sep=0.5pt, fill=white] at (0,0,3) {\footnotesize$\textcolor{red}{S}$};
        \node [inner sep=0.5pt, fill=white] at (0,1,2) {\footnotesize$\textcolor{red}{S}$};
        \node [inner sep=0.5pt, fill=white] at (0,-1,2) {\footnotesize$\textcolor{red}{S}$};

        \node [inner sep=0.5pt, fill=white] at (2,0,3) {\footnotesize$\textcolor{red}{S^\dagger}$};
        \node [inner sep=0.5pt, fill=white] at (3,0,2) {\footnotesize$\textcolor{red}{S}$};
        \node [inner sep=0.5pt, fill=white] at (2,-1,2) {\footnotesize$\textcolor{red}{S^\dagger}$};
    \end{tikzpicture}
    \hspace{5pt}
    \cdot 
    \hspace{5pt}
    \begin{tikzpicture}[scale = 1, baseline = -0.6cm]
        \draw (-1.5, 0, 0) -- (1.5, 0, 0);
        \draw[dotted] (0, -1.5, 0) -- (0, 0, 0);
        \draw (0, 1.5, 0) -- (0, 0, 0);
        \draw (0, 0, -1.5) -- (0, 0, 1.5);

        \draw (-1.5, 0, 2) -- (1.5, 0, 2);
        \draw (0, -1.5, 2) -- (0, 1.5, 2);
        \draw (0, 0, 0.5) -- (0, 0, 3.5);

        \draw (0.5, 0, 2) -- (3.5, 0, 2);
        \draw (2, -1.5, 2) -- (2, 1.5, 2);
        \draw (2, 0, 0.5) -- (2, 0, 3.5);

        \draw [blue] (0,0,0) -- (2,0,0) -- (2,0,2) -- (0,0,2) -- cycle;

        \node [fill, circle, red, inner sep = 1pt] at (0,1,0) {};
        \node [fill, circle, red, inner sep = 1pt] at (1,0,0) {};
        \node [fill, circle, red, inner sep = 1pt] at (0,0,-1) {};
        \node [fill, circle, red, inner sep = 1pt] at (-1,0,0) {};
        \node [fill, circle, red, inner sep = 1pt] at (0,0,1) {};
        \node [fill, circle, red, inner sep = 1pt] at (-1,0,2) {};
        \node [fill, circle, red, inner sep = 1pt] at (0,0,3) {};
        \node [fill, circle, red, inner sep = 1pt] at (0,-1,2) {};
        \node [fill, circle, red, inner sep = 1pt] at (2,0,1) {};
        \node [fill, circle, red, inner sep = 1pt] at (2,1,2) {};

        \draw [red] (0,0,-1) -- (1,0,0);
        \draw [red] (0,1,0) -- (1,0,0);
        \draw [red] (2,0,1) -- (1,0,0);

        \draw [red] (-1,0,0) -- (0,0,1);
        
        \draw [red] (-1,0,2) -- (0,-1,2);
        \draw [red] (-1,0,2) -- (0,0,3);
        \draw [red] (0,-1,2) -- (0,0,3);

        \draw [red] (2,1,2) -- (2,0,1);
    \end{tikzpicture}
\end{equation*}
\begin{equation*}
    \begin{tikzpicture}[scale = 1, baseline = 0.6cm]
        \draw [dotted] (-1.5, 0, 0) -- (0, 0, 0);
        \draw (1.5, 0, 0) -- (0, 0, 0);
        \draw (0, -1.5, 0) -- (0, 0, 0);
        \draw (0, 1.5, 0) -- (0, 0, 0);
        \draw (0, 0, -1.5) -- (0, 0, 1.5);

        \draw (-1.5, 2, 0) -- (1.5, 2, 0);
        \draw (0, 0.5, 0) -- (0, 3.5, 0);
        \draw (0, 2, -1.5) -- (0, 2, 1.5);

        \draw (-1.5, 2, 2) -- (1.5, 2, 2);
        \draw (0, 0.5, 2) -- (0, 3.5, 2);
        \draw (0, 2, 0.5) -- (0, 2, 3.5);

        \draw [blue] (0,0,0) -- (0,2,0) -- (0,2,2) -- (0,0,2) -- cycle;
        
        \node [inner sep=0.5pt, fill=white] at (0,1,0) {\footnotesize$X$};
        \node [inner sep=0.5pt, fill=white] at (0,0,1) {\footnotesize$X\textcolor{red}{S^\dagger}$};
        \node [inner sep=0.5pt, fill=white] at (0,1,2) {\footnotesize$X\textcolor{red}{S^\dagger}$};
        \node [inner sep=0.5pt, fill=white] at (0,2,1) {\footnotesize$X$};

        \node [inner sep=0.5pt, fill=white] at (0,0,-1) {\footnotesize$\textcolor{red}{S^\dagger}$};
        \node [inner sep=0.5pt, fill=white] at (1,0,0) {\footnotesize$\textcolor{red}{S^\dagger}$};
        \node [inner sep=0.5pt, fill=white] at (0,-1,0) {\footnotesize$\textcolor{red}{S}$};
        
        \node [inner sep=0.5pt, fill=white] at (0,3,0) {\footnotesize$\textcolor{red}{S}$};
        \node [inner sep=0.5pt, fill=white] at (0,2,-1) {\footnotesize$\textcolor{red}{S}$};
        \node [inner sep=0.5pt, fill=white] at (1,2,0) {\footnotesize$\textcolor{red}{S}$};
        \node [inner sep=0.5pt, fill=white] at (-1,2,0) {\footnotesize$\textcolor{red}{S}$};
        
        \node [inner sep=0.5pt, fill=white] at (0,3,2) {\footnotesize$\textcolor{red}{S^\dagger}$};
        \node [inner sep=0.5pt, fill=white] at (-1,2,2) {\footnotesize$\textcolor{red}{S}$};
        \node [inner sep=0.5pt, fill=white] at (0,2,3) {\footnotesize$\textcolor{red}{S}$};
    \end{tikzpicture}
    \hspace{5pt}
    \cdot
    \hspace{5pt}
    \begin{tikzpicture}[scale = 1, baseline = 0.6cm]
        \draw [dotted] (-1.5, 0, 0) -- (0, 0, 0);
        \draw (1.5, 0, 0) -- (0, 0, 0);
        \draw (0, -1.5, 0) -- (0, 0, 0);
        \draw (0, 1.5, 0) -- (0, 0, 0);
        \draw (0, 0, -1.5) -- (0, 0, 1.5);

        \draw (-1.5, 2, 0) -- (1.5, 2, 0);
        \draw (0, 0.5, 0) -- (0, 3.5, 0);
        \draw (0, 2, -1.5) -- (0, 2, 1.5);

        \draw (-1.5, 2, 2) -- (1.5, 2, 2);
        \draw (0, 0.5, 2) -- (0, 3.5, 2);
        \draw (0, 2, 0.5) -- (0, 2, 3.5);

        \draw [blue] (0,0,0) -- (0,2,0) -- (0,2,2) -- (0,0,2) -- cycle;

        \node [fill, circle, red, inner sep = 1pt] at (0,3,0) {};
        \node [fill, circle, red, inner sep = 1pt] at (0,2,-1) {};
        \node [fill, circle, red, inner sep = 1pt] at (1,2,0) {};
        \node [fill, circle, red, inner sep = 1pt] at (-1,0,0) {};
        \node [fill, circle, red, inner sep = 1pt] at (0,0,1) {};
        \node [fill, circle, red, inner sep = 1pt] at (0,1,2) {};
        \node [fill, circle, red, inner sep = 1pt] at (0,2,1) {};
        \node [fill, circle, red, inner sep = 1pt] at (1,2,2) {};
        \node [fill, circle, red, inner sep = 1pt] at (0,3,2) {};
        \node [fill, circle, red, inner sep = 1pt] at (-1,2,2) {};
        \node [fill, circle, red, inner sep = 1pt] at (0,2,3) {};
        
        \draw [red] (0,3,0) -- (0,2,-1);
        \draw [red] (1,2,0) -- (0,2,-1);
        \draw [red] (1,2,0) -- (0,3,0);

        \draw [red] (-1,0,0) -- (0,0,1);
        \draw [red] (0,1,2) -- (0,0,1);

        \draw [red] (0,2,1) -- (1,2,2);
        \draw [red] (0,3,2) -- (1,2,2);

        \draw [red] (-1,2,2) -- (0,2,3);
    \end{tikzpicture}
\end{equation*}
Together with the dual symmetry $B_v$ in~\eqref{eq:dual_1_form}, the Hamiltonian
\begin{align*}
    H_\s = -\sum_v B_v - \sum_p K_p
\end{align*}
gives the commuting projector Hamiltonian for semion WW on a cubic lattice. Note the plaquette operators are only hermitian up to multiplication of vertex terms. That is, they are hermitian in the symmetric subspace.

It is worth noting the building block operators Eq.~\eqref{eq:building_block} are not independent. Consider a cube with faces numbered as below
\begin{align}\label{eq:cube_numbering}
    \begin{tikzpicture}[scale = 1, baseline = 0.5cm]
    \coordinate (O) at (0,0,0); 
    \coordinate (A) at (2,0,0);
    \coordinate (B) at (2,2,0);
    \coordinate (C) at (0,2,0);
    \coordinate (D) at (0,0,2);
    \coordinate (E) at (2,0,2);
    \coordinate (F) at (2,2,2);
    \coordinate (G) at (0,2,2);
    %
    %
%
    \draw [dotted] (O) -- (A);
    \draw (A) -- (B) -- (C); 
    \draw [dotted] (O) -- (C);
    \draw (D) -- (E) -- (F) -- (G) -- cycle; 
    \draw [dotted] (O) -- (D); 
    \draw (A) -- (E);
    \draw (B) -- (F);
    \draw (C) -- (G);
    \node at (1, 1, 2) {$1$};
    \node at (1, 1, 0) {$2$};
    \node at (1, 2, 1) {$3$};
    \node at (1, 0, 1) {$4$};
    \node at (0, 1, 1) {$5$};\node at (2, 1, 1) {$6$};
    \end{tikzpicture}
\end{align}
Then we have the relation
\begin{align}
    G_{CZ}^{(1,4)}
    G_{CZ}^{(1,5)}
    G_{CZ}^{(4,5)}
    G_{CZ}^{(2,3)}
    G_{CZ}^{(2,6)}
    G_{CZ}^{(3,6)} = 
    G_S^{(1)}
    G_{S}^{(4)}
    G_{S}^{(5)}
    G_{S^\dagger}^{(2)}
    G_{S^\dagger}^{(3)}
    G_{S^\dagger}^{(6)}
\end{align}
which gives the vertex relation after gauging
\begin{align}\label{eq:ver_relation}
    CZ_{1,4}
    CZ_{1,5}
    CZ_{4,5}
    CZ_{2,3}
    CZ_{2,6}
    CZ_{3,6}
    =
    S_1
    S_4
    S_5
    S^\dagger_2
    S^\dagger_3
    S^\dagger_6
\end{align}

Similarly, the semion WW plaquette terms are also not independent. Consider the product of the plaquette terms $K_p$ around a cube. Although the stabilziers mutually commute, we fix an ordering for concreteness
\begin{align*}
\prod_{p\in c} K_p = K_{p_6}K_{p_5}K_{p_4}K_{p_3}K_{p_2}K_{p_1}
\end{align*}
where the ordering is according to Eq.~\eqref{eq:cube_numbering}. We organize the product by moving all Pauli $X$ to the left, which will cancel, leaving only $Z$, $S$, $S^\dagger$ and CZ operators. Specifically, $\prod_{p\in c} K_p = K_{p_6}K_{p_5}K_{p_4}K_{p_3}K_{p_2}K_{p_1}$ is given by
\begin{align*}
\begin{tikzpicture}[scale = 1, baseline = 0.5cm]
    \coordinate (O) at (0,0,0); 
    \coordinate (A) at (2,0,0);
    \coordinate (B) at (2,2,0);
    \coordinate (C) at (0,2,0);
    \coordinate (D) at (0,0,2);
    \coordinate (E) at (2,0,2);
    \coordinate (F) at (2,2,2);
    \coordinate (G) at (0,2,2);
    %
    %
%
    \draw [dotted] (O) -- (A);
    \draw (A) -- (B) -- (C); 
    \draw [dotted] (O) -- (C);
    \draw (D) -- (E) -- (F) -- (G) -- cycle; 
    \draw [dotted] (O) -- (D); 
    \draw (A) -- (E);
    \draw (B) -- (F);
    \draw (C) -- (G);
    \draw (-1.5, 2, 2) -- (1.5,2,2);
    \draw (0, 2, 3.5) -- (0,2,2);
    \draw (0, 2, 2) -- (0,3.5,2);
    %
    \draw (0, 2, -1.5) -- (0,2,0);
    \draw (0, 2, 0) -- (0,3.5,0);
    \draw[dotted] (-1.5, 0, 0) -- (1.5,0,0);
    \draw[dotted] (0, 0, -1.5) -- (0,0,0);
%
    \draw (2, 2, 2) -- (3.5,2,2);
    \node[inner sep=0.5pt, fill=white] at (-1, 0, 0) {\footnotesize$Z$};
    \node[inner sep=0.5pt, fill=white] at (0, 0, -1) {\footnotesize$Z$};
    \node[inner sep=0.5pt, fill=white] at (0, 3, 0) {\footnotesize$Z$};
    \node[inner sep=0.5pt, fill=white] at (0, 2, -1) {\footnotesize$Z$};
    \node[inner sep=0.5pt, fill=white] at (0, 2, 3) {\footnotesize$Z$};
    \node[inner sep=0.5pt, fill=white] at (-1, 2, 2) {\footnotesize$Z$};
    \node[inner sep=0.5pt, fill=white] at (1, 2, 2) {\footnotesize$Z$};
    \node[inner sep=0.5pt, fill=white] at (2, 1, 2) {\footnotesize$Z$};
    \node[inner sep=0.5pt, fill=white] at (2, 2, 1) {\footnotesize$Z$};
    \node[inner sep=0.5pt, fill=white] at (3, 2, 2) {\footnotesize$Z$};
    \end{tikzpicture}
\cdot
\begin{tikzpicture}[scale = 1, baseline = 0.5cm]
    \coordinate (O) at (0,0,0); 
    \coordinate (A) at (2,0,0);
    \coordinate (B) at (2,2,0);
    \coordinate (C) at (0,2,0);
    \coordinate (D) at (0,0,2);
    \coordinate (E) at (2,0,2);
    \coordinate (F) at (2,2,2);
    \coordinate (G) at (0,2,2);
    %
    %
%
    \draw [dotted] (O) -- (A);
    \draw (A) -- (B) -- (C); 
    \draw [dotted] (O) -- (C);
    \draw (D) -- (E) -- (F) -- (G) -- cycle; 
    \draw [dotted] (O) -- (D); 
    \draw (A) -- (E);
    \draw (B) -- (F);
    \draw (C) -- (G);
    \draw (-1.5, 0, 2) -- (1.5,0,2);
    \draw (0, 0, 3.5) -- (0,0,2);
    \draw (0, -1.5, 2) -- (0,0,2);
    \draw (-1.5, 2, 2) -- (1.5,2,2);
    \draw (0, 2, 3.5) -- (0,2,2);
    \draw (0, 2, 2) -- (0,3.5,2);
    \draw (0,2,0) -- (-1.5,2,0);
    \draw [dotted] (0,0,0) -- (0,-1.5,0);
    \draw (3.5, 2, 0) -- (2,2,0);
    \draw (2, 2, -1.5) -- (2,2,0);
    \draw (2, 2, 0) -- (2,3.5,0);
    \draw (3.5, 2, 2) -- (2,2,2);
    \draw (2, 2, 3.5) -- (2,2,2);
    \draw (2, 2, 2) -- (2,3.5,2);
    \draw (3.5, 0, 2) -- (2,0,2);
    \draw (2, 0, 3.5) -- (2,0,2);
    \draw (2, 0, 2) -- (2,-1.5,2);
    \node[inner sep=0.5pt, fill=white] at (0, 0, 1) {\footnotesize$S$};
    \node[inner sep=0.5pt, fill=white] at (1, 0, 0) {\footnotesize$Z$};
    \node[inner sep=0.5pt, fill=white] at (0, -1, 0) {\footnotesize$Z$};
    \node[inner sep=0.5pt, fill=white] at (-1, 0, 2) {\footnotesize$S$};
    \node[inner sep=0.5pt, fill=white] at (0, 0, 3) {\footnotesize$S$};
    \node[inner sep=0.5pt, fill=white] at (0, -1, 2) {\footnotesize$S$};
    \node[inner sep=0.5pt, fill=white] at (1, 0, 2) {\footnotesize$S$};
    \node[inner sep=0.5pt, fill=white] at (1, 2, 2) {\footnotesize$Z$};
    \node[inner sep=0.5pt, fill=white] at (-1, 2, 2) {\footnotesize$S\dagger$};
    \node[inner sep=0.5pt, fill=white] at (0, 2, 3) {\footnotesize$S^\dagger$};
    \node[inner sep=0.5pt, fill=white] at (0, 3, 2) {\footnotesize$S^\dagger$};
    \node[inner sep=0.5pt, fill=white] at (0, 2, 1) {\footnotesize$S$};
    \node[inner sep=0.5pt, fill=white] at (-1, 2, 0) {\footnotesize$Z$};
    \node[inner sep=0.5pt, fill=white] at (1, 2, 0) {\footnotesize$S$};
    \node[inner sep=0.5pt, fill=white] at (3, 2, 0) {\footnotesize$S$};
    \node[inner sep=0.5pt, fill=white] at (2, 2, -1) {\footnotesize$S$};
    \node[inner sep=0.5pt, fill=white] at (2, 3, 0) {\footnotesize$S$};
    \node[inner sep=0.5pt, fill=white] at (2, 3, 2) {\footnotesize$S\dagger$};
    \node[inner sep=0.5pt, fill=white] at (3, 2, 2) {\footnotesize$S$};
    \node[inner sep=0.5pt, fill=white] at (2, 2, 3) {\footnotesize$S$};
    \node[inner sep=0.5pt, fill=white] at (2, 1, 2) {\footnotesize$S$};
    \node[inner sep=0.5pt, fill=white] at (2, 1, 0) {\footnotesize$S\dagger$};
    \node[inner sep=0.5pt, fill=white] at (2, 0, 1) {\footnotesize$Z$};
    \node[inner sep=0.5pt, fill=white] at (2, 0, 3) {\footnotesize$Z$};
    \node[inner sep=0.5pt, fill=white] at (2, -1, 2) {\footnotesize$Z$};
    \node[inner sep=0.5pt, fill=white] at (3, 0, 2) {\footnotesize$Z$};
    \end{tikzpicture}
\cdot
\begin{tikzpicture}[scale = 1, baseline = 0.5cm]
    \coordinate (O) at (0,0,0); 
    \coordinate (A) at (2,0,0);
    \coordinate (B) at (2,2,0);
    \coordinate (C) at (0,2,0);
    \coordinate (D) at (0,0,2);
    \coordinate (E) at (2,0,2);
    \coordinate (F) at (2,2,2);
    \coordinate (G) at (0,2,2);
    %
    %
%
    \draw [dotted] (O) -- (A);
    \draw (A) -- (B) -- (C); 
    \draw [dotted] (O) -- (C);
    \draw (D) -- (E) -- (F) -- (G) -- cycle; 
    \draw [dotted] (O) -- (D); 
    \draw (A) -- (E);
    \draw (B) -- (F);
    \draw (C) -- (G);
    \draw (-1.5, 0, 2) -- (1.5,0,2);
    \draw (0, 0, 3.5) -- (0,0,2);
    \draw (0, -1.5, 2) -- (0,0,2);
    \draw (-1.5, 2, 2) -- (1.5,2,2);
    \draw (0, 2, 3.5) -- (0,2,2);
    \draw (0, 2, 2) -- (0,3.5,2);
    \draw (3.5, 2, 0) -- (2,2,0);
    \draw (2, 2, -1.5) -- (2,2,0);
    \draw (2, 2, 0) -- (2,3.5,0);
    \draw (3.5, 2, 2) -- (2,2,2);
    \draw (2, 2, 3.5) -- (2,2,2);
    \draw (2, 2, 2) -- (2,3.5,2);
    \node (A) [fill, circle, red, inner sep = 1pt] at (0,3,2) {};
    \node (B) [fill, circle, red, inner sep = 1pt] at (0,2,1) {};
    \node (C) [fill, circle, red, inner sep = 1pt] at (1,2,2) {};
    \draw[red] (A) -- (B) -- (C)--(A);
    \node (A) [fill, circle, red, inner sep = 1pt] at (0,1,2) {};
    \node (B) [fill, circle, red, inner sep = 1pt] at (0,0,1) {};
    \node (C) [fill, circle, red, inner sep = 1pt] at (1,0,2) {};
    \draw[red] (A) -- (B) -- (C)--(A);
    \node (A) [fill, circle, red, inner sep = 1pt] at (-1,2,2) {};
    \node (B) [fill, circle, red, inner sep = 1pt] at (0,2,3) {};
    \node (C) [fill, circle, red, inner sep = 1pt] at (0,1,2) {};
    \draw[red] (A) -- (B) -- (C)--(A);
    \node (A) [fill, circle, red, inner sep = 1pt] at (-1,0,2) {};
    \node (B) [fill, circle, red, inner sep = 1pt] at (0,0,3) {};
    \node (C) [fill, circle, red, inner sep = 1pt] at (0,-1,2) {};
    \draw[red] (A) -- (B) -- (C)--(A);
    \node (A) [fill, circle, red, inner sep = 1pt] at (1,2,2) {};
    \node (B) [fill, circle, red, inner sep = 1pt] at (2,2,3) {};
    \node (C) [fill, circle, red, inner sep = 1pt] at (2,1,2) {};
    \draw[red] (A) -- (B) -- (C)--(A);
    \node (A) [fill, circle, red, inner sep = 1pt] at (1,2,0) {};
    \node (B) [fill, circle, red, inner sep = 1pt] at (2,2,1) {};
    \node (C) [fill, circle, red, inner sep = 1pt] at (2,1,0) {};
    \draw[red] (A) -- (B) -- (C)--(A);
    \node (A) [fill, circle, red, inner sep = 1pt] at (2,3,2) {};
    \node (B) [fill, circle, red, inner sep = 1pt] at (2,2,1) {};
    \node (C) [fill, circle, red, inner sep = 1pt] at (3,2,2) {};
    \draw[red] (A) -- (B) -- (C)--(A);
    \node (A) [fill, circle, red, inner sep = 1pt] at (2,3,0) {};
    \node (B) [fill, circle, red, inner sep = 1pt] at (2,2,-1) {};
    \node (C) [fill, circle, red, inner sep = 1pt] at (3,2,0) {};
    \draw[red] (A) -- (B) -- (C)--(A);
    \end{tikzpicture}
\end{align*}
The first term are the Pauli $Z$ operators by commuting $X$ through $S$, $S^\dagger$ and CZ. The second term is the product of all $S$, $S^\dagger$ and $Z$ operators in the six $K_p$. The last terms is the product of all CZ gates in the six $K_p$. Using the relation Eq.~\eqref{eq:ver_relation}, we check the product is equivalent to a product of vertex operators. Indeed, since the plaquette operators product to a symmetry operator before gauging, they must also product to a symmetry operator after gauging.

\subsection{Lagrangians and Measurements}\label{app:ta_lm}

In this section, we assume the Hamiltonians of a pair of Lagrangians $\mathcal{L}_1$ and $\mathcal{L}_2$ are given, and we would like to construct the theory $\mathcal{L}_1 + \mathcal{L}_2$. For simplicity we assume there is only one $\mathbb{Z}_2$ $n$-form gauge field giving $\mathcal{L}_1[A]$ and $\mathcal{L}_2[A]$. The Hilbert space of $\mathcal{L}_1[A]$ and $\mathcal{L}_2[A]$ are both consists of one qubit for each $n$-simplex. The Hamiltonians are given by commuting projectors $H_1 = -\sum_i K_i$ and $H_2 = -\sum_j P_j$ following the procedures in Appendix~\ref{app:ta}. Then the Hamiltonian of $\mathcal{L}_1 + \mathcal{L}_2$ can be constructed as follows. We prepare a copy for each $\mathcal{L}_1 $ and $ \mathcal{L}_2$. This enlarges the Hilbert space to two qubits per $n$-simplex, giving the Hamiltonian and Lagrangian
\begin{align*}
    H = -\sum_i K_i^{(1)} - \sum_j P_j^{(2)},
    \hspace{35pt}
    \mathcal{L}[A_1, A_2] = \mathcal{L}_1[A_1] + \mathcal{L}_2[A_2],
\end{align*}
where $K_i^{(1)}$ and $P_j^{(2)}$ acts only on the first and second set of qubits respectively. Next, we preform the projection $Z_1Z_2$ on each $n$-cell, that is
\begin{align*}
    H' = -\sum_i K_i^{(1)} - \sum_j P_j^{(2)} - \Lambda \sum_{\Delta \in \text{$n$-cell}} Z_{\Delta,1}Z_{\Delta,2},
    \hspace{25pt}
    \mbox{with}\,\Lambda \rightarrow \infty,
\end{align*}
where $Z_{\Delta,i}$ is the Pauli $Z$ acting on $i$th qubit on the $n$-cell $\Delta$. This identifies the two gauge fields, giving the desired
\begin{align*}
    \mathcal{L}'[A] = \mathcal{L}_1[A_1 = A] + \mathcal{L}_2[A_2 = A].
\end{align*}

For example, the Pontryagin square classifies the $\mathbb{Z}_4$ group of anomalous $\mathbb{Z}_2$ $1$-form symmetry in 2D, in which case the anomaly is canceled by the 3D term $\frac{k}{4}P(A)$, $k = 0,1,2,3$~\cite{hsin2019comments}. The 3D bulk Hamiltonian $H_\s$ of the generator $\frac{1}{4}P(A)$ was derived in Appendix~\ref{app:ta_s_ww}. We first introduce two decoupled copies of $H_\s$, and measure $Z_1Z_2 = 1$ on all the edges, this gives the 3D fermionic toric code. The details are presented in Sec.~\ref{sec:examples_sf}. If we introduce another copy of $H_\s$, and do the projection $Z_1Z_2 = 1$ again, then we obtain the anti-semionic Walker-Wang model, whose plaquette terms are exactly given by the plaquette terms of the semionic Walker-Wang model in Appendix~\ref{app:ta_s_ww}, but replacing all $S$ by $S^\dagger$, and all $S^\dagger$ by $S$.

\end{document}